\documentclass[12pt]{article}
\usepackage{amsmath} 
\usepackage{amssymb}
\usepackage{hhline}
\usepackage[scale={.8,.8}]{geometry}
\usepackage{url}
\usepackage{lscape}
\usepackage{caption}
\usepackage[numbers,sort&compress]{natbib}
\usepackage{epsfig}
\usepackage{multirow} 
\usepackage{color}
\usepackage{verbatim}
\usepackage{nicefrac}
\usepackage{upgreek}
\usepackage{bbm}
\usepackage{setspace}
\usepackage{axodraw4j}
\usepackage{pstricks}
\usepackage{psfrag}
\usepackage{color}
\usepackage{anysize}

\newcommand{\slashed}[1]{\displaystyle{\not}{#1}}

\usepackage{hyperref}

\definecolor{green}{rgb}{0,0.5,0}

\begin{document}
\date{}
\title{
\begin{flushright}
{\scriptsize \tt TUM-HEP-979/15}  
\end{flushright}
{\bf Combining Experimental and Cosmological Constraints on Heavy Neutrinos
}}
\author{Marco Drewes, Bj\"orn Garbrecht\\ 
\scriptsize{Physik Department T70, Technische Universit\"at M\"unchen, James Franck Stra\ss e 1, D-85748 Garching, Germany}
}
\maketitle
\begin{abstract}
  \noindent 
We study experimental and cosmological constraints on the extension of the Standard Model by three right handed neutrinos with masses between those of the pion and W boson. 
We combine for the first time direct, indirect and cosmological constraints in this mass range.
This includes experimental constraints from neutrino oscillation data, neutrinoless double $\beta$ decay, electroweak precision data, lepton universality, searches for rare lepton decays, tests of CKM unitarity and past direct searches at colliders or fixed target experiments.
On the cosmological side, big bang nucleosynthesis has the most pronounced impact. Our results can be used to evaluate the discovery potential of searches for heavy neutrinos at LHCb, BELLE II, SHiP, ATLAS, CMS or a future lepton collider. 
\end{abstract}

\newpage
\tableofcontents
\section{Introduction}

\paragraph{Motivation} - 
The Standard Model (SM) of particle physics and the theory of general relativity together can explain almost all phenomena observed in nature, covering scales that range from the extension of the observable universe down to the inner structure of the proton \cite{Agashe:2014kda}.
To date, neutrino flavour oscillations are the only confirmed experimental result which requires within the framework of renormalisable  quantum field theory the existence of new particles.

All fermions except neutrinos are known to exist with both, left-handed (LH) and right-handed (RH) chirality.
One reason to suspect that massive RH neutrinos exist is that these offer an explanation for the neutrino masses 
through the seesaw mechanism \cite{Minkowski:1977sc,GellMann:1980vs,Mohapatra:1979ia,Yanagida:1980xy,Schechter:1980gr,Schechter:1981cv},
hence they can explain the observed neutrino flavour oscillations. 
In addition, they could explain a number of cosmological problems as well as a few unconfirmed experimental anomalies; see e.g. Ref.~\cite{Drewes:2013gca,Drewes:2015jna} for a review.
Most noticeable, RH neutrinos can generate the observed baryon asymmetry of the universe (BAU) \cite{Canetti:2012zc} via leptogenesis \cite{Fukugita:1986hr}, and are a natural Dark Matter candidate \cite{Dodelson:1993je,Shi:1998km}, see \cite{Adhikari:2016bei} for a recent review. 
For a specific mass pattern they can even explain all of these phenomena simultaneously \cite{Asaka:2005pn,Canetti:2012vf}, see \cite{Boyarsky:2009ix,Canetti:2012kh} for a detailed summary.

As for any newly postulated particle, the experimentally most interesting properties of RH neutrinos $N_I$ are their masses $M_I$ and the interaction with other particles.
The latter is characterised by a matrix of Yukawa coupling constants $F_{\alpha I}$ with ordinary leptons of flavour $\alpha$ or, alternatively, mixing angles $\Theta_{\alpha I}$.
Unfortunately, the magnitude of the $M_I$ and $F_{\alpha I}$ is entirely unknown. 
Neutrino oscillation experiments are only sensitive to a particular combination of the masses and couplings that determines the physical neutrino masses at low energies (essentially the ratio between the square of the coupling constant and the mass).\footnote{To be specific, the light neutrino masses are given by the eigenvalues $m_i^2$ of the flavour matrix $m_\nu m_\nu^\dagger$, where $m_\nu$ in terms of $F_{\alpha I}$ and $M_I$ is given in Eq.~(\ref{activemass}).}  
In this work we combine the negative results of direct searches for heavy neutrinos in the past with observables that are indirectly affected by their existence and cosmological bounds from primordial nucleosynthesis to impose constraints on the interaction strength of heavy neutrinos with the individual families in the SM. To the best of our knowledge, this is the first analysis of this kind for the model with three right handed neutrinos. 

\begin{figure}
\includegraphics[width=\textwidth]{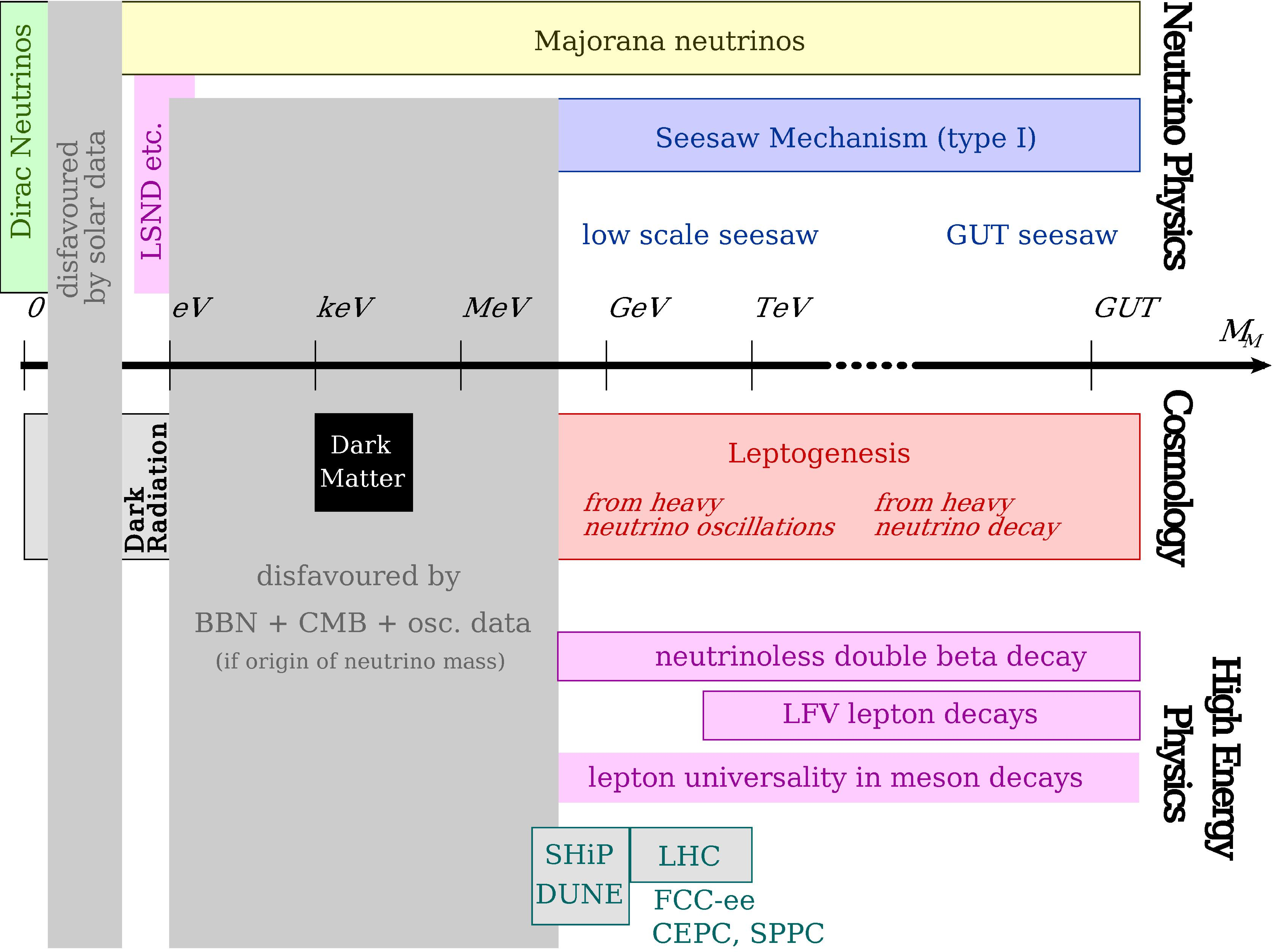}
\caption{
\label{SummaryPlot}
A schematic overview of the allowed range of values for the ``seesaw scale''
(i.e.\ the eigenvalues $M_I$ of $M_M$)
and their implications for neutrino physics, cosmology and high energy physics.
Constraints from neutrino physics are discussed in section \ref{sec:oscdata}.
The gray area below $M_I\simeq 100$ MeV is disfavoured by the considerations in section \ref{sec:cosmo}.
Some comments on leptogenesis can be found in section \ref{Sec:leptogenesis}, the possibility that heavy sterile neutrinos compose the Dark Matter is discussed in detail in \cite{Adhikari:2016bei}.
The indirect constraints indicated by the violet boxes are discussed in section \ref{Sec:IndirectSignatures}.
The range of existing of planned direct search experiments that could find heavy neutrinos is indicated in turquoise and discussed in section \ref{DirectSearchSubsection}.
}
\end{figure}

\paragraph{Allowed range of masses} - 
If Yukawa interactions of  the heavy neutrinos are to be described by perturbative quantum field theory, then the their masses $M_I$ should be at least 1-2 orders of magnitude below the Planck mass.\footnote{This can be estimated by inserting the observed neutrino mass differences into (\ref{activemass}).}
On the lower end they can have eV (or even sub-eV) \cite{deGouvea:2005er} masses, and any value in between is experimentally allowed if the number $n$ of heavy states $N_I$ exceeds 2 \cite{Hernandez:2014fha}.
Cosmological constraints can be used to push the lower bound up. If one requires that the observed BAU is generated by leptogenesis, then past studies suggest that $M_I$ has to be larger than a few MeV \cite{Canetti:2010aw,Canetti:2012kh}.
This bound can be raised to roughly 100 MeV if one combines the negative result of past direct searches  with the cosmological requirement that the $N_I$-lifetime is shorter than about $0.1$s \cite{Ruchayskiy:2011aa,Ruchayskiy:2012si,Hernandez:2014fha}. 
For longer lifetimes the energy released during the $N_I$ decay would affect the process of big bang nucleosynthesis (BBN) \cite{Ruchayskiy:2012si} and lead to an observable change in the abundance of light elements in the intergalactic medium.
If the $N_I$ decay after BBN, then this decay injects entropy into the primordial plasma. Moreover, if they are sufficiently light and abundant, they themselves also contribute to the effective number of relativistic degrees of freedom $N_{\rm eff}$, which is constrained by observations of the CMB and light element abundances.\footnote{The bound $M_I>100$ MeV can be circumvented within the minimal seesaw model if the $N_I$ are very long lived and decay at a negligible rate through the history of the universe. In this case they can in principle be viable Dark Matter candidates \cite{Adhikari:2016bei}. 
However, such longevity requires mixing angels $U_I^2$ that are so tiny that the contribution of these particles to the seesaw mechanism and active neutrino mass generation is negligible. We do not consider this case here in detail, but briefly discuss it in section \ref{sec:cosmo}.
If one does not require the $N_I$ to explain the observed light neutrino masses, then the constraints on their mass and mixing are generally much weaker \cite{deGouvea:2015euy}.
}
This still leaves  open a window $100\, {\rm MeV} < M_I < 10^{15}$~GeV of roughly sixteen orders of magnitude, see Fig.~\ref{SummaryPlot}.

Hence, theoretical prejudice is the only guideline when picking a mass scale.
In grand unified theories the RH neutrino mass scale is usually placed near (slightly below) the scale of grand unification. This choice is highly popular because it allows to explain neutrino masses with $F_{\alpha I}\sim 1$ and leads to standard thermal leptogenesis. The disadvantage of this scenario is that the $N_I$ will most likely never be observed directly, though indirect constraints can be derived, e.g. from neutrinoless double $\beta$-decay \cite{Blennow:2010th} and cosmological considerations \cite{Deppisch:2013jxa}.

A  direct discovery of the $N_I$ in the laboratory in the near future is only possible if the $M_I$ are at the TeV scale or below.
Apart from the usual ``lamp-post approach'' argument of experimental testability, an appeal of this choice is that more massive RH neutrinos would destabilise the Higgs mass~\cite{Vissani:1997ys}, leading to a severe fine tuning
problem in the absence of mechanisms that cancel radiative corrections such as supersymmetry.
Ideas that motivate a low scale seesaw include the possibility that the scale(s) $M_I$ and the electroweak scale have a common origin \cite{Iso:2009ss,Iso:2012jn,Khoze:2013oga,Khoze:2016zfi}, 
``no new scale''-considerations~\cite{Shaposhnikov:2007nj}, 
applying Ockham's razor to the number of new particles required to explain the known beyond the SM phenomena~\cite{Asaka:2005pn,Asaka:2005an},
minimal flavour violation  \cite{Cirigliano:2005ck,Gavela:2009cd}, 
 left-right-symmetric models \cite{Pati:1974yy,Mohapatra:1974hk,Senjanovic:1975rk,Wyler:1982dd} in which the complete breaking of GUT symmetry happens near the TeV scale and the possibility that $B-L$ is a spontaneously broken symmetry \cite{Chikashige:1980ui,Gelmini:1980re,GonzalezGarcia:1988rw} that is approximately conserved \cite{Branco:1988ex,Gluza:2002vs,Shaposhnikov:2006nn,Kersten:2007vk,Gavela:2009cd,Racker:2012vw}. Two of the most popular classes of scenarios are often referred to as ``inverse seesaw models'' \cite{Mohapatra:1986bd,Mohapatra:1986aw,GonzalezGarcia:1988rw} and ``linear seesaw models'' \cite{Akhmedov:1995vm,Akhmedov:1995ip,Barr:2003nn,Malinsky:2005bi} (see also \cite{Pilaftsis:1991ug,Abada:2007ux,Sierra:2012yy,Fong:2013gaa,Ballesteros:2016euj,Ballesteros:2016xej}).
Experimentally this mass range is very interesting because the heavy neutrinos can be found directly. The  strategy in direct searches strongly depends on $M_I$. For $M_I>5$ GeV, $N_I$-particles can only be produced in high energy collisions at the LHC \cite{Das:2012ze,Helo:2013esa,Dev:2013wba,Das:2014jxa,Ng:2015hba,Izaguirre:2015pga,Hung:2015hra,Peng:2015haa,Dev:2015kca,Gluza:2015goa,Gago:2015vma,Das:2015toa,Kang:2015uoc,Dib:2015oka,Degrande:2016aje}, either via vector boson fusion ($M> 500$ GeV), s-channel exchange of W-bosons ($500 {\rm GeV} > M > 80$ GeV) or in real gauge boson decays ($M<80$ GeV).
In the future, a higher sensitivity can be reached at a high energy lepton collider ILC \cite{Das:2012ze,Asaka:2015oia,Antusch:2015mia,Banerjee:2015gca,Hung:2015hra} FCC-ee \cite{Blondel:2014bra,Antusch:2015mia,Antusch:2015gjw,Abada:2015zea} or the CEPC \cite{Antusch:2015mia,Antusch:2015rma}.
Heavy neutrinos with $M<5$ GeV can be searched for in meson decays at b-factories \cite{Kobach:2014hea,Canetti:2014dka,Shuve:2016muy,Milanes:2016rzr,Asaka:2016rwd} or fixed target experiments \cite{Gorbunov:2007ak}, including NA62 \cite{Asaka:2012bb,Talk:Spadaro}, the SHiP experiment proposed at CERN \cite{Anelli:2015pba,Alekhin:2015byh,Graverini:2015dka} or a similar setup proposed at the DUNE beam at FNAL \cite{Akiri:2011dv,Adams:2013qkq}.
Cosmologically the low scale seesaw is interesting because it allows to explain the BAU via leptogenesis from flavour oscillations during the thermal production of the $N_I$ in the early universe \cite{Akhmedov:1998qx,Asaka:2005pn}. 
In this work we study existing experimental constraints on $N_I$ in the (type I) seesaw model with $M_I$ below the mass of the W boson. For a similar study with TeV masses we refer the reader to Refs.~\cite{Ibarra:2011xn,Bambhaniya:2014hla,Bambhaniya:2014kga,Antusch:2015mia,Deppisch:2015qwa}.

\paragraph{Goals of this work} - 
In principle the number $n$ of RH neutrinos is a free parameter, as they are SM gauge singlets and not subject to any anomaly cancellation requirement.
If the seesaw mechanism is the sole origin of neutrino masses, this implies $n\geq2$ because two non-zero mass differences between the light SM neutrinos have been observed, and the seesaw mechanism requires one RH neutrino per observed non-zero light neutrino mass.\footnote{ This may not be true in models with extended scalar sectors~\cite{Ibarra:2011gn,Ibarra:2014pfa,Tang:2014hna}.} 
Leptogenesis also requires $n\geq2$.
Previous studies of experimental constraints are mostly focused on the case $n=2$ (or even $n=1$), either for simplicity or because the third neutrino is assumed to be a Dark Matter candidate that interacts so feebly that its influence on neutrino masses and leptogenesis is negligible.
Known constraints for $n=2$ can e.g. be found in Ref.~\cite{Atre:2009rg,Boyarsky:2009ix,Drewes:2013gca,Drewes:2015jna} and references therein, recent works include \cite{Ruchayskiy:2011aa,Ruchayskiy:2012si,Asaka:2011pb,Asaka:2012hc,Asaka:2012bb,Canetti:2012vf,Canetti:2012kh,Asaka:2013jfa,Asaka:2014kia,Liventsev:2013zz,Aaij:2014aba,Drewes:2016jae}.
These studies have revealed that, for $n=2$, the combination of known direct and indirect constraints leads to much stronger bounds on the magnitude of the heavy neutrino couplings than one may expect from considering them individually, see e.g. Ref.~\cite{Drewes:2016jae} for a detailed discussion.

The motivation to extend the analysis from $n=2$ to $n=3$ is two-fold.
First, within the seesaw framework, we know that $n\geq3$ if the lightest neutrino turns out to be massive ($m_{\rm lightest}\neq0$).
Second, the range of $N_I$ parameters for which leptogenesis can be realised with $M_I\sim$ GeV turns out to be very different for $n=2$ and $n=3$.
For $n=2$ the two masses masses $M_1$ and $M_2$ have to be quasi-degenerate in order to explain the observed BAU, see \cite{Pilaftsis:1997jf,Pilaftsis:2003gt} and \cite{Asaka:2005pn,Canetti:2010aw,Asaka:2011wq,Canetti:2012kh,Canetti:2012vf,Shuve:2014zua,Gorbunov:2013dta}. This is also a necessary condition to measure the CP-violation in the sterile sector \cite{Cvetic:2014nla}, which is responsible for the BAU in most of the parameter space \cite{Canetti:2012kh}.
Moreover, leptogenesis then requires the Yukawa coupling constants $F_{\alpha I}$ to be too tiny to give measurable branching ratios in most existing experiments to find the $N_I$ \cite{Canetti:2012vf,Canetti:2012kh}.
With $n=3$ both of these restrictions can be overcome \cite{Drewes:2012ma,Canetti:2014dka,Garbrecht:2014bfa}; in this case leptogenesis does not require a mass degeneracy, and the parameter region where the BAU can be explained is within reach of existing experiments \cite{Canetti:2014dka}.
This provides strong motivation to re-analyse the known bounds to understand the parameter space for $n=3$.

The purpose of the present work is to combine all known direct and indirect experimental constraints as well as bounds from cosmology on RH neutrinos with masses between the pion and W boson mass. This allows to compare the allowed parameter space to the cosmologically motivated region where leptogenesis is possible and the sensitivity of future searches. 
Our analysis includes indirect constraints from neutrino oscillation data, searches for lepton flavour violating decays, neutrinoless double $\beta$ decay, CKM-unitarity tests, lepton universality tests, electroweak precision data and big bang nucleosynthesis. These are combined with constraints from direct searches at colliders and in beam dump experiments.

The present article is organised as follows.
In section~\ref{sec:seesaw} we recapitulate the seesaw model and introduce our notation.
In section~\ref{sec:scenarios} we define  
eight benchmark scenarios, characterised by the heavy and light neutrino mass spectrum, which we will use to explore the parameter space.
In the following sections we recapitulate different bounds on this parameter space, coming from indirect probes (section~\ref{Sec:IndirectSignatures}),
various direct search experiments (section~\ref{DirectSearchSubsection}) and cosmology (section~\ref{sec:cosmo}).
In section~\ref{numericssection} we numerically obtain combined constraints on the masses and mixings of the RH neutrinos in our benchmark scenarios.
We discuss the implications of these constraints for cosmology and future experiments in section~\ref{sec:discussion} and conclude in section~\ref{sec:conclusions}.

\section{The seesaw model}\label{sec:seesaw}
The well-known (type-I) seesaw model is defined by adding $n$ neutral fermions $\nu_R$ with RH chirality to the SM. These can couple to the SM neutrinos $\nu_L$ in the same way as the RH and LH components of the charged leptons are coupled together, hence the name RH neutrinos.
The Lagrangian reads  
\begin{eqnarray}
	\label{L}
	\mathcal{L} &=&\mathcal{L}_{SM}+ 
	i \overline{\nu_{R}}\slashed{\partial}\nu_{R}-
	\overline{\ell_{L}}F\nu_{R}\tilde{\Phi} -
	\tilde{\Phi}^{\dagger}\overline{\nu_{R}}F^{\dagger}\ell_L 
-{\rm \frac{1}{2}}(\overline{\nu_R^c}M_{M}\nu_{R} 
	+\overline{\nu_{R}}M_{M}^{\dagger}\nu^c_{R}). 
	\end{eqnarray}
Here flavour and isospin indices are suppressed.
$\mathcal{L}_{SM}$ is the SM Lagrangian, 
$\ell_{L}=(\nu_{L},e_{L})^{T}$ are the LH lepton doublets in the SM and $\Phi$ is the Higgs doublet with $\tilde{\Phi}=\epsilon\Phi^*$, where $\epsilon$ is the antisymmetric   
$SU(2)$-invariant tensor, 
$M_{M}$ a Majorana mass matrix for $\nu_{R}$ with $\nu_R^c=C\overline{\nu_R}^T$, and $F$ is a matrix of Yukawa couplings. 
The charge conjugation matrix is $C=i\gamma_2\gamma_0$.
We work in a flavour basis where $M_M={\rm diag}(M_1,M_2,M_3)$.

\subsection{Mass and interaction eigenstates}
If the eigenvalues of $M_M$ are far above the electroweak scale, the heavy neutrinos cannot be produced in experiments, and one can use an effective field theory \cite{Weinberg:1979sa,Broncano:2002rw}
\begin{eqnarray}
\label{Leff}
	\mathcal{L}_{\rm eff} &=&\mathcal{L}_{SM}+ \frac{1}{2}\bar{\ell_{L}}\tilde{\Phi}
F M_M^{-1}F^T
\tilde{\Phi}^{T}\ell_{L}^{c}
	\end{eqnarray}
obtained by integrating out the fields $\nu_R$ instead of (\ref{L}).
The Higgs mechanism generates the usual Majorana mass term $\overline{\nu_L}m_\nu^{\rm tree}  \nu_L^c$, where $m_\nu^{\rm tree}$ is given by 
\begin{equation}
m_\nu^{\rm tree}=-M_D M_M^{-1} M_D^T=-\theta M_M \theta^T = -v^2 F M_M^{-1}F^T\label{TreeLevelSeesaw}
\,.
\end{equation}
In the present work we are, however, interested in experimentally accessible heavy neutrino masses. 

\paragraph{Neutrino mass matrix} - 
There are different scenarios that predict eigenvalues of $M_M$ at or below the electroweak scale, including the inverse seesaw \cite{Mohapatra:1986bd} and linear seesaw \cite{Malinsky:2005bi}, the $\nu$MSM \cite{Shaposhnikov:2006nn,Araki:2011zg} or Coleman-Weinberg type models \cite{Khoze:2013oga}.
The full neutrino mass term after electroweak symmetry breaking can be written as
\begin{eqnarray}\label{neutrinomassfull}
\frac{1}{2}
(\overline{\nu_L} \  \overline{\nu_R^c})
\mathfrak{M}
\left(
\begin{tabular}{c}
$\nu_L^c$\\
$\nu_R$
\end{tabular}
\right) + {\rm h.c.}
\equiv
\frac{1}{2}
(\overline{\nu_L} \ \overline{\nu_R^c})
\left(
\begin{tabular}{c c}
$\delta m_\nu^{\rm 1loop}$ & $M_D$\\
$M_D^T$ & $M_M$
\end{tabular}
\right)
\left(
\begin{tabular}{c}
$\nu_L^c$\\
$\nu_R$
\end{tabular}
\right) + {\rm h.c.} 
\end{eqnarray}
where $M_D\equiv Fv$, $v$ is the temperature dependent Higgs field expectation value ($v=174$ GeV at temperature $T=0$), and h.c. denotes the hermitian conjugate term. $\delta m_\nu^{\rm 1loop}$ is the leading radiative correction to the light neutrino mass matrix $m_\nu$, which we consider later. 
For 
all $M_I\gg 1$  eV
one observes the hierarchy $M_D\ll M_M$ and there are two distinct sets of mass eigenstates, one set of ``active'' light neutrinos that are mostly SU(2) doublets and one set of ``sterile'' heavy neutrinos that are mostly gauge singlets.
Mixing between active and sterile mass states is suppressed by elements of the mixing matrix
\begin{equation}
\theta\equiv M_D M_M^{-1}.
\end{equation}  
It is straightforward to derive the well-known expressions for the neutrino mass and mixing matrices by expanding in the elements of $\theta$ to second order.\footnote{In principle it could be that terms beyond second order are important. However, in the $M_I$-range we are interested in, direct search constraints are sufficiently strong to ensure that the expansion in $\theta$ can be truncated at second order.
}
The matrix
\begin{equation}
\mathcal{U}=
\left[
\left(
\begin{tabular}{c c}
$\mathbbm{1}-\frac{1}{2}\theta\theta^\dagger$ & $\theta$\\
$-\theta^\dagger$ & $\mathbbm{1}-\frac{1}{2}\theta^\dagger\theta$
\end{tabular}
\right)  \ + \ \mathcal{O}[\theta^3]
\right]
\left(
\begin{tabular}{c c}
$U_\nu$ & $ $\\
$ $ & $U_N^*$
\end{tabular}
\right) 
.\end{equation}
diagonalises the full $6\times 6$ neutrino mass matrix $\mathfrak{M}$ at second order as
\begin{eqnarray}\label{diagonalisationofmathfrakM}
\mathcal{U}^\dagger \mathfrak{M}\mathcal{U}^*=
\left(
\begin{tabular}{c c}
$U_\nu^\dagger m_\nu U_\nu$ & $ $\\
$ $ & $U_N^T M_N U_N^*$
\end{tabular}
\right)
\,.
\end{eqnarray} 
Note that the loop correction $\delta m_\nu^{\rm 1loop}$ is of $\mathcal{O}[\theta^2]$.
The unitary matrices $U_\nu$ and $U_N$ diagonalise the mass matrices 
\begin{eqnarray}
m_\nu&=& m_\nu^{\rm tree} + \delta m_\nu^{\rm 1loop}\label{activemass} 
\end{eqnarray}
and 
\begin{eqnarray}
M_N&=&M_M + \frac{1}{2}\big(\theta^{\dagger} \theta M_M + M_M^T \theta^T \theta^{*}\big), \label{HeavyMassdef}
\end{eqnarray}
with\footnote{The additional complex conjugation for $U_N$ is to ensure that $U_N$ and $U_\nu$ relate mass and flavour eigenstates in the active and sterile sector in an analogous way within the present notation. 
}
\begin{eqnarray}
M_N=U_N^* M_N^{\rm diag} U_N^\dagger \ &,& \ m_\nu=U_\nu m_\nu^{\rm diag} U_\nu^T\,,\\
 M_N^{\rm diag} = {\rm diag}(M_1,M_2,M_3) \ &,& \ m_\nu^{\rm diag}={\rm diag}(m_1,m_2,m_3)
 \,.
\end{eqnarray}
Here $m_\nu^{\rm tree}$ is given by (\ref{TreeLevelSeesaw}).
In the flavour basis where the physical heavy neutrino mass matrix is diagonal, $\delta m_\nu^{\rm 1loop}$ is given by \cite{Pilaftsis:1991ug}
\begin{eqnarray}\label{RadiativeCorrectionExpression}
(\delta m_\nu^{\rm 1loop})_{\alpha\beta} = \sum_IF_{\alpha I}M_IF^T_{I \beta}
l(M_I)\,,
\end{eqnarray}
with
\begin{eqnarray}\label{ldef}
l(M_I)\equiv
\frac{1}{(4\pi)^2}
\left(
\frac{3\ln[(M_I/m_Z)^2]}{(M_I/m_Z)^2-1}
+\frac{\ln[(M_I/m_H)^2]}{(M_I/m_H)^2-1}
\right).
\end{eqnarray}
Inserting this into (\ref{activemass}) allows to rewrite it as
\begin{eqnarray}\label{activemassfull}
m_\nu=-\theta\tilde{M}\theta^T
\end{eqnarray}
with\footnote{In practice, we  assume in this context that this flavour basis coincides to a good approximation with that where $M_M$ is diagonal. If two of the $M_I$ are degenerate, then in principle there can be a significant misalignment between the bases, but in that case the relevant sub-matrix of $M_M$ is almost a unit matrix and one can assume that the transformation has no big effect. }
\begin{eqnarray}
\tilde{M}&=&U_N \tilde{M}^{\rm diag} U_N^T,\label{Mdiagforradcorr}\\
\tilde{M}^{\rm diag}_{IJ}&=&M_I\delta_{IJ}\left[
 1 - \frac{M_I^2}{v^2} l(M_I)
\right].
\end{eqnarray}
Note that under the present assumptions, the matrix $\mathcal{U}$ at second order in $\theta$ is not modified by radiative corrections except for the change in $m_\nu$. In the absence of radiative corrections, the matrix $\tilde{M}$ would coincide with $M_N$ at this order.

\paragraph{Neutrino mixing matrix} - 
All six mass eigenstates are Majorana fermions.
In the following we represent them by Majorana spinors instead of chiral spinors.
The elements $\upnu_i$ of flavour make up the vector 
\begin{equation}\label{LightMassEigenstates}
\upnu=V_\nu^{\dagger}\nu_L-U_\nu^{\dagger}\theta\nu_{R}^c
+ V_\nu^{T}\nu_L^c-U_\nu^{T}\theta\nu_{R} 
\end{equation}
and are mostly superpositions of the ``active'' SU(2) doublet states $\nu_L$, they have light masses 
$m_i\sim \theta^2 M_I$. 
The elements $N_I$ of 
\begin{equation}
N=V_N^\dagger\nu_R+\Theta^{T}\nu_{L}^{c}
+  V_N^T\nu_R^c+\Theta^{\dagger}\nu_{L}
\end{equation}
are mostly superpositions of the ``sterile'' singlet states $\nu_R$ and have masses of the order of $M_I$. 
The observed light mass eigenstates $\upnu_i$ are connected to the active flavour eigenstates by the matrix
\begin{equation}\label{VnuDef}
V_\nu\equiv (\mathbbm{1}-\frac{1}{2}\theta\theta^{\dagger})U_\nu. 
\end{equation}
That is, $V_\nu$ is the usual neutrino mixing matrix and $U_\nu$ its unitary part. $V_N$ and $U_N$ are their equivalents in the sterile sector,
\begin{equation}
V_N\equiv (\mathbbm{1}-\frac{1}{2}\theta^T\theta^*)U_N. 
\end{equation}
The deviation of $V_\nu$ from unitarity due to the $\theta\theta^\dagger$ comes from the small admixture of sterile states into the $\upnu_i$, which may lead to some observable effects \cite{Nelson:2010hz,Fan:2012ca,Antusch:2014woa,Escrihuela:2015wra}.\footnote{Unfortunately, unitarity violation due to heavy neutrinos does not appear to be able to resolve the long-standing issues of oscillation anomalies \cite{Kopp:2013vaa}.}
The mixing between active and sterile flavours is given by the matrix
\begin{equation}
\Theta\equiv\theta U_N^*.
\end{equation}
The only way how the $N_I$ interact with the SM at low energies is via this mixing with active neutrinos, which is characterised by the elements $|\Theta_{\alpha I}|\ll 1$. Hence, they are a type of \emph{heavy neutral lepton}.
Altogether, we can express the interaction eigenstates (singlets $\nu_R$ and doublet component $\nu_L$) as
\begin{eqnarray}\label{PMNSfull}
\left(
\begin{tabular}{c}
$\nu_L$\\
$\nu_R^c$
\end{tabular}
\right)
=
P_L\mathcal{U}
\left(
\begin{tabular}{c}
$\upnu$\\
$N$
\end{tabular}
\right)\,,
\end{eqnarray}
where $P_L$ is the left chiral projector.
Since the eigenvalues of $M_M$ and $M_N$ coincide in very good approximation, we will in the following not distinguish between these and use the notation $M_I$ for both.
The terms of order $\mathcal{O}[\theta^2]$ in the 
above expressions
are small, but in general cannot be neglected for two reasons.
First, they can lead to significant deviations of $U_N$ from unity if two of the $M_I$ are quasi-degenerate even if one starts in a basis where $M_M$ is diagonal.
Second, 
the factor $(\mathbbm{1}-\frac{1}{2}\theta\theta^{\dagger})$ in (\ref{VnuDef}) characterises the unitarity violation in the neutrino mixing matrix $V_\nu$ of the active flavours with each other.\footnote{The $3\times 3$ matrix $V_\nu$ is often called Pontecorvo-Maki-Nakagawa-Sakata (PMNS) matrix, though some authors use that term for the complete and unitary matrix $\mathcal{U}$.}

\subsection{Modification of the weak currents}\label{WeakCurrentModified}
The interactions of neutrinos in the SM are described by the Lagrangian term
\begin{equation}\label{WeakWW}
-\frac{g}{\sqrt{2}}\overline{\nu_L}\gamma^\mu e_L W^+_\mu
-\frac{g}{\sqrt{2}}\overline{e_L}\gamma^\mu \nu_L W^-_\mu  
- \frac{g}{2\cos\theta_W}\overline{\nu_L}\gamma^\mu\nu_L Z_\mu\,.
\end{equation}
Inserting $\nu_L=P_L(V_\nu \upnu + \Theta N)$ from (\ref{PMNSfull}) shows that the existence of sterile mass states modifies the interaction strength of the light neutrinos \cite{Shrock:1980ct,Shrock:1981wq}. Moreover, the heavy states $N_I$ do have a $\Theta$-suppressed weak interaction.
The $N_I$ also interact with Higgs particles and $\nu_L$ directly via the Yukawa coupling, but this term is not important for the mass range we have in mind, in which Higgs decays do not give dominant constraints.

Low energy experiments can be described in terms of the Fermi theory after integrating out the W and Z bosons. If the energy scale is much below $M_I$, one can also integrate out the $N_I$. 
The Fermi constant $G_F$ is measured via the lifetime of the muon, which is lighter than the $M_I$ under consideration in this work. Therefore, only the light states (\ref{LightMassEigenstates}) are allowed in the final state of the decay $\mu^-\rightarrow e^-\bar{\nu}_e\nu_\mu$. 
Hence, the weak interaction strength of the $\nu_L$ in this process is effectively suppressed by the factor $(1-\frac{1}{2}\theta^\dagger \theta)$ in (\ref{PMNSfull}).
This slightly suppresses the decay rate, leading to an ``incorrect'' result if one extracts $G_F$ from this measurement under the assumption that there is no physics beyond the SM.
Neglecting the $m_i$, one can relate the  Fermi constant measured in muon decays $G_\mu$ to the true Fermi constant $G_F$ as
\begin{eqnarray}\label{GmuDef}
G_\mu^2=G_F^2\big[1-\sum_I\theta_{\mu I}\theta^\dagger_{I\mu}-\sum_J\theta_{e J}\theta^\dagger_{J e}\big]
\simeq G_F^2 
(V_\nu V_\nu^\dagger)_{ee} (V_\nu V_\nu^\dagger)_{\mu\mu}\,.
\end{eqnarray}
The last step makes obvious that the deviation from the SM is due to the non-unitarity of $V_\nu$.

\subsection{Parameterisation}\label{parametrisationsubsec}
For $n=3$ the Lagrangian (\ref{L}) contains $18$ new physical parameters. 
Three of these are the masses $M_I$, the remaining 15 are hidden in the Yukawa matrices, for which we use the generalised Casas-Ibarra parametrisation for $\Theta$ that has been proposed in \cite{Lopez-Pavon:2015cga}
\begin{equation}\label{CasasIbarra2}
\Theta =i U_\nu\sqrt{m_\nu^{\rm diag}}\mathcal{R}\sqrt{\tilde{M}^{\rm diag}}^{-1}\,.
\end{equation}
This parametrisation can be derived from (\ref{activemassfull}) analogous to its tree level equivalent \cite{Casas:2001sr}, but takes into account the effect of $\delta m_\nu^{\rm 1loop}$. 
In terms of $F$, this reads
\begin{eqnarray}\label{CasasIbarra}
FU_N^*&=&\frac{i}{v}U_\nu\sqrt{m_\nu^{\rm diag}}\mathcal{R}\sqrt{\tilde{M}^{\rm diag}}^{-1}U_N^T M_MU_N^*\\
&\simeq&\frac{i}{v}U_\nu\sqrt{m_\nu^{\rm diag}}\mathcal{R}\sqrt{\tilde{M}^{\rm diag}}^{-1}M_N^{\rm diag}\nonumber.
\end{eqnarray}
In the second line we have again neglected the difference between the eigenvalues of $M_M$ and $M_N$.
$\mathcal{R}$ is a complex matrix with $\mathcal{R}^T\mathcal{R}=\mathbbm{1}$, which can be parametrised by complex angles $\omega_{ij}$ as 
\begin{equation}
\mathcal{R}=\mathcal{R}^{(23)}\mathcal{R}^{(13)}\mathcal{R}^{(12)}.
\end{equation}
The non-zero elements of the $\mathcal{R}^{(ij)}$ are
\begin{eqnarray}
\mathcal{R}^{(ij)}_{ii}&=&\mathcal{R}^{(ij)}_{jj}=\cos\omega_{ij} ,\nonumber\\
\mathcal{R}^{(ij)}_{ij}=\sin\omega_{ij} \ &,& \
\mathcal{R}^{(ij)}_{ji}=-\sin\omega_{ij} \ , \ 
\mathcal{R}^{(ij)}_{kk}\underset{k\not=i,j}{=}1.\nonumber
\end{eqnarray}
We work in the flavour basis where the charged lepton Yukawa couplings are diagonal, then the matrix $U_\nu$ can be parametrised as
\begin{equation}
U_\nu=V^{(23)}U_{\delta}V^{(13)}U_{-\delta}V^{(12)}{\rm diag}(e^{i\alpha_1 /2},e^{i\alpha_2 /2},1). \nonumber
\end{equation}
Here $U_{\pm\delta}={\rm diag}(e^{\mp i\delta/2},1,e^{\pm i\delta/2})$ and 
the non-zero entries of the matrices $V$ are given by
\begin{eqnarray}
V^{(ij)}_{ii}&=&V^{(ij)}_{jj}=\cos\uptheta_{ij} , \nonumber\\
V^{(ij)}_{ij}=\sin\uptheta_{ij} \ &,& \
V^{(ij)}_{ji}=-\sin\uptheta_{ij} \ , \ 
V^{(ij)}_ {kk}\underset{k\not=i,j}{=}1. \nonumber
\end{eqnarray}
Here $\uptheta_{ij}$ are the mixing angles
amongst the active neutrinos, and $\alpha_1$, $\alpha_2$ and $\delta$ are CP-violating phases.
This parametrisation allows to encode all constrains from neutrino oscillation experiments in $U_\nu$ and $m_\nu^{\rm diag}$. 
Another advantage is that (\ref{CasasIbarra}) nicely factorises into a part $\frac{1}{v}U_\nu\sqrt{m_\nu^{\rm diag}}$ that only contains parameters associated with the light active neutrinos and a  remainder that contains the parameters in the sterile sector.

\section{Benchmark scenarios}\label{sec:scenarios}
Only five of the $7n-3$ parameters contained in (\ref{L}) in addition to the SM have been determined experimentally (two mass differences $|m_i^2-m_j^2|$ and three mixing angles $\uptheta_{ij}$), see \cite{Gonzalez-Garcia:2014bfa} for the recent values used in our analysis. 
In the near future, the CP-violating Dirac phase $\delta$ is expected to be measured in neutrino oscillation experiments, and the absolute neutrino mass scale may be determined from cosmology. 
This leaves us with a large number of free parameters, making it very difficult to understand the parameter space analytically. This is in contrast to the case $n=2$, in which there are only four parameters in the sterile sector, and further simplifications arise if one assumes the mass degeneracy required for leptogenesis. 
Therefore most of our following analysis is numerical, and we define a number of representative benchmark scenarios.

\subsection{The relevant observables} 
The high dimensionality also raises the question of how to project the parameter space, i.e. how to plot our results. Since most of the parameters defined in the equations following~(\ref{CasasIbarra}) do not directly correspond to physical observables, it is not straightforward to identify physically meaningful exclusion regions in terms of these parameters. 
We therefore mainly express our results in terms of the physical masses $M_I$ and the mixings
\begin{equation}\label{UaIdef}
U_{\alpha I}^2\equiv |\Theta_{\alpha I}|^2
\end{equation}
or their combinations 
\begin{equation}
U_I^2\equiv U_{e I}^2 + U_{\mu I}^2 + U_{\tau I}^2 \ , \ U_\alpha^2\equiv \sum_I U_{\alpha I}^2.
\end{equation}
These are the quantities that enter branching ratios and rates observed in experiments.
In order to judge the potential of future searches for $M_I$, it is crucial to understand which range of values for $U_{\alpha I}^2$ is allowed for given $M_I$ after imposing the combination of all known constraints. This is the main goal of this paper.
We address this question with the underlying assumptions that there are $n=3$ heavy neutrinos which generate the light neutrino masses via the seesaw mechanism, and that all of them have masses between the pion and the W mass, which are discussed in some more detail in Sec.~\ref{dependenceonassumptions}.
In the following, we address several questions in the framework of the seesaw mechanism:
\begin{center}
\emph{Assuming a heavy neutrino $N_I$ with given mass $M_I$ exists, what are the largest values of $U_{e I}^2$, $U_{\mu I}^2$, $U_{\tau I}^2$ and $U_I^2$ that are consistent with all known experimental and observational data? And what is the smallest possible value of $U_I^2$?} 
\end{center}
Note that this approach does not provide a ``global fit'' to the properties of heavy and light neutrinos \cite{Antusch:2014woa,Fernandez-Martinez:2016lgt}. Instead of obtaining best fit values for the model parameters, we utilise the fact that no significant deviation from the SM predictions has been observed (apart from neutrino masses)\footnote{
The $1-2\sigma$ evidence for non-zero $U_{\alpha I}^2$ found in  
\cite{Akhmedov:2013hec,Antusch:2014woa,Fernandez-Martinez:2016lgt}
is not significant by our criteria defined in section  \ref{Sec:IndirectSignatures}.} to impose bounds on the mixing angles $U_{\alpha I}^2$ of heavy neutrinos. Our approach also differs from the global analysis of experimental bounds in \cite{deGouvea:2015euy}, which mostly ignored cosmological constraints and assumed that heavy neutrinos decay into unobservable particles.

\subsection{Choice of the Majorana mass scale}
To define our benchmark scenarios, we specify the total mass range $M_{\rm min}<M_I<M_{\rm max}$ within which we allow the $M_I$ to lie.  
That is, we assume that either there are no further states $N_I$ with masses outside this range, or they decouple to good approximation and do not affect the observables we study.
As pointed out in the introduction, there are essentially no experimental constraints on $M_{\rm min}$  and $M_{\rm max}$ unless one adds cosmological considerations discussed in section \ref{sec:cosmo}.
These roughly imply that $N_I$ should be heavier than 100 MeV.
For the sake of definiteness, in all that follows we assume that the $N_I$ are heavier than the pion,
\begin{equation}\label{MminDefEq}
M_{\rm min}=m_\pi.
\end{equation}

\paragraph{Scenario A: The $\nu$MSM scenario} - 
In this scenario one sterile neutrino $N_1$ is lighter than allowed by (\ref{MminDefEq}), but it can avoid the constraint described in section \ref{DRsection} due to its very feeble mixing angle. 
For most purposes (especially in experiments), $N_1$ can safely be ignored, and scenario A is effectively equivalent to the case $n=2$. Cosmologically $N_1$ can, however, be interesting because it is a DM candidate \cite{Adhikari:2016bei}.
The remaining two $N_2$ and $N_3$ are in the mass range between 100~MeV and 80~GeV considered here. This scenario has been 
studied in Ref.~\cite{Asaka:2005pn,Boyarsky:2009ix,Shaposhnikov:2008pf,Canetti:2012kh,Drewes:2016jae} and references therein.
We will not repeat or refine this analysis here.
Note that the phenomenology of $N_2$ and $N_3$ is the same as in the model with $n=2$ as far as neutrino masses, baryogenesis and direct searches are concerned because $N_1$ effectively decouples.
Two aspects of this specific scenario should be recalled in the present context.
First, baryogenesis requires a mass degeneracy between $M_2$ and $M_3$ at a level of $\sim 10^{-3}$ \cite{Canetti:2012vf,Shuve:2014zua}. 
The small mass splitting introduces a long oscillation time scale, which can give rise to resonant phenomena that require special treatment in the early universe \cite{Garny:2011hg,Garbrecht:2011aw,Garbrecht:2014aga,Iso:2013lba,Dev:2014wsa,Dev:2014laa,Biondini:2015gyw}  and possibly in the laboratory \cite{Boyanovsky:2014una,Boyanovsky:2014lqa,Cvetic:2015ura}.  For general $n\geq 3$ scenarios, successful Leptogenesis does not necessarily require a mass degeneracy;
in this case standard methods pertinent to short oscillation time-scales can be used to describe the propagation of the $N_I$ even in the early universe \cite{Anisimov:2010aq,Anisimov:2010dk,Drewes:2012qw,Drewes:2012ma}.
Second, due to the mass degeneracy, most experiments cannot measure the $U_{\alpha I}^2$ individually, but are only sensitive to $U_{\alpha 2}^2+U_{\alpha 3}^2\simeq U_\alpha^2$ instead. There exists a strict lower limit on the $U_{\alpha}^2$ with individual active neutrino flavours, which is a function of their mass \cite{Asaka:2011pb,Ruchayskiy:2011aa,Canetti:2012kh,Gorbunov:2013dta}.

\paragraph{Scenario B: The meson decay scenario
} - 
This scenario is defined by the choice
\begin{equation}\label{ScenarioB}
M_{\rm max}=m_B\,.
\end{equation}
For this choice the masses of all $N_I$ lie in the region where they can efficiently be produced in meson decays.
For $M_I<m_D$ the SHiP experiment will have its highest sensitivity to $U_{\alpha I}^2$, estimated to be $\sim 10^{-9}$ \cite{Alekhin:2015byh}. This parameter region is interesting for leptogenesis because the BAU can be explained without significant tuning in the ratio $U_{e I}^2/U_{\mu I}^2$ \cite{Canetti:2014dka}.
The sensitivity drops considerably for $M_I$ larger than the D meson mass $m_D=1,869.61 \pm 0.1$ MeV \cite{Agashe:2014kda} because the $N_I$ are mainly produced in decays of D mesons.
Other experiments (such as LHCb and BELLE) also have their main sensitivity in this mass range \cite{Milanes:2016rzr}
and may improve bounds from past searches \cite{Bernardi:1987ek,Vaitaitis:1999wq,Badier:1985wg,Vilain:1994vg,Abreu:1996pa,Artamonov:2014urb}, see also \cite{Atre:2009rg,Ruchayskiy:2011aa,Kobach:2014hea}.

For $m_D<M_I<m_B$ the $N_I$ can still be produced in B meson decays (with the charged B meson mass $m_B=5279.26\pm 0.17$ MeV).
This makes a detection of the $N_I$ in B meson decays at LHCb or BELLE II possible. While the current bounds from these experiments \cite{Aaij:2014aba,Liventsev:2013zz} in most of the mass range are still weaker than the old bounds from DELPHI \cite{Abreu:1996pa}, they will soon become the front line in the exploration of the cosmologically motivated parameter space \cite{Canetti:2014dka}.

\paragraph{Scenario C: The general sub-electroweak seesaw} -
Finally, we explore a general scenario in which the $N_I$ are lighter than the W boson,
\begin{equation}\label{ScenarioD}
M_{\rm max}= m_W.
\end{equation}
This mass range is currently accessible to ATLAS and CMS. 
Both collaborations have published constraints for $M_I>m_W$  \cite{ATLAS:2012ak,Khachatryan:2014dka,Aad:2015xaa,Khachatryan:2015gha,CMS:2015sea}.
Besides the minimal seesaw (\ref{L}), similar searches can be carried out for its left-right symmetric extension, see e.g.~\cite{Keung:1983uu,Nemevsek:2011hz,Mitra:2016kov,Chen:2013fna}.
For $M_I<m_W$ some bounds have been released \cite{Khachatryan:2015gha,CMS:2015sea}, and the experimental perspectives are thought to improve in the future because the $N_I$ can be produced in gauge boson decays. 
It has been claimed that CMS can probe mixing angles as small as $U_I^2\sim10^{-9}$  \cite{NicoElena}, though the decay products in this case are comparably soft and the results appear to have a rather strong dependence on the cuts that are imposed during the analysis \cite{Helo:2013esa}. A sensitivity to mixings $U_I^2\sim10^{-7}$, however, seems realistic \cite{Izaguirre:2015pga}, see also \cite{Gago:2015vma,Dib:2015oka,Dib:2016wge}. The reach of the LHC could further be extended with the MATHUSLA surface detector \cite{Chou:2016lxi}.
Even further improvement could be made at a future lepton collider \cite{Abada:2014cca,Blondel:2014bra,Graverini:2015dka,Antusch:2015mia,Asaka:2015oia,Abada:2015zea,Antusch:2015rma,Antusch:2016vyf}.

\subsection{The range of active-sterile mixing}\label{MixingScenarios}

\subsubsection{The ``naive'' seesaw}  
If all entries of the neutrino mass matrix are of the same order, then the Yukawa couplings $F_{\alpha I}$ are of the order of the "naive seesaw expectation" 
\begin{equation}
F_0^2\equiv \sqrt{m_{\rm atm}^2 + m_{\rm lightest}^2}M_M/v^2, \label{F0}
\end{equation}
i.e. the would-be value of the coupling constant in a world without flavour, $n=1$ and a neutrino mass of the order $m_{\rm atm}$. 
The naive seesaw usually predicts values that are close to the minimal possible $U_{\alpha I}^2$ for given $M_I$, which for $n=2$ have e.g. been studied in \cite{Asaka:2011pb,Ruchayskiy:2011aa}, and one can neglect the radiative correction $\delta m_\nu^{\rm 1loop}$ in the mass range we consider.
In the parametrisation (\ref{CasasIbarra2}) it can be realised for $|{\rm Im}\omega_{ij}|$ not much
larger than one.

\subsubsection{Approximate lepton number conservation}\label{BminusL}
It is possible that individual $F_{\alpha I}$ are much larger than $F_0$ and the smallness of the $m_i$ is not (only) due to the suppression $\sim 1/M_M$ in (\ref{activemass}), but also the result of a cancellation amongst individual contributions to the neutrino mass matrix. Such a cancellation could either be ``accidental'' or due a symmetry, such as an approximately conserved lepton number \cite{Wyler:1982dd,Mohapatra:1986bd,Branco:1988ex,GonzalezGarcia:1988rw,Barr:2003nn,Malinsky:2005bi,Shaposhnikov:2006nn}.

\paragraph{Accidental cancellations} - Such a case is usually considered to be ``fine tuned''. Moreover, if it is imposed at a given level in the perturbative expansion in $F$ (or $\theta$), there is in principle no reason why it should hold at higher order. In particular, if it is imposed onto the tree level estimate $m_\nu^{\rm tree}$, it usually does not occur in the radiative correction $\delta m_\nu^{\rm 1loop}$ given by (\ref{RadiativeCorrectionExpression}), which has a different flavour structure. This leads to large corrections to the active neutrino masses in the parameter region where the $U_{\alpha I}^2$ are large. Of course, one can always impose the accidental cancellation on the radiatively corrected formula (\ref{activemass}).

\paragraph{Symmetry protected scenario} - A cancellation can be explained and imposed at all orders if it is caused by a symmetry. This is, for instance, the case for the inverse seesaw \cite{Mohapatra:1986bd}, linear seesaw \cite{Barr:2003nn,Malinsky:2005bi} or possibly the $\nu$MSM \cite{Shaposhnikov:2006nn}.
This situation is often referred to as \emph{``approximate lepton number conservation''}.
More precisely, it is usually not the lepton number of the SM that is conserved, but the combination $B-L$ of baryon number $B$ and a generalised lepton number $L$ under which the $\nu_R$ are charged. 
Individually, $B$ and $L$ are both violated by a quantum anomaly in the SM \cite{Adler:1969gk,Bell:1969ts}, which leads to efficient conversion in the early universe \cite{Kuzmin:1985mm} at temperatures above $\sim 130$ GeV \cite{D'Onofrio:2014kta}, but respects the conservation of $B-L$.
The Majorana mass $M_M$ in general also violates $B-L$ conservation.
However, an approximate conservation is realised if two of the $\nu_R$ (let us call these $N_2$ and $N_3$) form a Dirac-spinor $\Psi_N\equiv (iN_2+N_3)/\sqrt{2}$
and the third one ($N_1$) ``decouples'' ($F_{\alpha 1}=0$), see e.g. appendix B of \cite{Canetti:2012kh}.
Then unitary transformations $N_I=U_{IJ}N_J'$ can be used to bring $F$ and $M_M$ into the form \cite{Abada:2007ux,Fernandez-Martinez:2015hxa}
\begin{eqnarray}
F=\left(
\begin{tabular}{c c c}
$F_e$ & $\epsilon_e$ & $\epsilon_e'$\\
$F_\mu$ & $\epsilon_\mu$ & $\epsilon_\mu'$\\
$F_\tau$ & $\epsilon_\tau$ & $\epsilon_\tau'$
\end{tabular}
\right)
\
,
\
M_M=\left(
\begin{tabular}{c c c}
$M'$ & $\mu_4$ & $\mu_3$\\
$\mu_4$ & $\mu_2$ & $M$\\
$\mu_3$ & $M$ & $\mu_1$
\end{tabular}
\right).\label{BmLcons}
\end{eqnarray}
Here $\epsilon_\alpha,\epsilon_\alpha'\ll F_\alpha$ and $\mu_i\ll M, M'$ are lepton number violation (LNV) parameters, which must vanish if $B-L$ is exactly conserved.
In this limit one finds 
\begin{eqnarray}
U_{\alpha 1}= 0 \ &,& \ U_{\alpha 2}^2=U_{\alpha 3}^2 \ {\rm } \ {\rm for} \ \alpha = e,\mu,\tau\,,\label{equalcouplings}
\end{eqnarray}
In the parametrisation (\ref{CasasIbarra2}) this can be realised for $M_2=M_3$, $\omega_{12}=\omega_{13}=0$, $\omega_{23}\rightarrow \pm i\infty$. 
An exact $B-L$ conservation would of course also require $m_i=0$.
The approximately $B-L$ conserving scenario is of crucial importance in the present context because it usually allows for the largest possible $U_{\alpha I}^2$ for a given $M_I$ that are experimentally allowed. 
The reason is that some of the strongest constraints on $U_{\alpha I}^2$ come from lepton number violating (LNV) observables, including light neutrino masses, neutrinoless double $\beta$ decay or same sign dilepton signatures in collider experiments. Such signals are suppressed by the small parameters in (\ref{BmLcons}) in the $B-L$-conserving limit.  
The relatively large $U_{\alpha}^2$ imply that the experimental perspectives for a discovery are particularly promising \cite{Kersten:2007vk,Ibarra:2011xn,Das:2012ze,Canetti:2012vf,Canetti:2012kh,Das:2014jxa,Dev:2013wba,Bambhaniya:2014hla,Bambhaniya:2014kga,Antusch:2015mia,Deppisch:2015qwa,Dev:2015kca}.

\section{Indirect signatures of heavy neutrinos}\label{Sec:IndirectSignatures}
Heavy neutrinos in the mass range $m_\uppi<M_I<m_W$ can be produced in experiments in different ways. 
In addition to such ``direct searches'', in which the $N_I$ appear as real particles, the properties of heavy neutrinos can also be constrained in ``indirect searches'' by observing processes in which they only appear as virtual particles.
In fact, the existence of $N_I$ would affect essentially all low energy observations of weak processes via the difference between $G_\mu$ and $G_F$ discussed in section \ref{WeakCurrentModified}. In the following we outline in some more detail how individual observables get affected beyond that before turning to direct searches in the next section.

\subsection{Neutrino oscillation data}\label{sec:oscdata}
Neutrino oscillation data constrains the differences between the neutrino masses $m_i$ as well as the mixing angles and phases in $U_\nu$.
It also imposes constraints on the mixings $U_{\alpha I}^2$. That is, once the $m_i$ and $U_\nu$ are fixed, not all values for the $U_{\alpha I}^2$ can be realised for two reasons.
First, at least some of the $N_I$ have to have a significant mixing with active neutrinos to yield eigenvalues for $m_\nu m_\nu^\dagger$ that are large enough to explain the observed $|m_i^2-m_j^2|$.
Second, the $U_{\alpha I}^2$ cannot be too large. Otherwise higher order corrections might lead to physical neutrino masses that are too large even if the tree level $m_i$ are tuned to be small due to some cancellations.

\subsubsection{Minimal mixing}\label{sec:minimalmixing}
It is clear from (\ref{activemass}) that some mixing between active and sterile neutrinos must exist in order to explain non-zero neutrino masses via the seesaw mechanism: If all elements of $\theta$ vanish, then all active neutrinos remain massless.
An obvious question to pose in any of the scenarios defined above would be
\begin{center}
\emph{
What is the minimal mixing $U_{e I}^2$ with $\nu_e$ that $N_I$ with $M_{\rm min}^{\rm exp}<M_I<M_{\rm max}^{\rm exp}$ can have?} 
\end{center}
Unfortunately the answer is $U_{e I}^2=0$ (and same for mixing with $\nu_{\mu}$ or $\nu_\tau$). 
This statement in some sense is trivial: Since only two neutrino mass differences have been observed, we can explain all observational data with $n=2$ (in which case the lightest neutrino is massless). This is e.g. effectively realised for $n=3$ with $U_{e3}^2=U_{\mu3}^2=U_{\tau3}^2=0$, hence this scenario must be allowed. 
For illustrative purposes we consider the special case of vanishing phases $\delta=\alpha_1=\alpha_2=0$ and $\omega_{23}=\omega_{13}=0$ and normal hierarchy. 
Then all mixings $\Theta_{\alpha 3}\propto m_1$, i.e. $N_3$ entirely decouples if the lightest neutrino is massless (irrespectively of the choice of the $M_I$-spectrum). 
For $\omega_{12}=-\uptheta_{12}$ one can in addition suppress $\Theta_{e 2}$ up to corrections of order $m_1/m_3$, i.e. the ratio between the lightest and heaviest neutrino.
Even if one requires that all light neutrinos are massive, we can still choose the remaining parameters such that individual $U_{\alpha I}^2$ vanish. 
We have checked analytically that there are choices for the $\omega_{ij}$ for which a given $\Theta_{e I}=0$ or $\Theta_{\mu I}=0$.
There exist analytic solutions for $\omega_{ij}$ that set the coupling of two of the $N_I$ to electrons exactly to zero, but the expressions are rather lengthy and not illuminating. Hence, we cannot impose a lower bound on the mixing of any individual $N_I$ with a specific active flavour.
These findings are consistent with what has been described in \cite{Gorbunov:2013dta}.

If we demand that all masses $M_I$ lie within the experimentally accessible region, then we can pose the question
\begin{center}
\emph{Assuming $N_1$ is the sterile neutrino with the largest mixing with $\nu_\alpha$ of a given flavour $\alpha$ (i.e. $U_{1\alpha}^2>U_{2\alpha}^2,U_{3\alpha}^2$), what is the smallest value that $U_{1\alpha}^2$ can take as a function of $M_1$?}\footnote{Here the choice of labelling $N_1$, $N_2$ and $N_3$ is arbitrary.
The parametrisation in Eq.~(\ref{CasasIbarra}) and following implies certain relations between the specific sterile flavour $N_1$ and observed parameters. 
However, the labelling of the $N_I$ is arbitrary, as there exist choices of the unknown parameters for which they exactly swap properties. Since we scan over all unknown parameters we can simply \emph{define} $N_1$ as the RH neutrino that has the largest mixing with $\nu_\alpha$.  
}  
\end{center}
The answer to this question is some non-zero $U_{\alpha {\rm min}}^2$, which is a function of $M_1$.
Its meaning is the following: Assume that we have an experiment that covers a mass range between $M_{\rm min}^{\rm exp}$ and $M_{\rm max}^{\rm exp}$, and that all three $N_I$ have masses in this interval. Then we are guaranteed to find at least one sterile neutrino if we push the sensitivity in $U_{\alpha I}^2$ below $U_{\alpha {\rm min}}^2$ across the whole mass range. Otherwise we have excluded the model.
Unfortunately also this approach is of limited use unless 
all of the $M_I$ are within reach of the experiment, as otherwise the $N_I$ with sizable mixing could simply be too heavy to be seen. Hence, any conclusion would unavoidably strongly depend on the choice of $M_{\rm max}$.

We can, however, put a lower bound on the sum $U_I^2$. This bound strongly depends on the absolute neutrino mass scale.
If all three $m_i$ are non-zero, then a lower bound on $U_I^2$ is enforced by neutrino oscillation data: To generate three non-zero neutrino masses, all three $N_I$ have to mix with the active neutrinos.\footnote{More precisely: There is no direction in flavour space (superposition of the $N_I$ mass states) that decouples.}
This means that each $N_I$ must mix with at least one active flavour, meaning that $U_I^2$ is bound from below for given $M_I$.
If the lightest neutrino is massless, then there is no such lower bound. 
Two non-zero $m_i$ can be generated by the seesaw mechanism with two sterile flavours; the third sterile flavour is allowed to have an arbitrarily small mixing (even zero). Hence, for given $M_I$ there is no lower bound on $U_I^2$.

\subsubsection{Maximal mixing}\label{oscdatamaximalmixing}
The the tree-level seesaw relation (\ref{TreeLevelSeesaw})
implies that large mixings $U_{\alpha I}^2$ generally lead to large neutrino masses $m_i$.
Therefore, the smallness of the observed neutrino masses allows to impose an upper bound on the $U_{\alpha I}^2$.
Small neutrino masses $m_i$ can be made consistent with relatively large $U_{\alpha I}^2$ if there are cancellations in $m_\nu^{\rm tree}$. 
Such cancellations can either be ``accidental'' or due to a symmetry in the Lagrangian (such as approximate lepton number conservation discussed in Sec.~\ref{BminusL}).

The active neutrino masses $m_i$ are generated at tree level, but also receive radiative corrections from loops with virtual $N_I$. At leading order, the radiative corrections are given by the term $\delta m_\nu^{\rm 1loop}$ in the seesaw relation (\ref{activemass}).
For sufficiently large $U_{\alpha I}^2$, radiative corrections to $m_\nu^{\rm tree}$ 
can be sizable \cite{AristizabalSierra:2011mn,Fernandez-Martinez:2015hxa}. 
If the cancellations in $m_\nu^{\rm tree}$ are accidental, then they (most likely) do not occur in the loop correction. Then the $m_i$ obtained from $m_\nu^{\rm tree}+\delta m_\nu^{\rm 1loop}$ in (\ref{activemass}) are generally too large even if $m_\nu^{\rm tree}$ suggests that a given $F$ and $M_M$ were consistent with neutrino oscillation data.
One may, of course, find other parameters that lead to accidental cancellations in the one-loop relation (\ref{activemass}).
Apart from the fact that this is generally thought to be ``fine tuned'', the cancellations will (most likely) not occur at two-loop order, and so on.
If the smallness of $m_i$ is protected by a symmetry, then the cancellations also happen at the quantum level at all loop orders, and individual $F_{\alpha I}\gg F_0$ can be made consistent with the observed  $\Delta m^2_{\rm sol}$ and $\Delta m^2_{\rm atm}$ without fine tuning.

One way to impose an upper bound on $U_{\alpha I}^2$ from neutrino oscillation data is to demand that  $\Delta m^2_{\rm sol}$ and $\Delta m^2_{\rm atm}$ are given by $m_\nu^{\rm tree}$, and that corrections due to $\delta m_\nu^{\rm 1loop}$ remain small. 
In this case, large $U_{\alpha I}^2$ can only be consistent with the observed $\Delta m^2_{\rm sol}$ and $\Delta m^2_{\rm atm}$ if cancellations occur \emph{independently} in both, the tree level term (\ref{TreeLevelSeesaw}) and the correction (\ref{RadiativeCorrectionExpression}). This is naturally the case if there is a protecting symmetry, but is very unlikely to happen accidentally. 
Demanding that radiative corrections are small therefore effectively suppresses accidental cancellations in the neutrino mass matrix.

In the present work we follow a more ``agnostic'' approach and use the parametrisation (\ref{CasasIbarra2}), which automatically guarantees that all randomly generated parameter values in our scans are consistent with neutrino oscillation data at one-loop level. 
As a consistency check, we confirm that the relation 
\begin{eqnarray}\label{SeesawConsistency}
\sum_i m_i (U_\nu)_{\alpha i}^2 + \sum_I M_I \Theta_{\alpha I}^2 = (\delta m_\nu^{\rm 1loop})_{\alpha \alpha}
\end{eqnarray}
holds, which directly follows from (\ref{diagonalisationofmathfrakM}).
By doing so, we do not penalise ``fine-tuned'' parameter choices in which cancellations in $m_\nu^{\rm tree}+\delta m_\nu^{\rm 1loop}$ are accidental. 
While the latter may seem theoretically unmotivated, we choose to allow such situations because we prefer to remain as agnostic as possible regarding possible embeddings of the low scale seesaw into a bigger theoretical framework.

\subsection{Lepton flavour violation}\label{sec:nutoegamma}
The violation of lepton flavour in the neutrino sector also mediates lepton flavour violation (LFV) amongst the charged leptons. Contributions can come from both, the exchange of light and heavy neutrinos.
Observational bounds on LFV in rare processes allow to constrain the parameter space for the $N_I$.
This is particularly useful in scenarios discussed in Sec.~\ref{BminusL} there total lepton number is approximately conserved. Then the rates for some of the most promising LNV signals (including neutrinoless double $\beta$ decay or same sign dilepton signals at colliders) are suppressed by the small parameters in (\ref{BmLcons}), while lepton flavour violation may still be observed.

\subsubsection{Rare lepton decays} 
The rate for a radiative rare charged lepton decay $\ell_\alpha\rightarrow \ell_\beta + \gamma$ is given by~\cite{Cheng:1980tp,Bilenky:1977du}
\begin{eqnarray}\label{rareleptondecayformula}
\Gamma_{\ell_\alpha\rightarrow \ell_\beta + \gamma}=\frac{\alpha_{EM} m_\alpha^5 G_\mu^2}{2048 \pi^4}|R|^2
\end{eqnarray}
with
\begin{eqnarray}\label{Rfunction}
R&=&\sum_{i} (V_\nu^*)_{\alpha i} (V_\nu)_{\beta i}
G\left(\frac{m_i^2}{M_W^2}\right)
+\sum_I\Theta^*_{\alpha I} \Theta_{\beta I}G\left(\frac{M_I^2}{M_W^2}\right)
\end{eqnarray}
and the loop function
\begin{equation}\label{Gloop}
G(x)=\frac{10-43x+78 x^2-49 x^3 + 4 x^4 + 18 x^3 \log(x)}{3 (x - 1)^4}.
\end{equation}
The function $G(x)$ is often approximated by $G(0)=10/3$, which is justified if only the (approximately massless) SM neutrinos appear in the loop. This is not justified in the low scale seesaw model (\ref{L}) because the $N_I$  can also contribute.
The experimentally most constrained process is the decay $\mu\rightarrow e\gamma$, with a branching ratio $B(\mu\rightarrow e \gamma)<5.7\times10^{-13}$ \cite{Adam:2013mnn}.
Using (\ref{rareleptondecayformula}), the  predicted branching ratio in the seesaw model (\ref{L}) is given by 
\begin{eqnarray}\label{BranchingRatio}
B(\mu\rightarrow e \gamma) = \frac{\Gamma(\mu\rightarrow e \gamma)}{\Gamma(\mu\rightarrow e \nu_\mu \bar{\nu}_e)} = \frac{3\alpha_{\rm em}}{32\pi}|R|^2\,.
\end{eqnarray}
In addition to the $\mu\rightarrow e \gamma$ decay, we also use constraints from searches for $\tau\rightarrow e \gamma$ and $\tau\rightarrow \mu \gamma$ decays to constrain the $U_{\alpha I}^2$.
More specifically, we impose the conditions on the branching ratios\\

\begin{center}
\begin{tabular}{c c}
$B(\mu\rightarrow e \gamma) < 5.7\times10^{-13}$ &  \cite{Adam:2013mnn}\,,\\
$B(\tau\rightarrow e \gamma) < 1.5\times10^{-8}$ &  \cite{Blankenburg:2012ex}\,,\\
$B(\mu\rightarrow e \gamma) < 1.8\times10^{-8}$  & \cite{Blankenburg:2012ex}\,.\\
\end{tabular}
\end{center}
We do not include bounds from the process $\mu\rightarrow eee$ \cite{Ilakovac:1994kj}, which at present appear to be subdominant \cite{Agashe:2014kda}.
The non-observation of $\mu\rightarrow e\gamma$ decays imposes rather strong constraints on $U_{\alpha I}^2$ in models with $M_I\sim$ TeV \cite{Ibarra:2011xn}, see also \cite{Abada:2014kba}. For GeV masses, however, the bounds are known to be much less important already for $n=2$ \cite{Gorbunov:2014ypa}, as there exist stronger bounds from other sources. Since $n=3$ offers more freedom, it can be expected that the experimental limits are less constraining.

\subsubsection{$\mu\rightarrow e$ conversion in nuclei}
The rate of $\mu\rightarrow e$ conversion in nuclei is in general very small in the seesaw model (\ref{L}) because it scales as $\sim F^4/M_I^4$.
If two $M_I$ are degenerate, as it is the case in approximately lepton number conserving scenarios from section~\ref{BminusL},  
it may nevertheless be observed in the future.
In particular, in low scale seesaw models under consideration here, it may be a more sensitive probe of active-sterile mixing than the rate of $\mu\rightarrow e \gamma$. Therefore future $\mu\rightarrow e$ conversion searches could be one of the most powerful tools to look for $N_I$ indirectly \cite{Alonso:2012ji}. 
However, the present experimental bounds are generally weaker than direct search constraints \cite{Canetti:2013qna}.
Within the mass range we consider here, there is only a small window near $M_I\sim M_W$ in which $\mu\rightarrow e$ conversion bounds are competitive with DELPHI bounds \cite{Alonso:2012ji}. We therefore do not include  these in our present analysis.

\subsection{Neutrinoless double $\beta$ decay}\label{sec:0nubb}
In the seesaw model (\ref{L}) neutrinos are Majorana particles, and the Majorana mass term $M_M$ does not only violate lepton flavour, but also total lepton number. This makes neutrinoless double $\beta$ decays possible.  
The experimental constraints on neutrino properties are conventionally expressed in terms of the  \emph{effective Majorana mass}, which is defined as  
\begin{equation}\label{mee}
m_{ee}=\left|
\sum_i (U_\nu)_{ei}^2m_i + \sum_I \Theta_{eI}^2M_I f_A(M_I)
\right|.
\end{equation}
Here we follow Ref.~\cite{Ibarra:2011xn} and use $f_A=(M_A/M_I)^2 F_A$, $M_A\simeq 0.9$ GeV and $F_A=0.079$ for $^{76}\rm{Ge}$, based on previous estimates in Refs.~\cite{Vergados:1982wr,Haxton:1985am,Blennow:2010th}.\footnote{Strictly speaking the estimate for $f_A$ is only valid if $M_I$ is much larger than the exchanged momentum $\sim 100 $ MeV, see e.g. \cite{Faessler:2014kka} for some recent calculations beyond this approximation. 
This affects our estimate of the nuclear matrix element at the lower end $M_I\sim M_\pi$ of the mass range considered here, which introduces a small error. However, in this mass regime direct search constraints are much stronger than those from neutrinoless double $\beta$ decay.}
We can impose upper bounds on the $U_{e I}^2$ by requiring that the rate for neutrinoless double $\beta$ decay is consistent with the experimental upper bound $m_{ee}<0.2$~eV~ from the GERDA experiment\cite{Agostini:2013mzu}.\footnote{
After we had completed our numerical study, an updated constraint $m_{ee}<(0.06-0.161)$ eV has been published by the KamLAND-Zen experiment \cite{KamLAND-Zen:2016pfg}.}

\subsection{Lepton universality}\label{UniversalitySection}
The decay rates of pseudoscalar mesons $M$ into leptons depend on hadronic effects that suffer from considerable uncertainties. However, to a large degree, these cancel out if one considers ratios of the form \cite{Shrock:1980ct,Asaka:2014kia}
\begin{equation}\label{RM}
R^M_{\alpha\beta}\equiv\frac{\Gamma(M^+\rightarrow \ell_\alpha^+ \nu_\alpha )}{\Gamma(M^+\rightarrow \ell_\beta^+ \nu_\beta )}.
\end{equation}
This makes the deviation from the SM prediction,
\begin{equation}
\Delta r^M_{\alpha\beta}\equiv\frac{R^M_{\alpha\beta}}{R^{M}_{\alpha\beta {\rm SM}}}-1
\end{equation}
a sensitive observable.
The effect of $N_I$ on $R^M_{\alpha\beta}$ becomes clear if one decomposes (\ref{RM}) into mass eigenstates,
\begin{eqnarray}
  R^M_{\alpha\beta} = 
  \frac{
    \sum_{i=1,2,3} \Gamma (M^+ \rightarrow \ell_\alpha^+ \upnu_i)
    + 
    \sum_{I=1,2,3} \Gamma (M^+ \rightarrow \ell_\alpha^+ N_I)
  }{
    \sum_{i=1,2,3} \Gamma (M^+ \rightarrow \ell_\beta^+ \upnu_i)
    + 
    \sum_{I=1,2,3} \Gamma (M^+ \rightarrow \ell_\beta^+ N_I)
  }  \,.
\end{eqnarray}
\paragraph{Kinematically accessible $N_I$} - 
One can decompose \cite{Shrock:1980ct}
\begin{eqnarray}\label{DeltarWithN}
 \Delta r^M_{\alpha\beta}
  &=&
  \frac{
    \sum_{i=1,2,3} |(V_\nu)_{\alpha i}|^2 +
    \sum_{I=1,2,3} U_{\alpha I}^2 G_{\alpha I}
  }{
    \sum_{i=1,2,3} |(V_\nu)_{\beta i}|^2 +
    \sum_{I=1,2,3} U_{\beta I}^2 G_{\beta I}
  }
  - 1 \nonumber\\
&=& \frac{
    1 + 
     \sum_I U_{\alpha I}^2 
    \left[ G_{\alpha I} - 1 \right]
  }{
    1 +
    \sum U_{\beta I}^2 
    \left[ G_{\beta I} - 1 \right]
  }
  - 1
  \,,
\label{Deltar}
\end{eqnarray}
with
\begin{eqnarray}
  G_{\alpha I} =\vartheta(m_X - m_{l_\alpha}-M_I)
  \frac{ r_\alpha + r_I - (r_\alpha - r_I)^2 }
  { r_\alpha ( 1 - r_\alpha)^2}
  \sqrt{ 1 - 2 (r_\alpha + r_I) +(r_\alpha - r_I)^2 } ,
\end{eqnarray}
$r_\alpha = m_{l_\alpha}^2/m_X^2$, $r_I = M_I^2 /m_X^2$ and the Heavyside step function $\vartheta$.
In the second step in (\ref{Deltar}) we have used the unitarity of $\mathcal{U}$.
We require that 
\begin{eqnarray}
R^M_{\alpha\beta}=R^M_{\alpha\beta {\rm SM}}(1 + \Delta r^M_{\alpha\beta})
\end{eqnarray}
is within the 3$\sigma$ error bars of experimental measurements.
For kaon decay the SM prediction \cite{Cirigliano:2007xi} and result from NA62 \cite{Lazzeroni:2012cx} are
\begin{eqnarray}
R^K_{e\mu {\rm SM}}=(2.477 \pm 0.001)\times 10^{-5},\\
R^K_{e\mu }= (2.488 \pm 0.010)\times 10^{-5}.
\end{eqnarray}
We neglect the smaller error on the SM prediction.
For pion decays the theory prediction \cite{Cirigliano:2007xi} and experimental result \cite{Czapek:1993kc}
\begin{eqnarray}
R^\uppi_{e\mu {\rm SM}}=(1.2354 \pm 0.0002)\times 10^{-4},\\
R^\uppi_{e\mu }= (1.230 \pm 0.004)\times 10^{-4}.\label{ruppiexp}
\end{eqnarray}
Here we use kaon and pion decays. 
In principle we can also take heavy mesons, but the experimental constraints appear to be weaker \cite{Abada:2013aba}, see also \cite{Abada:2012mc,Basso:2013jka,Endo:2014hza,Asaka:2014kia}. The same applies to tests of lepton universality in $\tau$-decays. Here we take these bounds into account only in the regime where they are ``indirect''. 
\paragraph{Heavier $N_I$} - 
Heavy neutrinos can affect lepton universality observables ``directly'' (i.e. by appearing as real final state particles) only if they are lighter than the decaying particles. 
If they are heavier, then they still affect the decay ``indirectly'' due to the violation of unitarity in $V_\nu$.  
Then (\ref{DeltarWithN}) reduces to \cite{Shrock:1980vy,Shrock:1980ct,Shrock:1981wq}
\begin{equation}
 \Delta r^M_{\alpha\beta}
=
\frac{
    1 - 
    \sum_{I=1,2,3} U_{\alpha I}^2 
  }{
    1 -
    \sum_{I=1,2,3} U_{\beta I}^2 
  }
  - 1
\simeq  -U_\alpha^2+U_\beta^2.
\end{equation}
In the regime $M_I>m_\tau$ we also consider lepton universality violation in $\tau$-decays, using 
\begin{equation}
 \Delta r^\tau_{e\mu}
=
\frac{
    1 - 
    \sum_{I=1,2,3} U_{e I}^2 
  }{
    1 -
    \sum_{I=1,2,3} U_{\mu I}^2 
  }
  - 1
\simeq   \sum_{M_I>m_\tau}-U_{e I}^2+U_{\mu I}^2.
\end{equation}
The theory prediction \cite{Pich:2009zza}
and experimental result \cite{Beringer:1900zz} are
\begin{eqnarray}
R^\tau_{e\mu {\rm SM}}=0.973,\\
R^\tau_{e\mu }= 0.9764 \pm 0.0030.
\end{eqnarray}

\subsection{Electroweak precision data}
The Z-boson mass $m_Z$, fine structure constant $\alpha(m_Z)$ and Fermi constant have been measured at high accuracy \cite{Baak:2014ora}.
These parameters allow to predict the weak mixing angle $s_W={\rm sin}\theta_W$ and W boson mass $m_W$ via the relations
\begin{eqnarray}
s_W^2c_W^2&=&\frac{\alpha_Z\pi}{\sqrt{2}G_F m_Z^2}(1 + \delta r)\,,\label{sWSM}\\
m_W^2&=&c_W^2m_Z^2\label{mWSM}\,,
\end{eqnarray}
with $c_W={\rm cos}\theta_W$.
Here $\delta r$ are radiative corrections \cite{Hollik:1988ii,Awramik:2003rn}.
While $m_Z$, $\alpha(m_Z)$ and $\delta r$ are not significantly affected by the existence of the $N_I$, electroweak precision data is sensitive to these~\cite{delAguila:2008pw,Akhmedov:2013hec,Blas:2013ana} because
measurements of the Fermi constant would in fact have measured $G_\mu$ defined in  (\ref{GmuDef}) if $N_I$ with $M_I>m_\mu$ exist. 
The relations (\ref{sWSM}) and (\ref{mWSM}) are modified to \cite{Antusch:2014woa}
\begin{eqnarray}
s_W^2&=&\frac{1}{2}\left[
1-
\sqrt{
1-
\frac{2\sqrt{2}\alpha\pi(1 + \delta r)}{G_\mu m_Z^2}\sqrt{1-(\theta\theta^\dagger)_{ee}-(\theta\theta^\dagger)_{\mu\mu}}
}
\right]\label{sW}\,,\\
m_W^2&=&m_Z^2\frac{1}{2}\left[
1+
\sqrt{
1-
\frac{2\sqrt{2}\alpha\pi(1 + \delta r)}{G_\mu m_Z^2}\sqrt{1-(\theta\theta^\dagger)_{ee}-(\theta\theta^\dagger)_{\mu\mu}}
}
\right]\,.\label{mW}
\end{eqnarray}
We use the central values of the measured quantities \cite{Beringer:1900zz}
\begin{eqnarray}
\alpha(m_Z)^{-1}&=& 127.944(14)\,,\\
G_\mu &=& 1.1663787(6) \times 10^{−5} {\rm GeV}^{-2}\,,\\
m_Z&=&91.1875(21) {\rm GeV}\,,
\end{eqnarray}
and insert them into (\ref{sW}) and (\ref{mWSM}).
The results in \cite{Baak:2014ora} suggest $\delta r=-0.03244$. We ignore the small experimental errors on our input parameters and the theoretical error in $\delta r$.
Then we require that the prediction of (\ref{mWSM})  is consistent with the direct measurement 
\begin{eqnarray}
m_W&=& 80.385   \pm 0.015 {\rm GeV}
\end{eqnarray}
used in \cite{Baak:2014ora}\footnote{ Note that there is some discrepancy between the measurements for $s_W$ \cite{ALEPH:2005ab,Group:2012gb,Ferroglia:2012ir} 
\begin{eqnarray}
(s_W^\ell)^2&=&0.23113(21) \ {\rm  measured \  in \ leptons}\,,\notag\\
(s_W^\ell)^2&=&0.23222(27) \ {\rm  measured \  in \ hadrons}\,.\notag
\end{eqnarray}
We require to be either consistent with the one in hadrons or with the one in leptons.
}
at 3$\sigma$ level.

\subsection{CKM unitarity}
The non-unitarity in $V_\nu$ affects leptonic decays $\ell_\alpha\rightarrow\nu_\alpha \ell_\beta\bar{\nu}_\beta$. 
This change is effectively described by the difference between $G_F$ and $G_\mu$ defined in (\ref{GmuDef}).
It affects measurements of the CKM matrix $V_{\rm CKM}$ in the quark sector because the experimental values $(V_{\rm CKM}^{\rm exp})_{ab}^{(i)}$ are inferred from the direct experimental observables assuming that $\Theta\equiv0$.
The true CKM matrix is unitary in the model (\ref{L}), hence
\begin{equation}\label{unitarityrelationinCKM}
1 = |(V_{\rm CKM})_{ud}|^2 + |(V_{\rm CKM})_{us}|^2 + |(V_{\rm CKM})_{ub}|^2 \simeq |(V_{\rm CKM})_{ud}|^2 + |(V_{\rm CKM})_{us}|^2.
\end{equation} 
We will neglect the small $|(V_{\rm CKM})_{ub}|^2$ in the following.
For a given matrix $\Theta$, one can use differently measured values $(V_{\rm CKM}^{\rm exp})_{ab}^{(i)}$ to obtain a best fit value for the true $(V_{\rm CKM})_{ud}$ and $(V_{\rm CKM})_{us}$.  Using the CKM unitarity (\ref{unitarityrelationinCKM}), this becomes a one-parameter fit for $(V_{\rm CKM})_{ud}$.
The quality of the best fit, as characterised by the $\chi^2$, depends on $\Theta$. The relation between the true CKM elements and those extracted from experiment depends on $\Theta$ and is different for each experimental setup. 
The relation between $(V_{\rm CKM}^{\rm exp})_{us}^{(i)}$ extracted from experiment and the true $(V_{\rm CKM})_{ud}$ is given by
\begin{eqnarray}
|(V_{\rm CKM}^{\rm exp})_{us}^{(i)}|^2&=&|(V_{\rm CKM})_{us}|^2
\left[1 + f^{(i)}[\Theta]\right]\\
&\simeq& \left[1 - |(V_{\rm CKM})_{ud}|^2\right]
\times
\left[1 + f^{(i)}[\Theta]\right].\nonumber
\end{eqnarray}
Here we have factored out the ``error'' coming from the difference between the true $G_F$ and the measured $G_\mu$, which affects all experiments. $f^{(i)}[\Theta]$ is a function of the $\Theta_{\alpha I}$ that is specific to the experimental setup, and the index $^{(i)}$ labels different experiments. The quantity
$(V_{\rm CKM}^{\rm exp})_{ud}^{(i)}$ is then understood as the prediction for the direct observation
of this matrix element in a weak decay given the value of $(V_{\rm CKM}^{\rm exp})_{us}^{(i)}$ and assuming
CKM unitarity.
We use the same processes as the authors of \cite{Antusch:2014woa}, where a detailed description of the derivation of the formulae below is given.
The kaon decays into electrons
\begin{eqnarray}
(1) \ K_L\rightarrow \pi^+ e^- \bar{\nu}_e \ &;& \ 1+f^{(1)}[\Theta]=\frac{G_F^2}{G_\mu^2}\left[1-(\theta\theta^\dagger)_{ee}\right]\simeq 1+(\theta\theta^\dagger)_{\mu\mu}\\
(2) \ K_S\rightarrow \pi^+ e^- \bar{\nu}_e \ &;& \ f^{(2)}[\Theta]=f^{(1)}[\Theta]\\
(3) \ K^-\rightarrow \pi^0 e^-\bar{\nu}_e  \ &;& \ f^{(3)}[\Theta]=f^{(1)}[\Theta]
\end{eqnarray}
and their charge conjugates all have the same $f^{(i)}_{ab}[\Theta]$. The decays
\begin{eqnarray}
(4) \ K_L\rightarrow \pi^+ \mu^- \bar{\nu}_\mu \ &;& \ 1+f^{(4)}[\Theta]=\frac{G_F^2}{G_\mu^2}\left[1-(\theta\theta^\dagger)_{\mu\mu}\right]\simeq 1+(\theta\theta^\dagger)_{ee}\\
(5) \ K^-\rightarrow \pi^0 \mu^- \bar{\nu}_\mu \ &;& \ f^{(5)}[\Theta]=f^{(4)}[\Theta]
\end{eqnarray}
behave equivalently. 
We also consider the decay of $\tau$ leptons. For the ratio
\begin{eqnarray}
\frac{\tau^-\rightarrow K^- \nu_\tau}{\tau^-\rightarrow \pi^- \nu_\tau} \propto 
\frac{|(V_{\rm CKM})_{us}|^2}{|(V_{\rm CKM})_{ud}|^2}
\end{eqnarray}
one finds 
\begin{eqnarray} 
(6) \frac{\tau^-\rightarrow K^- \nu_\tau}{\tau^-\rightarrow \pi^- \nu_\tau}
\ &;& \ 1+f^{(6)}[\Theta]=1+(\theta\theta^\dagger)_{\mu\mu}
\end{eqnarray}
because in this measurement $|(V_{\rm CKM}^{\rm exp})_{us}^{(6)}|^2/|(V_{\rm CKM}^{\rm exp})_{ud}^{(6)}|^2=|(V_{\rm CKM})_{us}|^2/|(V_{\rm CKM})_{ud}|^2$.
For the $\tau$-decay into kaons, we use
\begin{eqnarray}
(7) \ \tau^-\rightarrow \pi^- \bar{\nu}_\tau \ &;& \ 1 + f^{(7)}[\Theta]=
1+(\theta\theta^\dagger)_{ee}+(\theta\theta^\dagger)_{\mu\mu}-(\theta\theta^\dagger)_{\tau\tau}
\end{eqnarray}
Finally, one could also include the branching ratio into strange mesons $\tau\rightarrow s/\tau\rightarrow {\rm anything}$, 
\begin{eqnarray}
(8) \ \tau\rightarrow s \ ; \ 1 + f^{(8)}[\Theta]=
1+0.2(\theta\theta^\dagger)_{ee}-0.9(\theta\theta^\dagger)_{\mu\mu}-0.2(\theta\theta^\dagger)_{\tau\tau}.\label{channel8}
\end{eqnarray}
Including this value tends to give a bad fit even in the SM configuration.
It is tempting to interpret this as a deviation from the SM that may be explained by RH neutrinos, but in the present work we prefer to be conservative and simply disregard the channel (8).
The experimental values for $(V_{\rm CKM})_{us}$ are given in table \ref{CKMtable}.
\begin{table}
\begin{minipage}{0.34\textwidth}
\centering
\begin{tabular}{|l|c|}
\hline
Process                & $(V_{\rm CKM})_{us} f_+(0)$     \\
\hline\hline
$K_L \rightarrow \pi e \nu$     & 0.2163(6)   \\
$K_L \rightarrow \pi \mu \nu$   & 0.2166(6)   \\
$K_S \rightarrow \pi e \nu$     & 0.2155(13)  \\
$K^\pm \rightarrow \pi e \nu$   & 0.2160(11)  \\
$K^\pm \rightarrow \pi \mu \nu$ & 0.2158(14)  \\
\hline
average                & 0.2163(5)   \\ 
\hline
\end{tabular}
\end{minipage}
\begin{minipage}{0.65\textwidth}
\begin{tabular}{|c|c|c|}
\hline
Process & $f^{(i)}(\Theta)$  & $|(V_{\rm CKM})_{us}|$  \\
\hline
$\frac{B(\tau \to K \nu)}{B(\tau \rightarrow \pi \nu)}$ & $(\theta\theta^\dagger)_{\mu\mu}$ & 0.2262(13) \\
$\tau \rightarrow K \nu	$				& $(\theta\theta^\dagger)_{ee} + (\theta\theta^\dagger)_{\mu\mu}-(\theta\theta^\dagger)_{\tau\tau}$	& 0.2214(22) \\
$\tau \rightarrow \ell,\,\tau \rightarrow s$ 			& $0.2(\theta\theta^\dagger)_{ee} - 0.9 (\theta\theta^\dagger)_{\mu\mu} - 0.2(\theta\theta^\dagger)_{\tau\tau}$	& 0.2173(22) \\
\hline
\end{tabular}
\end{minipage}
\caption{
We used the same data as the authors of \cite{Antusch:2014woa}.
The left panel shows values of $(V_{\rm CKM})_{us}$, determined from different kaon decay processes as listed in \cite{Antonelli:2010yf}, the quantity $f_+(0)=0.959(5)$ is taken from \cite{Aoki:2013ldr}. 
The right panel shows values of $(V_{\rm CKM})_{us}$, determined from different tau decay modes in \cite{Follana:2007uv,Amhis:2012bh}.}
\label{CKMtable}
\end{table}
We use this data and the world average $(V_{\rm CKM}^{\rm exp})_{ud}^{(0)}=0.97427(15)$ \cite{Beringer:1900zz} to perform a fit for $(V_{\rm CKM})_{ud}$ for a given choice of $\Theta$,
\begin{eqnarray}
(V_{\rm CKM}^{\rm exp})_{ud}=\frac{\sum_i(V_{\rm CKM}^{\rm exp})_{ud}^{(i)} \ \sigma_{(i)}^{-2}}{\sum_i\sigma_{(i)}^{-2}} \ , \ \chi^2=\sum_i\left[(V_{\rm CKM}^{\rm exp})_{ud}^{(i)} - (V_{\rm CKM}^{\rm exp})_{ud}\right]^2\sigma_{(i)}^{-2}
\end{eqnarray}
In order to accept a given parameter choice, we demand to remain  within $3\sigma$.
This test relies on the assumption that the actual quark mixing matrix is unitary.
It would of course be better to directly test the unitarity of $V_\nu$, but the data at present time is insufficient to impose significant constraints on $\theta$.

\subsection{Dipole moments}
\paragraph{Charged lepton electric dipole moment} -
Recently it has also been proposed that the impact of heavy neutrinos on charged lepton electric dipole moments can be used to constrain there properties. We have not included these observables in our analysis because they are not expected to give any significant constraints for $M_I<m_W$ \cite{Abada:2015trh}.
\paragraph{Neutrino transition dipole moment} -
Heavy neutrinos can radiatively decay into an active neutrino and a photon.
This process, which occurs at a rate \cite{Pal:1981rm,Barger:1995ty}
\begin{eqnarray}
 \Gamma(N_I \rightarrow \nu_\alpha \gamma)= \frac{9\alpha_{EM}G_F^2}{1024\pi^2}\sin^2(2U_{\alpha I})M_I^5, 
\end{eqnarray}
can be described by a neutrino transition dipole moment \cite{Beg:1977xz}.
In principle this dipole moment can be probed in neutrino nucleon scatterings. However, this requires a large neutrino flux and does not seem to be observable in the near future \cite{Gorbunov:2014ypa}.



\section{Direct searches for heavy neutrinos}\label{DirectSearchSubsection}

In the following we summarise constraints on the $N_I$-properties from direct searches at colliders and in fixed target experiments.
We display a summary of the existing direct search bounds in figures \ref{UePlot}-\ref{UtauPlot}.

\subsection{Production in meson and $\tau$ decays}
For masses $M_I<m_B$ the $N_I$ can be produced in meson decays. 
Bounds from the decays of gauge bosons or Higgs particles into RH neutrinos are sub-dominant in this regime.
\subsubsection{Beam dump experiments}
Searches for the decay of $N_I$ produced in meson decays been performed at the beam dump experiments PS191 \cite{Bernardi:1987ek}, NuTeV \cite{Vaitaitis:1999wq}, CHARM \cite{Bergsma:1985is}, CHARM II \cite{Vilain:1994vg}, NA3 \cite{Badier:1985wg},  E949 \cite{Artamonov:2014urb}, IHEP-JINR \cite{Baranov:1992vq}, BEBC \cite{CooperSarkar:1985nh}, FMMF \cite{Gallas:1994xp} and NOMAD \cite{Astier:2001ck}. In beam dump experiments, in which a proton beam is ``dumped'' into a fixed target, it is often not possible to identify any charged particles that are produced along with the $N_I$ in the meson decay because the background near the target is too strong. 
However,  heavy neutrinos with masses $M_I<m_D$ are rather long lived and travel macroscopic distances. One can therefore look for their charged decay products in a detector that is placed at some distance from the target.
The expected signals are proportional to combinations of $U_{\alpha I}^2 U_{\beta I}^2$, i.e. of fourth power in $\Theta$, because production and decay of the $N_I$ are suppressed by one mixing square each. 
The details depend on the specific production and decay channel and are e.g. discussed in \cite{Ruchayskiy:2011aa}. 
In \cite{Ruchayskiy:2011aa} it has been pointed out that the
constraints from PS191 \cite{Bernardi:1987ek} and CHARM \cite{Bergsma:1985is} were derived under the assumption that the heavy neutrinos only interact via the charged current. To take into account the neutral current contribution in (\ref{WeakWW}) in the $N_I$ decay, the constraints on $U_{e I}^4$ should be re-interpreted as constraints on the combination\footnote{The $N_I$ in the "WBB experiment" would be created in a scattering of muon neutrinos off a nucleus (rather than the decay of a D meson) and extend above $m_D$.
This channel only involves production via neutral current and decay via charged current, which was taken into account in \cite{Bergsma:1985is}, and no re-interpretation in terms of (\ref{ArtemReinterpret}) is needed for this channel.}
\begin{eqnarray}
\frac{1 + 4\sin^2\theta_W+8\sin^4\theta_W}{4}U_{e I}^2 
+ \frac{1 - 4\sin^2\theta_W+8\sin^4\theta_W}{4}(U_{\mu I}^2 + U_{\tau I}^2)\label{ArtemReinterpret}.
\end{eqnarray}
We use these re-interpreted bounds in our analysis.

\subsubsection{Peak searches}
\begin{figure}
\begin{center}
\includegraphics[width=12cm]{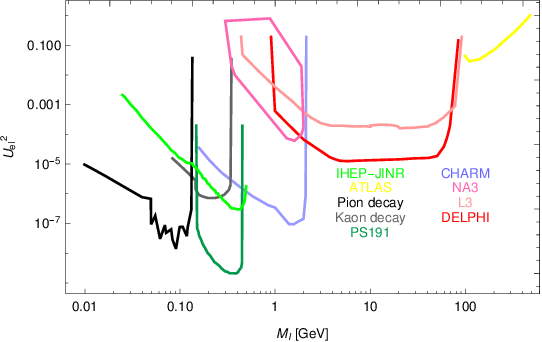}
\caption{\label{UePlot}
Constraints on $U_{e I}^2$ from the experiments 
DELPHI \cite{Abreu:1996pa}, L3 \cite{Adriani:1992pq}, ATLAS \cite{Aad:2015xaa},
PIENU \cite{PIENU:2011aa},
TINA \cite{Britton:1992xv},
PS191 \cite{Bernardi:1987ek},
CHARM \cite{Bergsma:1985is},
NA3  \cite{Badier:1985wg}, 
and kaon decays \cite{Yamazaki:1984sj}. 
For peak searches below the kaon mass we use the summary given in \cite{Atre:2009rg}, for PS191 we use the re-interpretation given in \cite{Ruchayskiy:2011aa}.
}
\end{center}
\end{figure}

\begin{figure}
\begin{center}
\includegraphics[width=12cm]{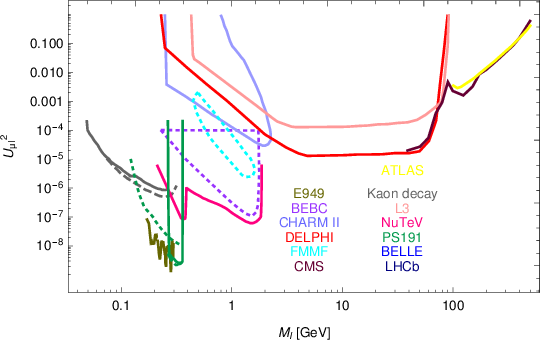}
\caption{\label{UmuPlot}
Constraints on $U_{\mu I}^2$ from the experiments 
DELPHI \cite{Abreu:1996pa}, 
L3 \cite{Adriani:1992pq}, 
LHCb \cite{Aaij:2014aba},
ATLAS \cite{Aad:2015xaa},
CMS \cite{Khachatryan:2015gha}, 
BELLE \cite{Liventsev:2013zz}, 
BEBC \cite{CooperSarkar:1985nh}, 
FMMF \cite{Gallas:1994xp}, 
E949 \cite{Artamonov:2014urb},
PIENU \cite{PIENU:2011aa}, 
TINA \cite{Britton:1992xv},
PS191 \cite{Bernardi:1987ek},  
CHARMII \cite{Vilain:1994vg},
NuTeV \cite{Vaitaitis:1999wq}, 
NA3 \cite{Badier:1985wg}
and kaon decays in \cite{Yamazaki:1984sj,Hayano:1982wu}.
For the bounds from kaon decays we use the interpretation given in \cite{Kusenko:2004qc,Atre:2009rg}.
For NA3, BEBC and FMMF we use the estimates from \cite{Atre:2009rg}.
For PS191 we compare the results according to the interpretation in Ref.~\cite{Ruchayskiy:2011aa} (solid line) as well as \cite{Artamonov:2014urb} for two different channels (dashed and dotted line). 
}
\end{center}
\end{figure}

\begin{figure}
\begin{center}
\includegraphics[width=12cm]{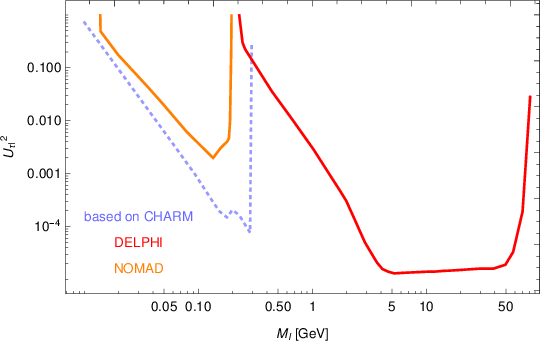}
\caption{\label{UtauPlot}
Bounds on $U_{\tau I}^2$ based on the interpretation of CHARM data in \cite{Orloff:2002de}, NOMAD \cite{Astier:2001ck}, L3 \cite{Adriani:1992pq} and DELPHI \cite{Abreu:1996pa}.
}
\end{center}
\end{figure}

Heavy neutrinos produced in meson decays can leave a signature in experiments even if their own decay is not observed.
When a meson with mass $m_X$ decays into a heavy neutrino $N_I$ and an electron or muon with mass $m_{\ell_\alpha}$, then the lepton spectrum shows a peak at an energy
\begin{equation}
E_{\rm peak}\simeq \frac{m_X^2+m_{\ell_\alpha}^2-M_I^2}{2m_X}\,.
\end{equation}
The branching ratio of this decay is $\propto U_{\alpha I}^2$, and the search for the peak in the lepton spectrum can be used to impose a bound on this mixing.
Various experiments have been performed, using kaons and pions as initial mesons \cite{PIENU:2011aa,Britton:1992pg,Britton:1992xv,Bryman:1996xd,Abela:1981nf,Daum:1987bg,Yamazaki:1984sj,Hayano:1982wu}, so far without any sign of $N_I$.
We plot the currently strongest bounds in figures~\ref{UePlot}-\ref{UtauPlot}.

\subsubsection{Lepton number violating decays}
LNV decays are ``smoking gun'' signals for the existence of heavy Majorana neutrinos and allow to impose some of the strongest bounds. They can, however, be avoided in the approximately lepton number conserving scenarios discussed in Sec.~\ref{BminusL}.
\paragraph{Meson decays} - 
Searches for LNV meson decays have been performed at CLEO \cite{He:2005iz}, Babar, BELLE \cite{Liventsev:2013zz} and LHCb \cite{Aaij:2014aba} (see also \cite{Helo:2010cw,Quintero:2011yh,Castro:2013jsn}).
The bounds given in \cite{Eidelman:2004wy} have been converted into bounds on the $U_{\alpha I}^2$ in \cite{Atre:2009rg}. These would be extremely strong if one could assume a vanishing lifetime for the $N_I$, which is, however,  not realistic. If one includes the lifetime, they appear to be sub-dominant. 
We therefore do not systematically re-derive these bounds to include more recent data \cite{Rubin:2010cq}. However, to verify that they are sub-dominant, we do include the bounds from LNV rare meson decays  shown in \cite{Atre:2009rg} and by hand strengthen them by a factor 3 to take account of the new data.
In contrast to that, we directly impose the more recent bounds from LHCb, which, however, turn out to be sub-dominant.

\paragraph{Rare $\tau$-decays} -
Heavy neutrinos with $M_I<m_\tau$ can also be produced in the decay of $\tau$ leptons \cite{Atre:2009rg,Castro:2012gi}.
The non-observation of LNV rare $\tau$-decays at Babar \cite{Aubert:2005tp} has been translated into bounds on the $U_{\tau I}^2$ in \cite{Atre:2009rg}.
We treat these with the same procedure as the bounds from LNV meson decays given in \cite{Atre:2009rg}, i.e. we strengthen them by a factor 3 to take account of new data  \cite{Miyazaki:2012mx,Miyazaki:2013yaa,Amhis:2014hma} to show that they remain sub-dominant. Some comments on the interpretation of this data can also be found in \cite{Gorbunov:2014ypa}.

\subsection{Production in gauge boson decays}
For $m_B<M_I<m_W$, the most efficient production channel is the decay of gauge bosons. 
In principle, the $N_I$ are also produced in Higgs decays \cite{Arganda:2004bz,Arganda:2014dta}.
Here we focus on the LEP bounds from $Z$-decays, which in the mass range we consider are stronger than LHC searches in W~decays \cite{Khachatryan:2015gha,CMS:2015sea,Khachatryan:2016olu}, lepton number or flavor violating
Higgs decays \cite{Graesser:2007yj,BhupalDev:2012zg,Cely:2012bz,Maiezza:2015lza,Khachatryan:2015kon} and Tevatron searches for LNV \cite{Abulencia:2007rd}.


\subsubsection{Bounds from LEP}

The experiments DELPHI \cite{Abreu:1996pa} and L3 \cite{Adriani:1992pq} have looked for the leptonic and semi-leptonic decays of heavy neutrinos produced in Z decays. The bounds from these searches remain to be the strongest to date in the mass range $m_B<M_I<m_W$. 
The existence of $N_I$ affects the decays of Z bosons also if the $N_I$ do not decay within the detector.
The decay rate into light neutrinos $\upnu_i$ is reduced due to their singlet component, which leads to the suppression of their weak interaction by a factor $\mathcal{O}[1-\theta]$.
For $M_I>m_Z$ this leads to an apparent violation of unitarity, which can be identified with the non-unitarity in $V_\nu$.
The decay rate $\Gamma_{Z\rightarrow \nu\nu}$ is (to date) not directly observable, but contributes to the total invisible decay width, which is experimentally constrained (in the SM it is the only contribution).
For the case $M_I<m_Z$ under consideration here, the decay into $N_I$ also contributes to the invisible decay width, so in principle unitarity is restored. This suggests that $Z$-decay parameters are not very sensitive to heavy neutrinos with $M_I\ll m_Z$.

In principle the modifications (\ref{GmuDef}) and (\ref{sW})  still lead to a deviation from the SM. Moreover, the SM prediction assumes that the decay products $\upnu_i$ have negligible masses, while the $M_I$ are kinematically not negligible if they are comparable to $m_Z$. Finally, the $N_I$ can decay within the detector, in which case these events might have been mis-identified in past searches.
In spite of this, we assume that Z-decay parameters do not impose significant additional bounds on $N_I$ with GeV masses for the following reasons. 
Regarding the last point, $N_I$ that are sufficiently short lived to decay in the detector in significant numbers would most likely have been observed at DELPHI and L3. We therefore cover this possibility by including the constraints from these experiments. 
Regarding the second point, the $M_I$ are kinematically negligible in almost the entire mass range we consider. 
Hence, the only remaining change is the modification of the rate due to (\ref{GmuDef}) and (\ref{sW}). In the often-quoted ratio $\Gamma_{Z \rightarrow {\rm invisible}}/\Gamma_{Z\rightarrow ll}$ the dependence on the Fermi constant cancels, and only a subtle (and weak) dependence due to $s_W$ remains. The same applies to the hadronic cross section $\Gamma_{Z \rightarrow ee}\Gamma_{Z \rightarrow {\rm hadrons}}/\Gamma_{Z}^2$. 
A comparison of the absolute observed value of $\Gamma_{Z \rightarrow{\rm invisible}}$ to the SM prediction would be sensitive to the change (\ref{GmuDef}) in the measured Fermi constant. However, it probes the same combination of mixings as the CKM unitarity test on (\ref{sW}) and (\ref{mW}), where the experimental error bars are smaller. 
We therefore conclude that Z-boson decay parameters only yield sub-dominant bounds and do not include them. This conclusion is consistent with the results of Ref.~\cite{Antusch:2015mia}, cf. also \cite{Abada:2014cca}.
Similar arguments can be made for the effect of $N_I$ on the decay of Higgs bosons, which has been studied in \cite{Arganda:2014dta}.

\subsubsection{Searches for heavier $N_I$}
For masses $M_I>m_Z$, the Z-decay parameters currently provide one of the strongest constraints on $U_{\alpha I}^2$. Indeed, in the global scan \cite{Antusch:2014woa} it was found that the invisible decay width and the hadronic branching ratio in Z decays slightly prefer a value of $U_{e I}^2\neq 0$, which is also supported by the present value of $s_W$. 
Direct search constraints in this region can be obtained from L3 \cite{Acciarri:1999qj,Achard:2001qv} and LHC \cite{ATLAS:2012ak,Khachatryan:2014dka,Aad:2015xaa,Khachatryan:2015gha,CMS:2015sea}. The latter could be improved by using additional production channels other than the s-channel exchange of W bosons \cite{Dev:2013wba,Alva:2014gxa,Das:2016hof}.

\subsection{Future searches}
In the low mass region $M_I<m_K$ even small improvements are significant, especially for the $n=2$ scenario or scenario A (which effectively is $n=2$). The reason is that in this case BBN can impose a lower bound on $U_{\mu I}^2$ (see section \ref{sec:bbn}), which is not much below the upper bound from direct searches, such that the remaining window may be closed in the near future. 
Further improvement will be possible with NA62 \cite{Moulson:2013oga} and T2K \cite{Asaka:2012bb}.  
The most significant improvement for $M_I<m_B$ could be made with the proposed SHiP experiment \cite{Bonivento:2013jag,Alekhin:2015byh,Anelli:2015pba}, or with a SHiP-like experiment at DUNE \cite{Akiri:2011dv,Adams:2013qkq}. 
If two $M_I$ are degenerate, one may even be able to measure the sterile sector CP-violation in $\mathcal{R}$ in $N_I$ decays \cite{Cvetic:2014nla,Cvetic:2015naa}, which can be responsible for the baryon asymmetry of the universe \cite{Canetti:2012kh}.

For larger masses, the LHC is the only existing collider that can improve the existing bounds. 
The dominant channel for $N_I$ production is via s-channel exchange of a W boson.\footnote{For masses above $\sim 600$ GeV, beyond the mass range we consider here, collinearly enhanced t-channel processes become more efficient \cite{Dev:2013wba,Alva:2014gxa}.}
For $M_I<15$ GeV, the $N_I$ lifetime is long enough to look for displaced vertices \cite{Izaguirre:2015pga,Helo:2013esa}, in particular a displaced lepton jet in combination with a prompt lepton from the $N_I$ production \cite{Izaguirre:2015pga}.
For masses $10\, {\rm GeV} < M_I < m_W$ it is more promising to search for LNV processes \cite{Izaguirre:2015pga,Dib:2015oka,NicoElena}.
However, a discovery seems only realistic (and compatible with bounds from neutrinoless double $\beta$ decay) in models where individual $F_{\alpha I}$ are much bigger than $F_0$ in (\ref{F0}) and neutrino masses are kept small by an approximate (fundamental or accidental) symmetry in $F$ and $M_M$ \cite{Kersten:2007vk}.
In view of this, the most promising machine would be a future lepton collider \cite{Blondel:2014bra,Antusch:2015mia,Antusch:2015rma,Banerjee:2015gca}, which would allow to look for signatures that do not rely on LNV, such as displaced vertices.

\subsection{Interpretation of experimental results}\label{Sec:interpretation}

\subsubsection{Dependence on the choice of scenario}\label{dependenceonassumptions}
All considerations in this paper are based on the minimal type-I seesaw described by the Lagrangian (\ref{L}). 
We assume that no other new physics affects the observables we consider significantly. Moreover, we assume that there are $n=3$ heavy neutrinos which generate the light neutrino masses via the seesaw mechanism, and that all of them have masses between the pion and the W mass.

``Direct'' observables related to particle kinematics and lifetime (such as peak searches, beam dump experiments, searches for missing energy and the constraints from nucleosynthesis) are more or less independent of the number and mass spectrum of heavy neutrinos
because they are sensitive to the properties of each $N_I$ individually,
and quantum interferences between different amplitudes have no strong effect on the cross section for these processes.
That is, they constrain the properties of each heavy neutrino $N_I$ individually, irrespectively of the properties of its siblings $N_J$ with ${J\neq I}$.

Most ``indirect'' observables, on the other hand, rely on this assumption (and the more subtle assumption that the effect of the $N_I$ is not accidentally cancelled by that of some other new physics). This includes neutrino oscillation data, rare lepton decays, lepton universality electroweak precision data and unitarity considerations.
This is obvious, as one cannot exclude the possibility that the effect of $N_I$ on any individual observable is minimised by other new particles if the $N_I$ are not observed directly.

A special situation is realised when the masses of two heavy neutrinos are too similar to be distinguished experimentally.\footnote{We have generally assumed that this is the case if the splitting is smaller than 10 MeV.} 
In this case even kinematic observables do not probe the different heavy neutrino flavours individually because they cannot be distinguished.
A specific realisation of this situation is given by the approximately $B-L$ conserving limit discussed in Sec.~\ref{MixingScenarios}.
In this case also LNV signatures (such as same sign dileptons) at colliders, which have been classified as ``direct observables'' here because the $N_I$ appear as on-shell particles, can be suppressed by interference terms. 

\subsubsection{Dirac and Majorana particles}
In several of the past experimental searches for heavy neutral leptons, it has been assumed that these are Dirac particles. Whether or not the constraints can directly be applied to the Majorana particles $N_I$ depends on the mass spectrum and type of experiment.

\paragraph{One sterile flavour or several non-degenerate flavours} -
In this case there is no difference between Dirac or Majorana as far as 
$N_I$ production is concerned. A Dirac neutrino has twice as many internal degrees of freedom (particle, antiparticle with spin up, down), but only half of these can be produced in the decay of a charged meson. 
This can be seen  by looking at the effective interaction Lagrangian between $N_I$ and charged pions $\uppi$:
\begin{equation}
\mathcal{L}^{\rm eff}_\pi=\frac{2 G_F}{\sqrt{2}}\Theta_{\alpha I}V_{ud}f_\pi\bar{e}_\alpha\left(P_R-\frac{m_\alpha}{M_I}P_L\right)N_I\uppi +{\rm h.c.}
\end{equation}
It is clear that only either the above term or its conjugate can give a non-zero contribution to $\langle e_\alpha^\pm N_I|\mathcal{L}^{\rm eff}_\pi|\uppi^\pm\rangle$, because either only $\bar{e}_\alpha$ or only $e_\alpha$ can be contracted with the final state lepton, regardless of whether $N$ is a Dirac or Majorana spinor. 

For the decay, on the other hand, there is  a difference between Dirac and Majorana fields; A Majorana particle can decay into twice as many final states because charge conjugate decays are also allowed. 
Peak searches only involve the production of $N_I$, so there is no difference between Dirac and Majorana. In beam dump experiments looking for $N_I$ decays one  expects twice as many events if $N_I$ is Majorana. Therefore, exclusion bounds that have been obtained under the assumption of Dirac particles are a factor 2 stronger for Majorana. 

Bounds from searches for LNV signals (e.g. same-sign dileptons) only apply to Majorana particles. 

\paragraph{Two mass-degenerate $N_I$} -
If two of the $M_I$ are very similar, then it might not be possible to resolve them experimentally. One has to distinguish two qualitatively different cases. 

If the mass splitting is smaller than the experimental resolution, but still much bigger than the widths of the $N_I$ resonances, then they are two 
well-separated mass eigenstates. The $N_I$ are, however, not produced as mass eigenstates, but in the flavour basis where $F$ is diagonal.
These two states can undergo coherent oscillations within the experimental setup. The phase of these oscillations as a function of distance $d$ can be estimated as $\phi(d)=d\frac{M_I^2-M_J^2}{2E}$, where $E$ is the particle energy. 
If $l$ is the detector length, then coherent effects can be neglected if $\phi=l \frac{ \delta M^2}{4 E} \gg1$, which is the case for most of the parameter space. The observed rate in peak searches would be simply the sum of rates for all $N_I$ within a given mass bin, i.e.\ one would constrain $\sum_I U_{\alpha I}^2$ (sum over all  $N_I$ in that mass bin) instead of $U_{\alpha I}^2$.
If the limit that two $N_I$ form a pseudo-Dirac spinor, then this amounts simply to a factor $2$ because $U_{\alpha I}^2=U_{\alpha J}^2$ in that limit, see (\ref{equalcouplings}). In addition to $M_I-M_J\rightarrow 0$, this requires a symmetry in $F$ that can be realised by sending some $|{\rm Im}\omega_{ij}|\rightarrow \infty$. This limit is technically natural because it leads to a conserved lepton number. 
The question to which degree one can distinguish Dirac, Majorana and pseudo-Dirac particles can be distinguished experimentally has e.g. been studied in Refs.~\cite{Dib:2016wge,Dib:2015oka,Anamiati:2016uxp}. However, in these works the contribution to $M_N$ from the coupling to the Higgs field has been neglected \cite{Drewes:2016jae}.

If the mass splitting is comparable to the width or smaller, then interference terms are important. In this case we are not aware of any simple way to interpret the results from beam dump experiments derived under the assumption $n=1$. In general, one would expect a suppression of lepton number violating signals due to the existence of an approximately conserved lepton number. We do not treat this case in the present work.

\subsubsection{Error bars}
Another issue is that different collaborations quote exclusion bounds with different confidence levels, which are partly summarised in~\cite{Atre:2009rg}. In principle this should be modelled with a smooth likelihood function. We entirely ignore this issue to simplify the present analysis and treat all bounds as ``hard'', i.e. as if all forbidden points are forbidden with absolute certainty and all allowed points can be realised with exactly the same probability. This procedure though serves the present purpose of charting the viable parameter space which is vast compared
to the size of the error bands.

\section{Cosmological constraints}\label{sec:cosmo}

\subsection{Big Bang nucleosynthesis}\label{sec:bbn}

If the $N_I$ decay during or shortly before BBN, then the decay products have energies of the order $M_I$, much  above the temperature of the primordial plasma at that time. This would affect the abundances of light elements that are produced, either by directly dissociating nuclei that have already formed or indirectly by causing a deviation from thermal equilibrium in the plasma that modifies the rates for the processes involved in nucleosynthesis. 
This can be used to impose a lower bound on $U_I^2$.

Here we adopt a simple (but rather conservative) criterion and demand that the lifetimes $\tau_I$ of all $N_I$ are shorter than $0.1$s. To calculate the lifetime, we use the decay rates for various decay channels given in Refs.~\cite{Gorbunov:2007ak} and \cite{Canetti:2012kh}.
This imposes a lower bound on $U_I^2$.\footnote{It can be softened in extensions of (\ref{L}) if there are decay channels for $N_I$ other than via mixing with active neutrinos.}
For $M_I< 140 $ MeV, where this effect is most important because the $N_I$-lifetime scales as $\tau_I\propto M_I^{-5}$, it has been studied in detail in Refs.~\cite{Dolgov:2000jw,Dolgov:2000pj,Ruchayskiy:2012si}. The results show that the actual bounds for $M_I$ below the pion mass are weaker than our criterion, while for larger masses our approach 
appears
justified as a first approximation.

In general, BBN cannot impose bounds on the mixings $U_{\alpha I}^2$ of $N_I$ with individual active flavours. A given $U_{\alpha I}^2$ can be arbitrarily small (even zero) if another $U_{\beta\neq\alpha I}^2$ is sufficiently large to ensure a lifetime below 0.1s. 
The situation is different in scenario A if one assumes a mass degeneracy that makes it impossible to measure the $U_{\alpha I}^2$ invididually. In this case there are far less free parameters in $F$; in particular, there is only one complex ``Euler angle'' $\omega$ in $\mathcal{R}$ that governs the mixing of $N_I$ with all active flavours, hence   $U_{e}^2$, $U_{\mu }^2$ and $U_{\tau }^2$ cannot be too different in size. This allows to translate the BBN bounds into constraints on individual $U_{\alpha }^2$ and to combine the upper bounds on the individual $U_{\alpha }^2$ to obtain a stronger upper bound on $U_I^2$.
For example, in the mass range $M_I<m_K$ and for $n=2$ the lower bounds on $U_{e}^2$ and $U_{\mu}^2$ also strongly constrain the poorly measured $U_{\tau}^2$, allowing to impose an upper bound on $U_I^2$ \cite{Asaka:2014kia,Drewes:2016jae}. The window between these lower bounds and the upper bounds from experiments may soon be closed.
For the $n=3$ scenario under investigation here  the three angles $\omega_{ij}$ offer more freedom to change the size of the different  $U_{\alpha I}^2$. This is also the reason why leptogenesis with $n=3$ does not  necessarily  rely on a mass degeneracy \cite{Drewes:2012ma} and can be achieved for much larger $U_{\alpha I}^2$ \cite{Canetti:2014dka}.

\subsection{$N_{\rm eff}$ and "Dark Radiation"}\label{DRsection}
If sterile neutrinos are present in the primordial plasma during or after BBN, they affect the effective number of relativistic degrees of freedom $N_{\rm eff}$ in the primordial plasma, which is constrained at the epoch of photon decoupling (from CMB observations \cite{Ade:2013zuv}) and BBN (from observations of light elements \cite{Steigman:2010zz}).
If they come into thermal equilibrium while being relativistic, they themselves directly contribute to $N_{\rm eff}$. 
If the $N_I$ decay prior to photon decoupling, then this process injects entropy into the primordial plasma and leads to a deviation from thermal equilibrium, which leads to a deviation of $N_{\rm eff}$ from the SM prediction. 
If they decay during BBN, their energetic decay products dissociate some of the formed nuclei.
The $N_I$-equilibration can in principle be avoided if the mixing $U_I^2$ is sufficiently small. However, if one at the same time wants to explain the active neutrino masses via the seesaw mechanism, then at least some sterile species must have a sizable mixing with active neutrinos.

In the $n=2$ scenario these considerations impose a strict constraint on the $M_I$ \cite{Hernandez:2013lza}. 
The logic is as follows: If the $N_I$ generate active neutrino masses, then this imposes a lower bound on their mixing $U_I^2$ with active neutrinos. It turns out that his means that they necessarily come into thermal equilibrium in the early universe. To avoid an effect on $N_{\rm eff}$, they must have decayed before BBN.

In the $n=3$ model, there is no lower bound on the smallest $U_I^2$ because only two active neutrino mass differences have been measured, and the lightest $m_i$ could in principle be zero.
In that case, only two $N_I$ are required for the seesaw mechanism. The third one can have an arbitrarily small mixing, hence avoid equilibration in the early universe and any bounds on its mass from $N_{\rm eff}$. 
A quantitative analysis \cite{Hernandez:2014fha} reveals that this particle could have a mass of less than 100 MeV if 
the lightest active neutrino is lighter than $\sim 10^{-3}$ eV.
At least two of the $N_I$, however, must be heavier than 100 MeV.\footnote{Obviously this bound can be avoided if there is another source of active neutrino masses.} 
This is e.g. realised in the scenario A, in which only two heavy neutrinos $N_2$ and $N_3$ contribute significantly to the measured neutrino mass differences (and leptogenesis), while the $N_1$ is lighter ($M_1\sim$ keV), very feebly coupled and has a lifetime that is comparable to the age of the universe. Then $N_1$ would leave no trace in light element abundances or the cosmic microwave background because of its longevity and low number density, but can be a viable DM candidate \cite{Asaka:2005pn}. 
The feeble interaction required to achieve this implies that the light $N_I$ would effectively decouple from the system and can be ignored as far as leptogenesis and the seesaw mechanism are concerned. 

Finally, in spite of strong constraints \cite{Vincent:2014rja}, cosmological data still leaves some room for very light sterile neutrinos with masses at or below the eV scale \cite{Hernandez:2014fha}. The reason is essentially that CMB data is consistent with the SM prediction $N_{\rm eff}= 3.046$ \cite{Mangano:2005cc}, but cannot exclude $N_{\rm eff}=4$. We do not  consider the case of very light sterile neutrinos here, see  instead \cite{Gariazzo:2015rra} for a recent review.

\subsection{Leptogenesis}\label{Sec:leptogenesis}
If one requires the $N_I$ not only to explain active neutrino masses, but also to generate the BAU from their CP-violating interactions in the early universe, then this imposes further constraints onto their mass and mixing.
In scenarios with superheavy $N_I$, the BAU is generated via thermal leptogenesis during $N_I$ freezeout and decay \cite{Fukugita:1986hr}.
In the simplest scenario, this mechanism only works if $M_I>10^9$ GeV \cite{Davidson:2002qv,Hambye:2003rt}. Flavour effects can reduce this bound by 1-2 orders of magnitude \cite{Antusch:2009gn,Racker:2012vw}, which is still far beyond the mass range considered here.
Thermal leptogenesis can be achieved with TeV Majorana masses only if they are degenerate \cite{Pilaftsis:2003gt}. 
Sterile neutrinos with smaller $M_I$ can  generate the BAU from oscillations during their thermal production (rather than freezeout and decay) \cite{Akhmedov:1998qx,Asaka:2005pn}. 
A detailed analysis of this mechanism in the $n=2$ model has been performed in \cite{Canetti:2012kh,Canetti:2014dka}. It was found that the BAU can only be explained if the masses $M_I$ are degenerate at least at a level of $10^{-3}$, and lower and upper bounds on $U_I^2$ for successful leptogenesis were identified. 
Interestingly, leptogenesis in this scenario can be realised with the CP-violation in $U_\nu$ alone \cite{Canetti:2012kh}.
In the future, a circular collider \cite{Blondel:2014bra} or the SHiP experiment \cite{Alekhin:2015byh} could be used to search for  these particles and the CP-violation.
In the $n=3$ scenario there is no need for a mass degeneracy for successful leptogenesis \cite{Drewes:2012ma}, and the mixing angles $U_I^2$ can be large enough to be probed by existing experiments \cite{Canetti:2014dka}.
Recently these studies have been refined and extended by a number of authors 
\cite{Khoze:2013oga,Frigerio:2014ifa,Shuve:2014zua,Garbrecht:2014bfa,Abada:2015rta,Hernandez:2015wna,Kartavtsev:2015vto,Drewes:2016gmt,Hernandez:2016kel,Drewes:2016lqo,Hernandez:2016kel,Drewes:2016gmt,Hambye:2016sby,Drewes:2016jae}. 
Though the connection to leptogenesis is very interesting, we do not include any bounds from the baryon asymmetry of the Universe into the present analysis because the search for RH neutrinos in the GeV-mass range is of interest in its own right.

\section{Numerical analysis and results}\label{numericssection}

\subsection{Numerical method}\label{methodsubsec}  

\paragraph{Scanning strategy} -
In our analysis, we fix the mixing angles $\uptheta_{ij}$ in $U_\nu$ and the mass splittings $m_i^2-m_j^2$ to the best fit values of the global fit presented in \cite{Gonzalez-Garcia:2014bfa}.
For each of the benchmark scenarios we independently study the cases of ``normal hierarchy'' ($m_1^2\lesssim m_2^2<m_3^2$) and ``inverted hierarchy'' ($m_3^2<m_1^2\lesssim m_2^2 $).
Since the absolute scale of neutrino masses is not known, we focus on the two extreme cases that the lightest neutrino is massless or has a mass of $0.23$ eV, near the upper limit from cosmology~\cite{Ade:2013zuv}. We study both situations independently.
This allows to define eight independent benchmark scenarios that are distinguished by a combination of the choice of $M_{\rm max}$ ($M_{\rm max}=m_B$ or $M_{\rm max}=m_W$), the hierarchy of light neutrino masses and the absolute scale of neutrinos masses ($m_{\rm lightest}=0$ or $m_{\rm lightest}=0.23$ eV).
For each of these scenarios we randomise all remaining parameters\footnote{Present neutrino data already allows to identify a preferred region for $\delta$ \cite{Gonzalez-Garcia:2014bfa}. However, since the significance of this preference is still low, we treat all values of $\delta$ as equally likely.} in order to determine the largest and smallest values of $U_I^2$, $U_{e I}^2$, $U_{\mu I}^2$ and $U_{\tau I}^2$ that are consistent with all constraints that are specified in the previous sections.

\paragraph{Scan I: Agnostic main scan} - In our main scan, we generate $6\times 10^7$ random parameter sets for each of the eight benchmark scenarios.
We use flat probability distributions in the parametrisation defined in Section~\ref{sec:seesaw}, where
we restrict ${\rm Im}\omega_{ij}$ to the interval $-8< {\rm Im}\omega_{ij} < 8$.
The dependence on the other parameters $\delta$, $\alpha_1$, $\alpha_2$ and ${\rm Re}\omega_{ij}$ is periodic. 
The use of flat probability distributions allows to interpret the distribution (or ``scatter'') of extremal $U_{\alpha I}^2$ in nearby mass bins, which characterises the degree of convergence of the scan towards some combined upper bound,
as a measure for the size of the allowed parameter regions in which $U_{\alpha I}^2$ of a given magnitude can be realised. 
This is an interesting piece of information, which otherwise could not be extracted from the projection of a high dimensional parameter space onto the mass-mixing plane. The interpretation in terms of parameter space volumes should of course be taken with a grain of salt because it depends on the parametrisation, which can be chosen arbitrarily.
Here we use the Casas-Ibarra parametrisation, which is amongst the most frequently used ones in the literature.

If the difference $\delta M$ between two masses $M_I$ and $M_J$ is smaller than the resolution of a direct search experiment, then the signals from $N_I$ and $N_J$ cannot be distinguished.
To take account of this, we apply the experimental bound to the sum of the signals whenever this happens.
For simplicity, we generally estimate the experimental mass resolution as 10 MeV. 
During the propagation within the detector, quantum coherences can be neglected as long as $l \frac{ \delta M^2}{4 E} \gg1$, where $l$ is the detector length and $E$ the particle energy. 
That is $N_I$ and $N_J$ behave like two well-defined individual on-shell particles.
The highly mass-degenerate case for which this not satisfied can lead to the interesting resonant leptogenesis scenarios \cite{Pilaftsis:2003gt,Garny:2011hg,Garbrecht:2011aw,Garbrecht:2014aga,Iso:2013lba,Dev:2014wsa,Hohenegger:2014cpa}, but the propagation in the laboratory may require a more sophisticated treatment \cite{Gorbunov:2007ak,Boyanovsky:2014una,Boyanovsky:2014lqa}.
In the extreme case that $\delta M$ is smaller than the intrinsic width of the heavy neutrino resonances, 
one cannot define two individual particles $N_I$ and $N_J$.
We exclude this case by requiring that all quantities $\Gamma_I\equiv(F^\dagger F)_{II}M_I/(16\pi)$ are smaller than the smallest mass splitting amongst the heavy neutrinos. One can loosely interpret $\Gamma_I$ as the width of $N_I$.\footnote{Note, however, that the lifetimes (and hence widths) of the quantum states should in principle be defined in the flavour basis where $F$ is diagonal, while physical particles correspond to the basis vectors in the basis where $M_M$ is diagonal.}
Amongst the points that are consistent with all these constraints we identify those that maximise the mixing angles $U_{e I}^2$, $U_{\mu I}^2$, $U_{\tau I}^2$ and their sum $U_I^2=U_{e I}^2+U_{\mu I}^2+U_{\tau I}^2$ in order
to determine upper bounds for these quantities. In addition, we identify the lower bound on $U_I^2$. As discussed in Section~\ref{sec:minimalmixing}, it is not possible to impose a lower bound on the individual $U_{\alpha I}^2$.

\paragraph{Scan II: The $B-L$ conserving parameter region} - In an additional scan we choose the probability distributions in a way that focuses on the $B-L$-conserving region described in Sec~\ref{BminusL}. 
We randomise $M_1$ and $M_2$ with a flat probability distribution between $M_{\rm min}$ and $M_{\rm max}$.
To determine $M_3$, we use a flat probability distribution in $\log_{10}|M_2-M_3|$ between $0$ and $-24$.
We randomise all  other parameters introduced in Sec.~\ref{parametrisationsubsec} in a two-step process. 
In the first step, we generate random values for these in the same way as in the main scan. We then multiply all parameters except $\omega_{23}$ by $x^3$, where $x$ is a random number between 0 and 1. This prefers parameter regions with ${\rm Im}\omega_{12}<1$ and ${\rm Im}\omega_{13}<1$, in which $N_1$ has little mixing with other flavours. 
On the other hand, it prefers degenerate $M_2$ and $M_3$ and relatively large ${\rm Im}\omega_{23}$ leading to large mixing $(U_N)_{23}$ amongst the states $\nu_{R 2}$ and $\nu_{R 3}$ in the degenerate mass eigenstates $N_2$ and $N_3$ along with comparably large $U_2^2$ and $U_3^2$, cf. (\ref{BmLcons})-(\ref{equalcouplings}).  
This which includes the point where $N_2$ and $N_3$ can be combined into a pseudo-Dirac spinor $\Psi_N\equiv (iN_2+N_3)/\sqrt{2}$ and $N_1$ decouples.


\paragraph{Validity of the  numerical approach} - Since our approach is based on a stochastic scan, it is not possible to prove that the extremal points we obtain are indeed the global extrema of the active-sterile mixings for given $M_I$ that are consistent with experiment.
We use the numerical convergence as a criterion to judge whether there the maximal and minimal mixings we find are close to the global extrema and therefore approximately trace strict upper/lower bounds, or whether the allowed parameter space volume is simply shrinking so rapidly for large $U_{\alpha I}^2$ that the scan fails to find the extrema.\footnote{We use a scanning strategy in which the parameters are randomly chosen without correlation to each other or the previous choice. In order to explore narrow ``funnels'', an adaptive Monte Carlo approach would certainly be more promising. On the other hand, such methods might be more likely to miss out on isolated regions. A future comparison of different scanning techniques would be of interest.}
In this context, it is helpful to keep dependencies of the indirect bounds on the parameters in mind.

\paragraph{Parametric dependencies} - The parameter space for $n=3$ is rather high-dimensional, which makes an analytic understanding of the constraints difficult. 
There are a few simple rules to understand the parameter dependence of the mixings $U_{\alpha I}^2$ for given $M_I$. For $n=2$ these have e.g. been studied in detail in \cite{Ruchayskiy:2011aa,Asaka:2011pb,Ibarra:2011xn,LopezPavon:2012zg}. For $n=3$ they are more complicated, but general tendencies can still be identified easily. 
\begin{itemize}
\item 
 In general, the dependence of the $U_{\alpha I}^2$ upon the parameters $\delta$, $\alpha_1$, $\alpha_2$ and the ${\rm Re}\omega_{ij}$ is relatively weak (though their precise value does matter if one aims to construct very special points, e.g.  a vanishing mixing $U_{\alpha I}^2$). 
\item As already pointed out in \cite{Asaka:2011pb,Shuve:2014zua,Ruchayskiy:2011aa,Canetti:2014dka}, the magnitude of the active-sterile mixing is mostly controlled by the ${\rm Im}\omega_{ij}$ because the elements of $F_{\alpha I}$ depend exponentially upon these parameters. Small $|{\rm Im}\omega_{ij}|<1$ tend to give values for $F_{\alpha I}$ near the ``naive seesaw value'' $F_0$, typically leading to small $U_{\alpha I}^2$ and small branching ratios in experiments. This situation is considered ``natural'' by many authors, though this statement of course depends on the parametrisation and an arbitrary definition of ``naturalness''.
\item Large  $|{\rm Im}\omega_{ij}|>1$ lead to comparably large individual $U_{\alpha I}^2$. This makes it easier to find the $N_I$ experimentally \cite{Gorbunov:2007ak,Kersten:2007vk}. The smallness of the $m_i$ in this case is not (only) due to the smallness of individual elements of $m_\nu$, but requires cancellations amongst them that lead to  small eigenvalues $m_i^2$ of $m_\nu m_\nu^\dagger$.
The use of the loop-corrected seesaw formula (\ref{activemass}) ensures that this cancellation holds at 1-loop level.
A particularly interesting and technically natural realisation of the large mixing scenario is provided by the approximately $B-L$ conserving models discussed in \ref{BminusL}.
\end{itemize}

\paragraph{Most saturated bounds} - 
It is interesting to see which indirect bound is most saturated for the parameter choice that leads to the largest allowed active-sterile mixing for given $M_I$.
To address this question, we introduce an (arbitrarily chosen, but well motivated) measure of saturation for each indirect constraint (labelled by an index $i$), which is quantified by a number $S_i$ with $0\leq S_i \leq 1$
for each of the allowed points.
For an observable $X_i$ that is predicted to vanish in the SM and for which an upper bound exists from experiment (e.g.\ $0\nu\beta\beta$, LFV lepton decays), $S_i$ is defined as
\begin{equation}\label{SiDef1}
S_i = \frac{X_{i,N}^2}{X_{i,\rm{exp}}^2},
\end{equation}
where $X_{i,\rm{exp}}$ is the experimental limit and $X_{i,N}$ the predicted value in the presence of heavy neutrinos $N$ with a given set $\{F,M_M\}$ of masses and couplings. 
For those observables that are non-zero in the SM, $S_i$ is defined via the deviation from the experimental value
\begin{equation}
S_i=\frac{(X-X_{i,\rm{exp}})^2}{(3
\sigma_{i,\rm{exp}})^2}
\end{equation}
Here $\sigma_{\rm exp}$ is the experimental error bar.

\subsection{Results}\label{ResultsSubSec}
The figures \ref{ClNH-1:fig}-\ref{EsIH-2:fig} show the results of our main scan. The upper panels in each figure show extremal values of the active-sterile mixing ($U_I^2$ or individual $U_{\alpha I}^2$), as explained in the following paragraphs. 
The colour of the points indicates which bound is the most saturated one for the displayed point.
The brightness is determined by the value corresponding $S_i$ and thus indicates how saturated that bound is (the brighter, the more saturated). The colour code is
\begin{itemize}
\item blue: LFV lepton decays,
\item red: neutrinoless double $\beta$ decay,
\item turquoise: EW precision data,
\item yellow: CKM unitarity,
\item green: lepton universality in pion, kaon and tauon decays.
\end{itemize}
The type of LFV decay and the process from which lepton universality is constrained (pion, kaon or tauon decays) are not distinguished in the upper panels. 

The lower panels show the degree of the saturation as quantified by the value of $S_i$ for the parameter choice that maximises the quantity plotted in the corresponding upper panel ($U_I^2$ or individual $U_{\alpha I}^2$). The colour code is the same as in the upper panels, but here the brightness is not related to the degree of saturation. Instead, it indicates type of LFV decay and the process from which lepton universality is constrained (pion, kaon or tauon decays),
\begin{itemize}
\item light blue: $\mu\rightarrow e\gamma$,
\item medium blue: $\tau\rightarrow\mu\gamma$,
\item dark blue: $\tau\rightarrow e\gamma$,
\item light green: lepton universality in pion decays,
\item medium green: lepton universality in kaon decays,
\item dark green: lepton universality in tauon decays.
\end{itemize}

\paragraph{Scenario B ($m_\pi<M_I<m_B$)} -  
We first focus on the scenario where $N_I$ can be produced in meson decays.
The dots in the upper panels in Fig.~\ref{ClNH-1:fig} show the extremal values of $U_I^2$ and $U_{eI}^2$ for given $M_I$ that are consistent with all constraints in the scenario with $m_{\rm lightest}=0.23$ eV, normal hierarchy and $m_\pi<M_I<m_B$. 
The colourful dots indicate the largest active-sterile mixing in each mass bin. The thick black line represents the upper bound from direct searches. 
The black dots show the smallest value found for $U_I^2$. 
The lower panels show the corresponding $S_i$, as explained in the previous paragraph.
Fig.~\ref{ClNH-2:fig} shows the maximal $U_{\mu I}^2$ and $U_{\tau I}^2$ we found and should be read in the same way.
The following figures display the same quantities for $m_{\rm lightest}=0.23\,{\rm eV}$ with inverted hierarchy (figures~\ref{ClIH-1:fig} and \ref{ClIH-2:fig}), $m_{\rm lightest}=0$ with normal hierarchy (figures~\ref{CsNH-1:fig} and \ref{CsNH-2:fig}) and
$m_{\rm lightest}=0$ with inverted hierarchy (figures~\ref{CsIH-1:fig} and \ref{CsIH-2:fig}).

\paragraph{Scenario C ($m_\pi < M_I < m_W$)} - 
The remaining figures show the same quantities (extremal $U_I^2$, $U_{e I}^2$, $U_{\mu I}^2$, $U_{\tau I}^2$ and the corresponding $S_i$) we found in the mass range $m_\pi < M_I < m_W$  for $m_{\rm lightest}=0.23\,{\rm eV}$ with normal hierarchy (Figs.~\ref{ElNH-1:fig} and \ref{ElNH-2:fig}), $m_{\rm lightest}=0.23\,{\rm eV}$ with inverted hierarchy (Figs.~\ref{ElIH-1:fig} and \ref{ElIH-2:fig}), $m_{\rm lightest}=0$ with normal hierarchy (Figs.~\ref{EsNH-1:fig} and \ref{EsNH-2:fig}) and
$m_{\rm lightest}=0$ with inverted hierarchy (Figs.~\ref{EsIH-1:fig} and \ref{EsIH-2:fig}).

\begin{landscape}
\begin{figure}
\begin{center}
\begin{tabular}{c c}
\includegraphics[width=10cm]{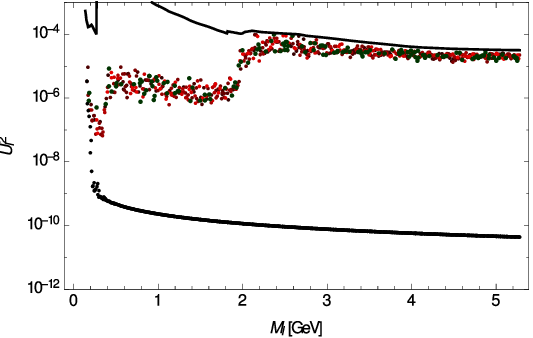}
&
\includegraphics[width=10cm]{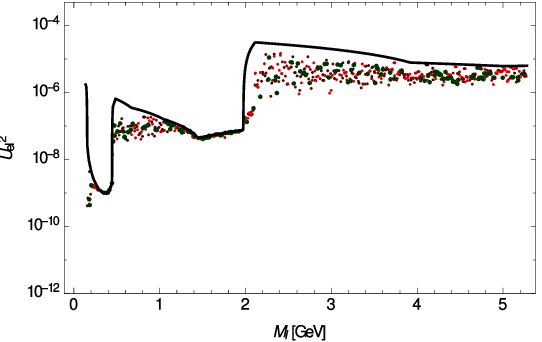}
\\
\includegraphics[width=10cm]{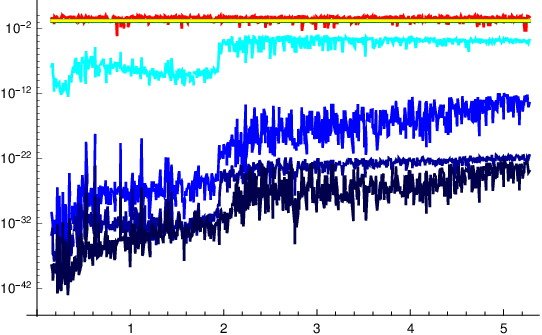}
&
\includegraphics[width=10cm]{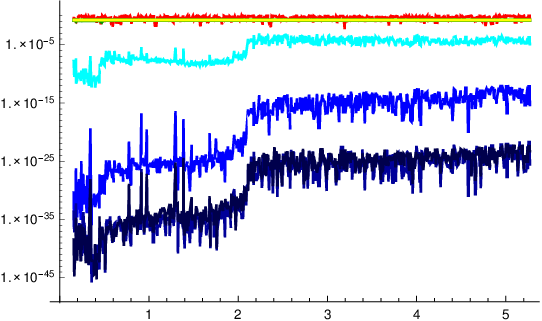}
\end{tabular}
\caption{\emph{Left panels}: The upper panel shows the largest allowed $U_I^2$ (colourful dots) and smallest allowed $U_I^2$ (black dots) found in our main scan for $m_{\rm lightest}=0.23\,{\rm eV}$, normal hierarchy and $m_\uppi<M_I<m_B$.
The black line represents the upper bound from direct searches. The lower panel shows the saturation measures $S_i$, as explained in the main text. 
\emph{Right panels}: The upper panel shows the largest allowed $U_{e I}^2$ (colourful dots) found in our main scan for $m_{\rm lightest}=0.23\,{\rm eV}$ and normal hierarchy. The black line represents the upper bound from direct searches.
 The lower panel shows the saturation measures $S_i$, as explained in the main text. 
\label{ClNH-1:fig}}
\end{center}
\end{figure}
\begin{figure}
\begin{center}
\begin{tabular}{c c}
\includegraphics[width=10cm]{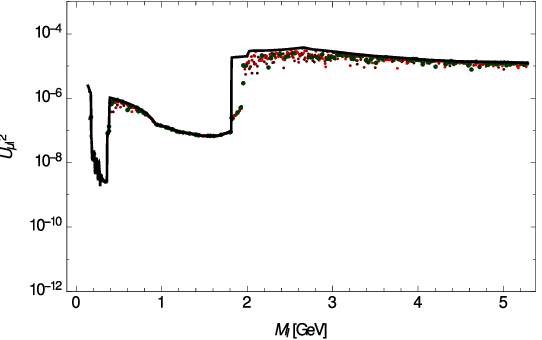}
&
\includegraphics[width=10cm]{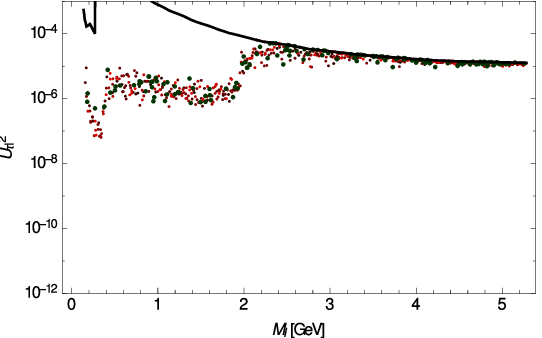}
\\
\includegraphics[width=10cm]{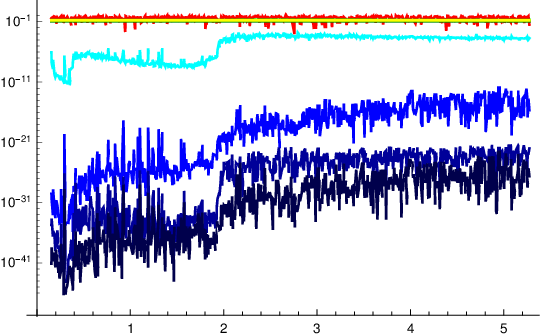}
&
\includegraphics[width=10cm]{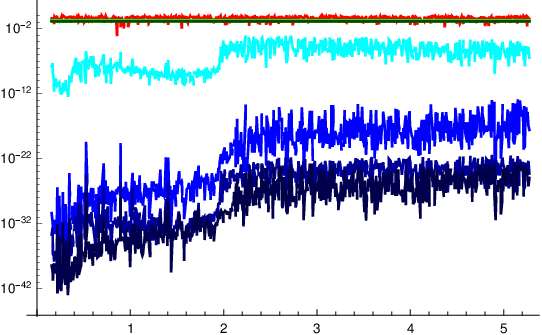}
\end{tabular}
\caption{\emph{Left panels}: The upper panel shows the largest allowed $U_{\mu I}^2$ (colourful dots) found in our main scan for $m_{\rm lightest}=0.23\,{\rm eV}$ and normal hierarchy.
The black line represents the upper bound from direct searches. The lower panel shows the saturation measures $S_i$, as explained in the main text. 
\emph{Right panels}: The upper panel shows the largest allowed $U_{\tau I}^2$ (colourful dots) found in our main scan for $m_{\rm lightest}=0.23\,{\rm eV}$, normal hierarchy and $m_\uppi<M_I<m_B$. The black line represents the upper bound from direct searches.
 The lower panel shows the saturation measures $S_i$, as explained in the main text.
\label{ClNH-2:fig}}
\end{center}
\end{figure}

\begin{figure}
\begin{center}
\begin{tabular}{c c}
\includegraphics[width=10cm]{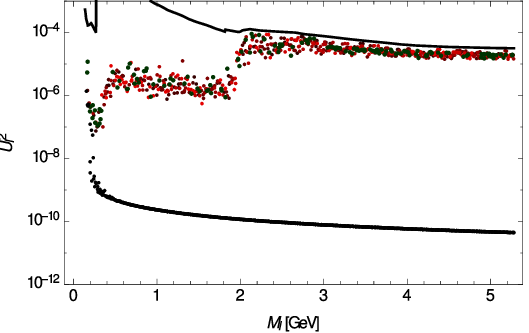}
&
\includegraphics[width=10cm]{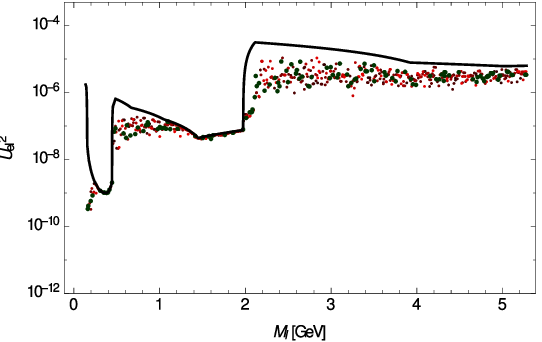}
\\
\includegraphics[width=10cm]{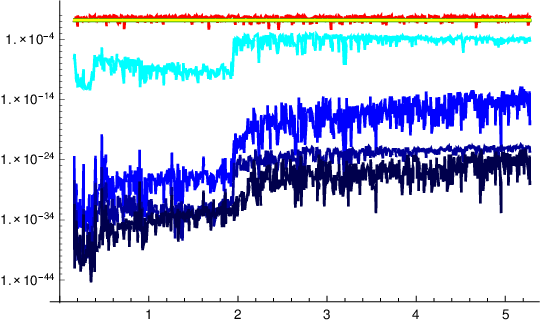}
&
\includegraphics[width=10cm]{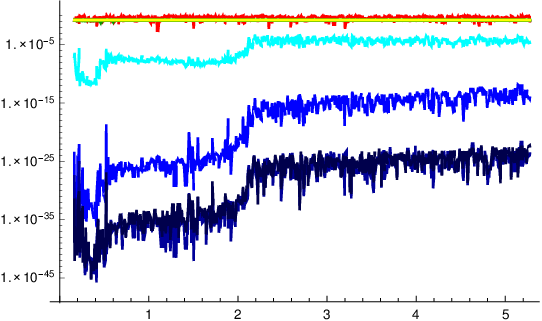}
\end{tabular}
\caption{Same as Fig.~\ref{ClNH-1:fig}, but for $m_{\rm lightest}=0.23\,{\rm eV}$ and inverted hierarchy.\label{ClIH-1:fig}}
\end{center}
\end{figure}
\begin{figure}
\begin{center}
\begin{tabular}{c c}
\includegraphics[width=10cm]{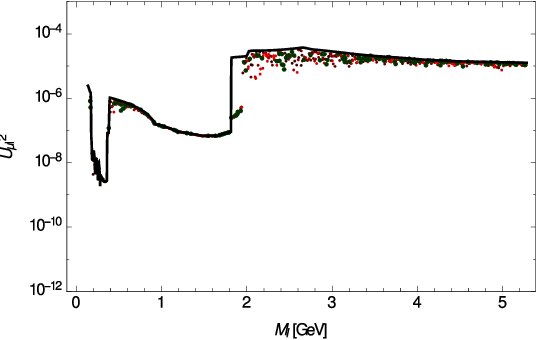}
&
\includegraphics[width=10cm]{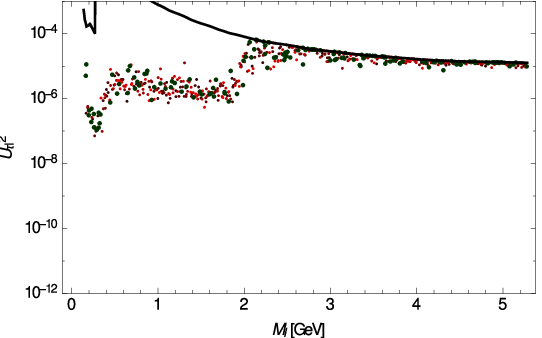}
\\
\includegraphics[width=10cm]{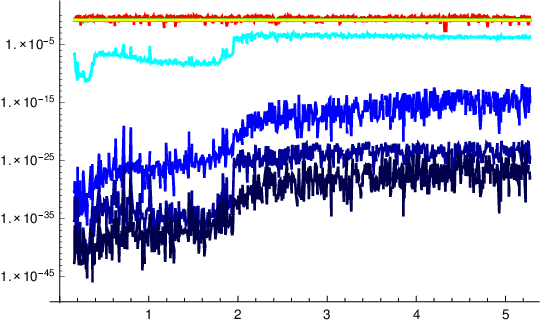}
&
\includegraphics[width=10cm]{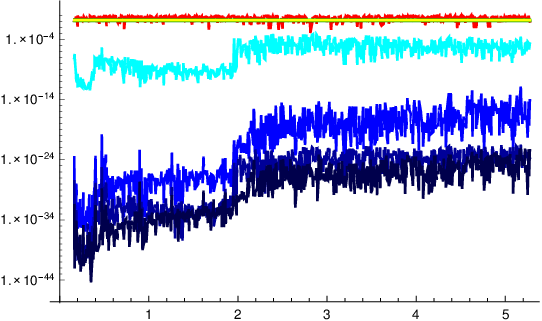}
\end{tabular}
\caption{Same as Fig.~\ref{ClNH-2:fig}, but for $m_{\rm lightest}=0.23\,{\rm eV}$ and inverted hierarchy.\label{ClIH-2:fig}}
\end{center}
\end{figure}

\begin{figure}
\begin{center}
\begin{tabular}{c c}
\includegraphics[width=10cm]{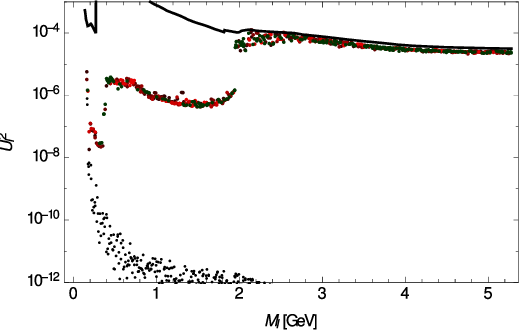}
&
\includegraphics[width=10cm]{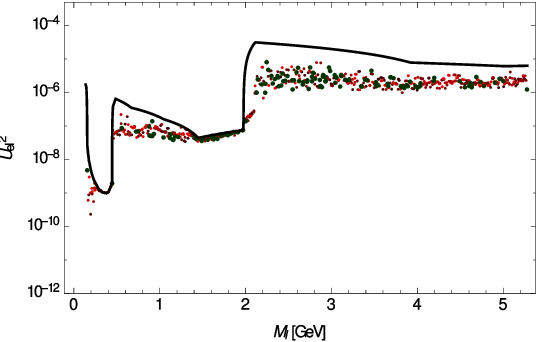}
\\
\includegraphics[width=10cm]{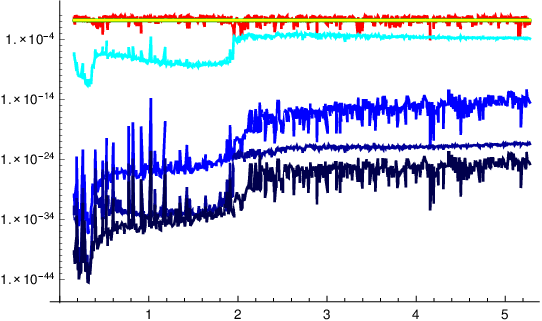}
&
\includegraphics[width=10cm]{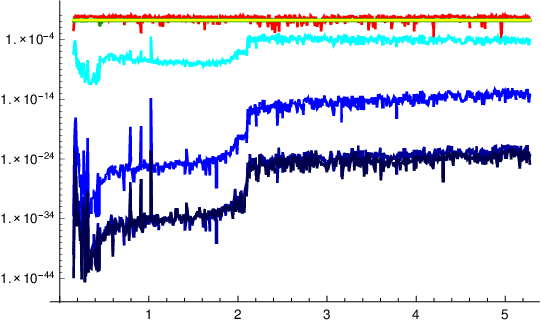}
\end{tabular}
\caption{Same as Fig.~\ref{ClNH-1:fig}, but for
$m_{\rm lightest}=0$ and normal hierarchy\label{CsNH-1:fig}.}
\end{center}
\end{figure}
\begin{figure}
\begin{center}
\begin{tabular}{c c}
\includegraphics[width=10cm]{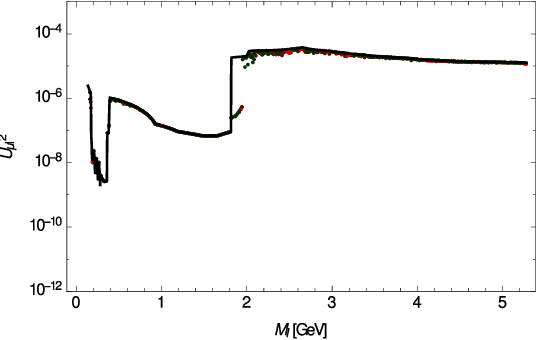}
&
\includegraphics[width=10cm]{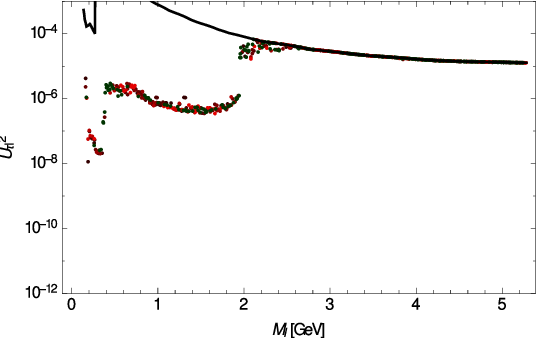}
\\
\includegraphics[width=10cm]{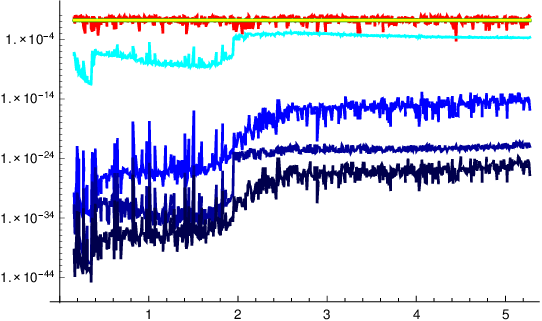}
&
\includegraphics[width=10cm]{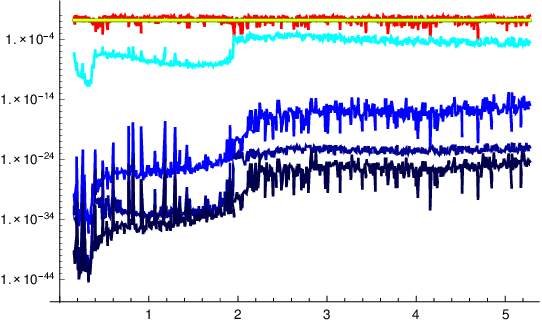}
\end{tabular}
\caption{Same as Fig.~\ref{ClNH-2:fig}, but for
$m_{\rm lightest}=0$ and normal hierarchy\label{CsNH-2:fig}.}
\end{center}
\end{figure}

\begin{figure}
\begin{center}
\begin{tabular}{c c}
\includegraphics[width=10cm]{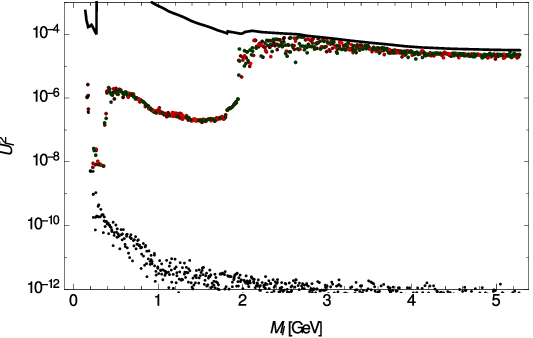}
&
\includegraphics[width=10cm]{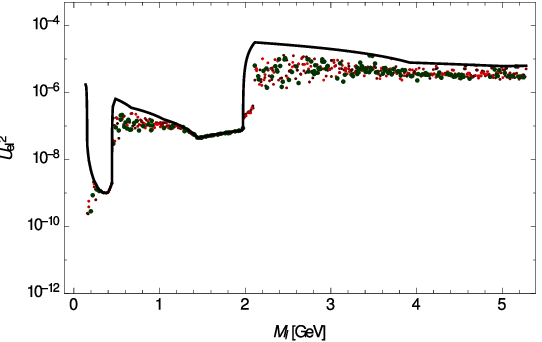}
\\
\includegraphics[width=10cm]{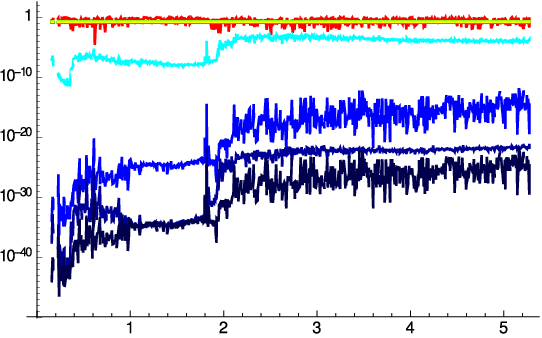}
&
\includegraphics[width=10cm]{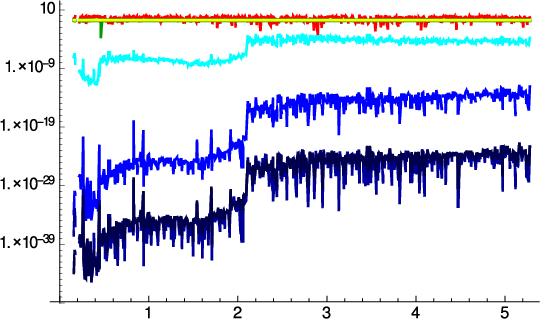}
\end{tabular}
\caption{Same as Fig.~\ref{ClNH-1:fig}, but for
$m_{\rm lightest}=0$ and inverted hierarchy\label{CsIH-1:fig}.}
\end{center}
\end{figure}
\begin{figure}
\begin{center}
\begin{tabular}{c c}
\includegraphics[width=10cm]{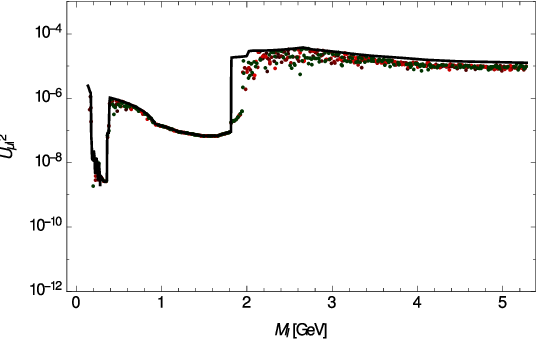}
&
\includegraphics[width=10cm]{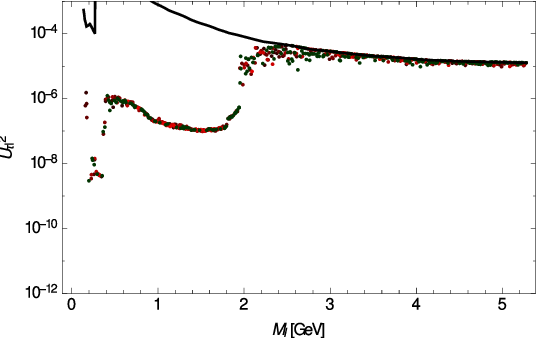}
\\
\includegraphics[width=10cm]{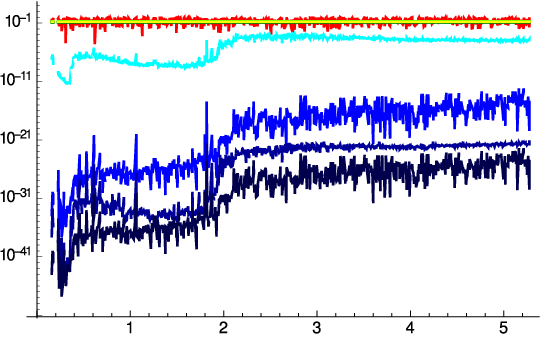}
&
\includegraphics[width=10cm]{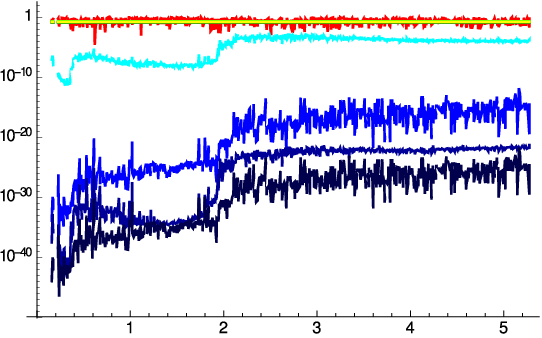}
\end{tabular}
\caption{Same as Fig.~\ref{ClNH-2:fig}, but for
$m_{\rm lightest}=0$ and inverted hierarchy\label{CsIH-2:fig}.}
\end{center}
\end{figure}


\begin{figure}
\begin{center}
\begin{tabular}{c c}
\includegraphics[width=10cm]{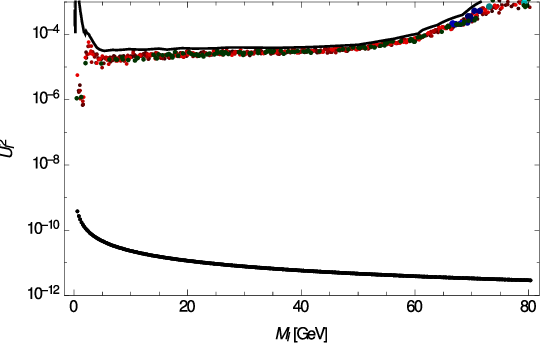}
&
\includegraphics[width=10cm]{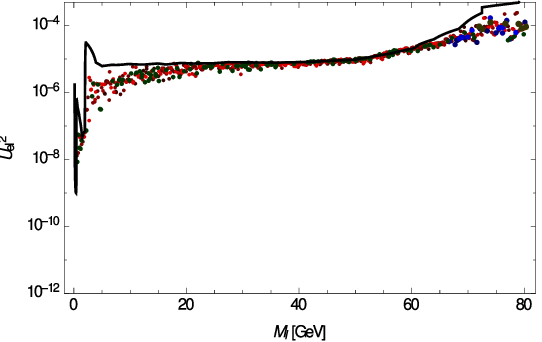}
\\
\includegraphics[width=10cm]{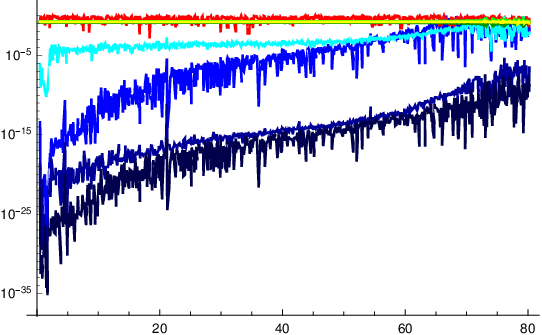}
&
\includegraphics[width=10cm]{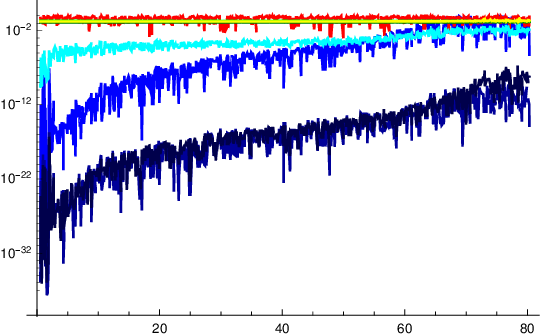}
\end{tabular}
\caption{Same as Fig.~\ref{ClNH-1:fig} ($m_{\rm lightest}=0.23\,{\rm eV}$, normal hierarchy), but with $m_\uppi<M_I<m_W$.
\label{ElNH-1:fig}}
\end{center}
\end{figure}
\begin{figure}
\begin{center}
\begin{tabular}{c c}
\includegraphics[width=10cm]{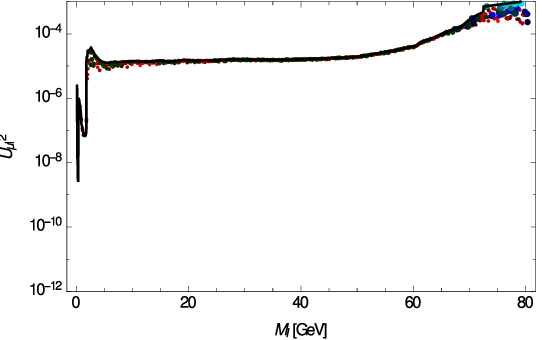}
&
\includegraphics[width=10cm]{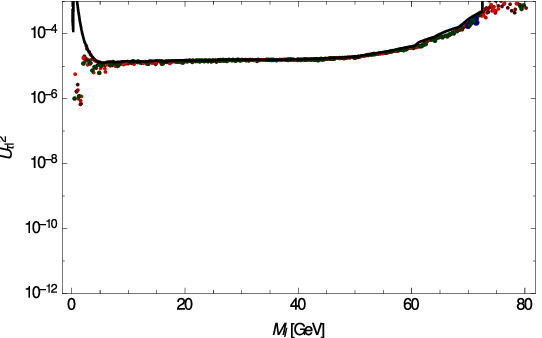}
\\
\includegraphics[width=10cm]{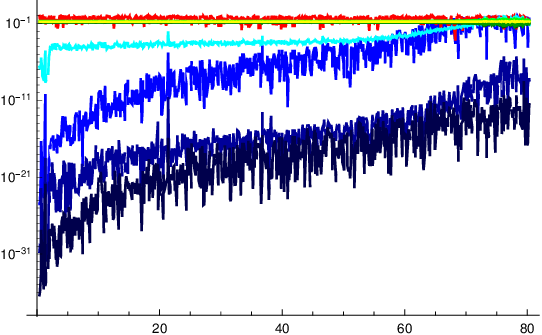}
&
\includegraphics[width=10cm]{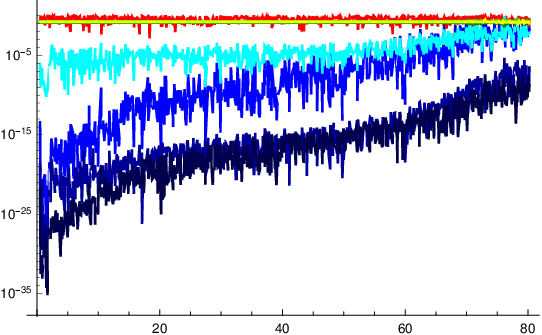}
\end{tabular}
\caption{Same as Fig.~\ref{ClNH-2:fig} ($m_{\rm lightest}=0.23\,{\rm eV}$, normal hierarchy), but with $m_\uppi<M_I<m_W$..
\label{ElNH-2:fig}}
\end{center}
\end{figure}


\begin{figure}
\begin{center}
\begin{tabular}{c c}
\includegraphics[width=10cm]{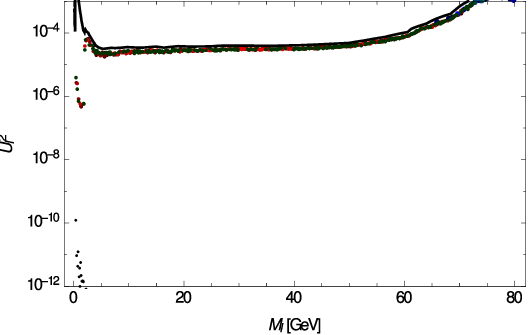}
&
\includegraphics[width=10cm]{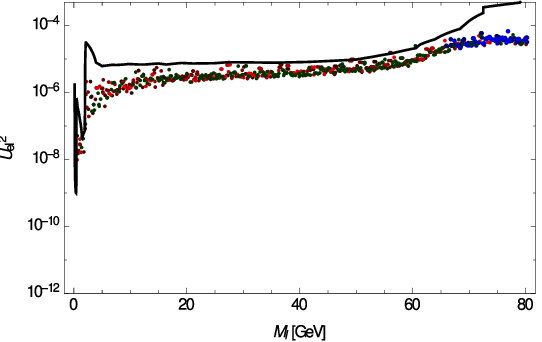}
\\
\includegraphics[width=10cm]{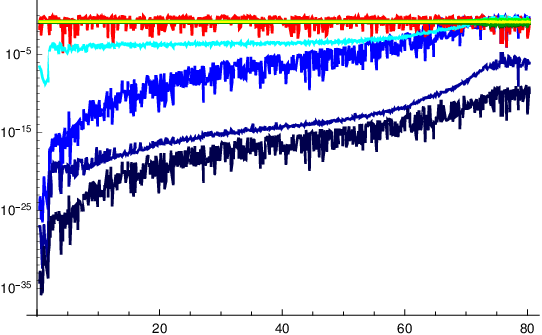}
&
\includegraphics[width=10cm]{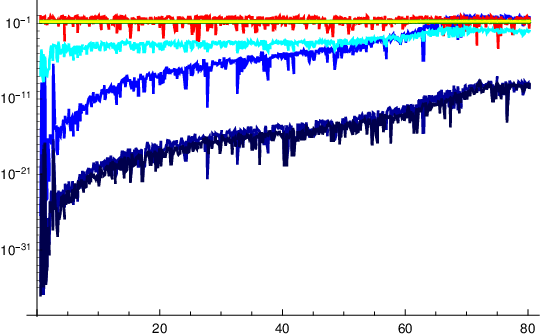}
\end{tabular}
\caption{Same as Fig.~\ref{ClNH-1:fig}, but for
$m_{\rm lightest}=0$, normal hierarchy and $m_\uppi<M_I<m_W$.
\label{EsNH-1:fig}}
\end{center}
\end{figure}
\begin{figure}
\begin{center}
\begin{tabular}{c c}
\includegraphics[width=10cm]{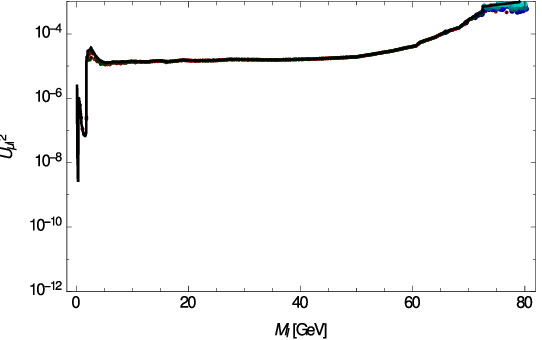}
&
\includegraphics[width=10cm]{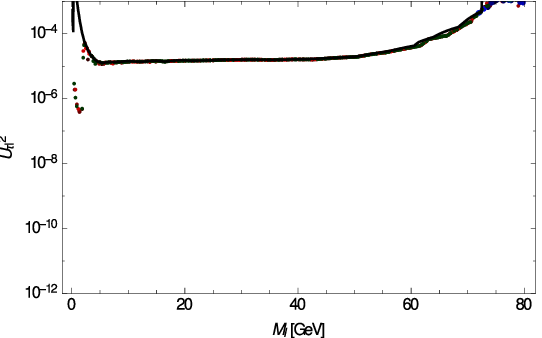}
\\
\includegraphics[width=10cm]{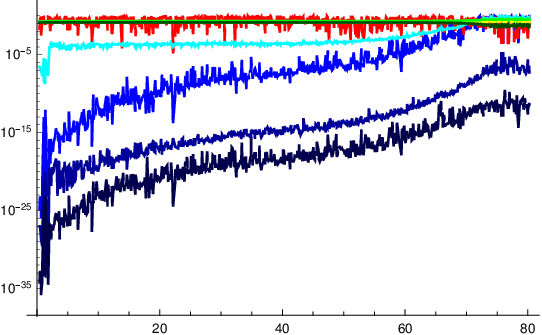}
&
\includegraphics[width=10cm]{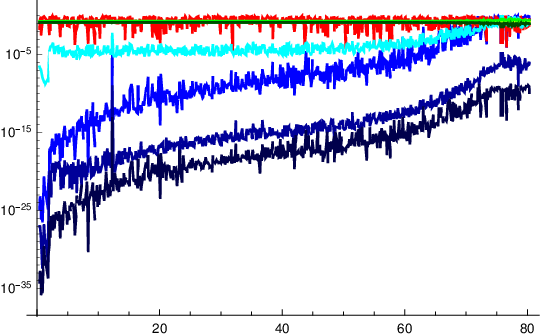}
\end{tabular}
\caption{Same as Fig.~\ref{ClNH-2:fig}, but for
$m_{\rm lightest}=0$, normal hierarchy and $m_\uppi<M_I<m_W$.\label{EsNH-2:fig}}
\end{center}
\end{figure}


\begin{figure}
\begin{center}
\begin{tabular}{c c}
\includegraphics[width=10cm]{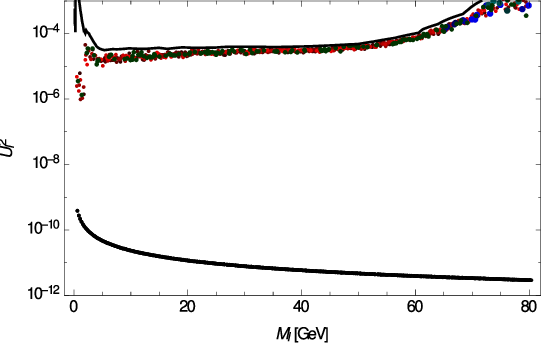}
&
\includegraphics[width=10cm]{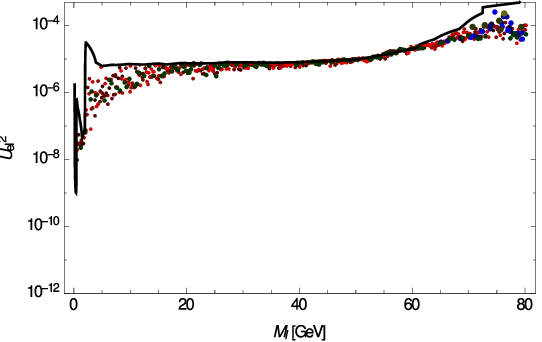}
\\
\includegraphics[width=10cm]{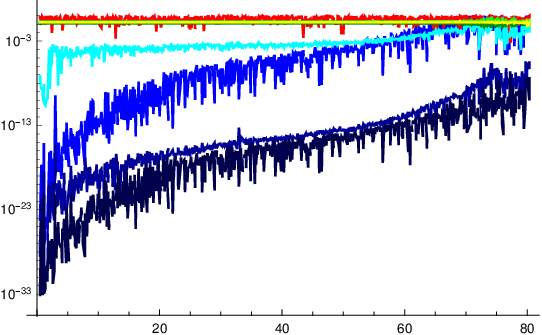}
&
\includegraphics[width=10cm]{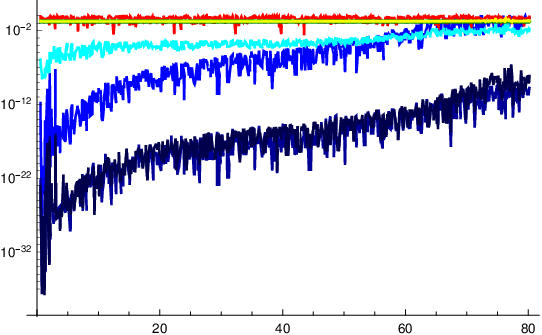}
\end{tabular}
\caption{Same as Fig.~\ref{ClNH-1:fig}, but for
$m_{\rm lightest}=0.23\,{\rm eV}$, inverted hierarchy  and $m_\uppi<M_I<m_W$.\label{ElIH-1:fig}}
\end{center}
\end{figure}
\begin{figure}
\begin{center}
\begin{tabular}{c c}
\includegraphics[width=10cm]{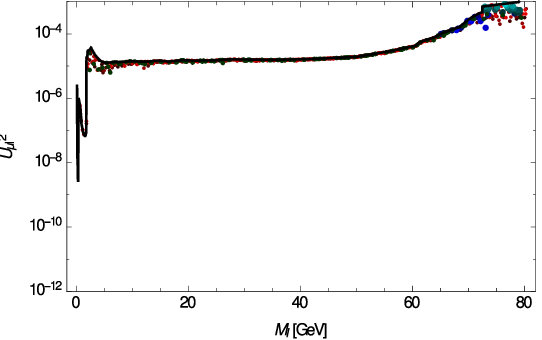}
&
\includegraphics[width=10cm]{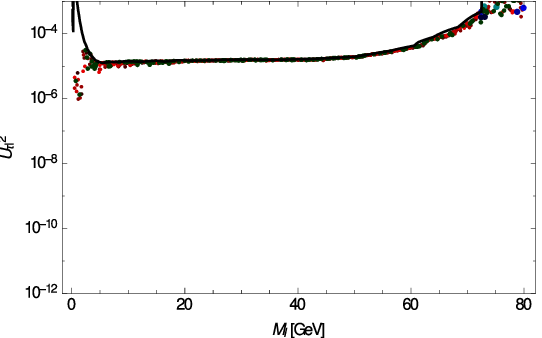}
\\
\includegraphics[width=10cm]{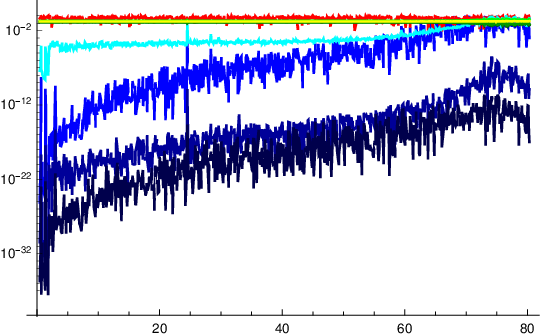}
&
\includegraphics[width=10cm]{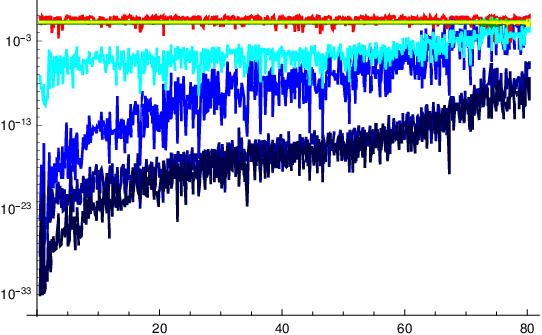}
\end{tabular}
\caption{Same as Fig.~\ref{ClNH-2:fig}, but for
$m_{\rm lightest}=0.23\,{\rm eV}$, inverted hierarchy and $m_\uppi<M_I<m_W$.\label{ElIH-2:fig}}
\end{center}
\end{figure}

\begin{figure}
\begin{center}
\begin{tabular}{c c}
\includegraphics[width=10cm]{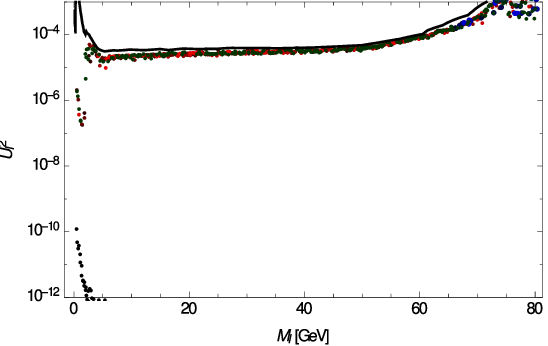}
&
\includegraphics[width=10cm]{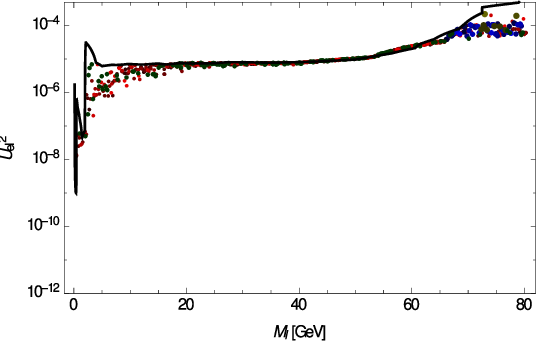}
\\
\includegraphics[width=10cm]{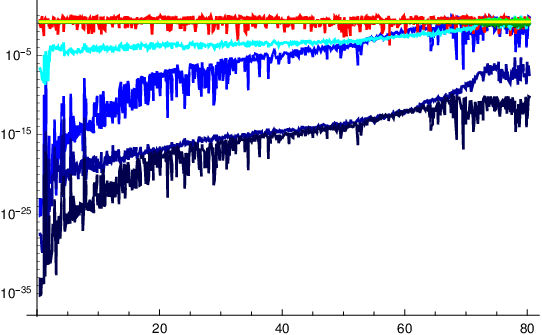}
&
\includegraphics[width=10cm]{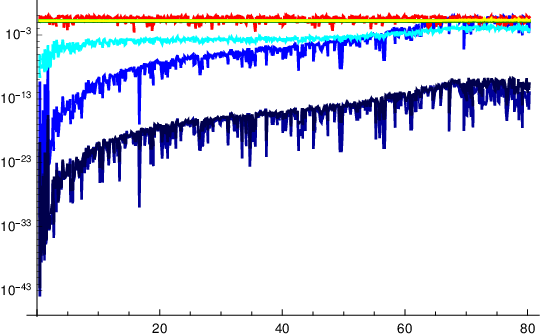}
\end{tabular}
\caption{Same as Fig.~\ref{ClNH-1:fig}, but for
$m_{\rm lightest}=0$, inverted hierarchy and $m_\uppi<M_I<m_W$.\label{EsIH-1:fig}}
\end{center}
\end{figure}
\begin{figure}
\begin{center}
\begin{tabular}{c c}
\includegraphics[width=10cm]{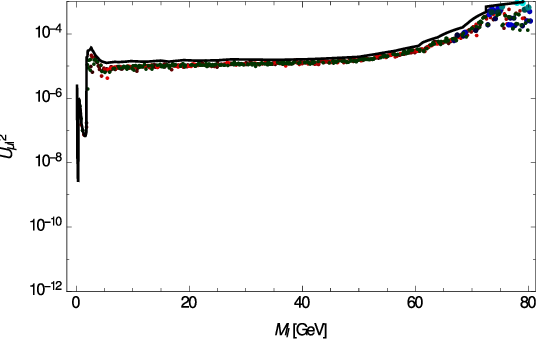}
&
\includegraphics[width=10cm]{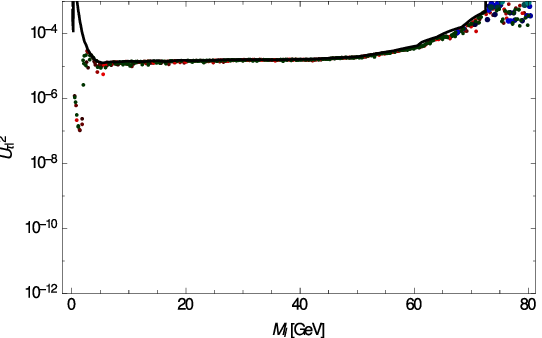}
\\
\includegraphics[width=10cm]{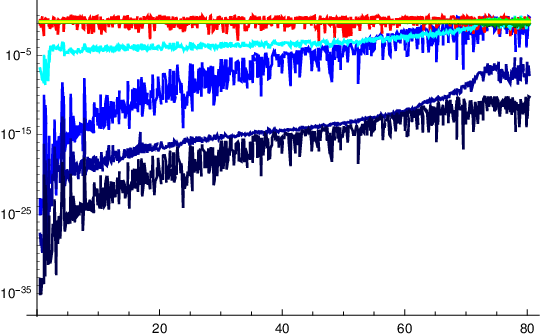}
&
\includegraphics[width=10cm]{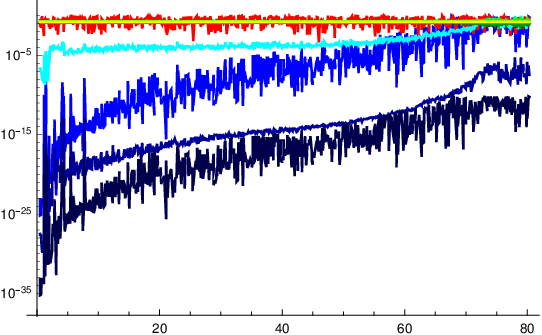}
\end{tabular}
\caption{Same as Fig.~\ref{ClNH-2:fig}, but for
$m_{\rm lightest}=0$, inverted hierarchy  and $m_\uppi<M_I<m_W$.\label{EsIH-2:fig}}
\end{center}
\end{figure}
\end{landscape}

\section{Discussion}\label{sec:discussion}
In the following we discuss the results of our scans and their physical interpretation.
\subsection{Smallest and largest active-sterile mixing}\label{DiscussionSmallLargeMixing} 

\paragraph{Smallest mixing} - Several factors conspire to impose a lower bound on $U_I^2$ in the seesaw model.
\begin{itemize}
\item The BBN-motivated requirement of a lifetime $\tau_I<0.1\,$s always imposes a lower bound on $U_I^2$. Due to the shorter lifetime $\tau_I$ of heavier $N_I$ ($\tau_I\propto M_I^{-5}$), this lower bound falls off rapidly and is practically irrelevant for any near future experiment if $N_I$ is heavier than the kaon.
This bound applies under the assumption that the $N_I$ came into thermal equilibrium in the early universe, or were at least sufficiently abundant that their decay would affect BBN. 
This is usually the case if $m_{\rm lightest}>10^{-3}$ eV \cite{Hernandez:2014fha}.
It can be circumvented in the scenario with one massless active neutrino ($m_{\rm lightest}=0$) and one decoupled sterile neutrino discussed in Sec.~\ref{DiscussionSmallLargeMixing}. 
\item For $m_{\rm lightest}\neq 0$ there is also a lower bound on $U_I^2$ from neutrino oscillation data, which roughly scales as $\propto m_{\rm lightest}$ and $\propto M_I^{-1}$.
For relatively large $m_{\rm lightest} \sim 0.23$ eV this lower bound is interesting because it might be reached by a realistic improvement of the SHiP experimental proposal \cite{Anelli:2015pba}. 
In contrast to $n=2$, there is no strict lower bound on individual $U_{\alpha I}^2$ from neutrino oscillation data for $n>2$.
\item If one required the $N_I$ not only to explain neutrino masses, but also the observed BAU by leptogenesis, then this imposes an upper as well as a lower bound on $U_I^2$. These have been studied in detail for Scenario A \cite{Canetti:2010aw,Canetti:2012vf,Canetti:2012kh,Hernandez:2015wna,Drewes:2016lqo,Drewes:2016gmt,Hernandez:2016kel,Drewes:2016jae} and partly been investigated for Scenario C \cite{Canetti:2014dka,Hernandez:2015wna}. A detailed investigation of the lower bound would be particularly interesting for the case that the lightest neutrino is very light and no lower bound can be obtained from neutrino oscillation data.
\end{itemize}

\paragraph{Largest mixing} -
Below we summarise the mass regions in which the combination of all constraints imposes a stronger upper bound on $U_{\alpha I}^2$ than the direct searches alone. 
\begin{itemize}
\item In several scenarios a small number of points near  $M_I\sim 100$ MeV suggests an upper limit on $U_{e I}^2$ or $U_{\mu I}^2$ that is much stronger than direct search bounds. 
However, these do not correspond to valid upper bounds, i.e.  there should exist allowed points with mixings  $U_{e I}^2$ and $U_{\mu I}^2$ all the way up to the direct search bounds (black line in Figs.~\ref{ClNH-1:fig}-\ref{EsIH-2:fig}) in this mass region. They are simply not found by our scan because they correspond to small volumes in parameter space which are squeezed between lower bounds from BBN and upper bounds form experiments. 
\item For masses $M_I<m_D$, the direct search bounds on $U_{e I}^2$ and $U_{\mu I}^2$ are very strong. Combining them with the indirect bounds allows to impose constraints on $U_{\tau I}^2$ that are much stronger than the direct search bounds on $U_{\tau I}^2$. This also leads to much stronger bounds on $U_I^2$. 
\item  The direct search bounds 
from experiments where $N_I$ are produced in D meson decays fall off rapidly once $M_I$ approaches $m_D$ because the phase space for this decay closes. 
However, bounds from the CHARM experiment \cite{Bergsma:1985is} extend to larger values of $M_I$ slightly above $m_D$.
This is because
in combination with the indirect bounds, the constraint on the quantity (\ref{ArtemReinterpret}) in this region leads to stronger limits on the individual $U_{\alpha I}^2$ than the direct search bounds.
\item In the region $m_D< M_I < m_B$, the largest values of $U_{e I}^2$ found in the main scan lie below the direct search bounds, and there is a huge scatter amongst them.
A similar situation is realised in the mass range between $m_K$ and 1.5 GeV. 
A smaller scatter can also be observed in $U_{\mu I}^2$ and $U_{\tau I}^2$.
The scatter is a result of the fact that the $0\nu\beta\beta$-constraint allows for large $U_{e I}^2$ only in very narrow funnels in parameter space where $B-L$ is conserved, see Sec.~\ref{LNVundLFVsec}. It reflects that fact that our scanning method has difficulties finding these funnels. 
From Figs.~\ref{ClNH-1:fig}, \ref{CsNH-1:fig}, \ref{ClIH-1:fig} and \ref{CsIH-1:fig} alone it is not clear whether
values of $U_{e I}^2$ all the way up the direct search bound can be made consistent with all indirect constraints for specific parameter choices that our scan failed to find. 
To answer this question, we perform a separate scan that is focused on the $B-L$-conserving region (scan II).
We cannot identify any strict upper bound that is stronger than the direct search bounds from combining all data. Allowed values of $U_{e I}^2$ up to the direct search bound can be found in all almost mass bins below $m_B$ (except in the small region discussed in the previous point). However, avoiding the constraints on $m_{ee}$ requires a mass degeneracy between two of the $N_I$, and experiments might not be able to resolve the signals from the two mass-degenerate heavy neutrinos. In this case the direct search bound should be applied to $U_{\alpha I}^2+U_{\alpha J}^2\simeq 2U_{\alpha I}^2$. That is, in combination with the constraint on $m_{ee}$, the direct search bound in the regime effectively 
becomes by a factor 2 stronger in the seesaw model. This is illustrated in Fig.~\ref{B-L_scan}.
 \begin{figure}
\begin{center}
\includegraphics[width=12cm]{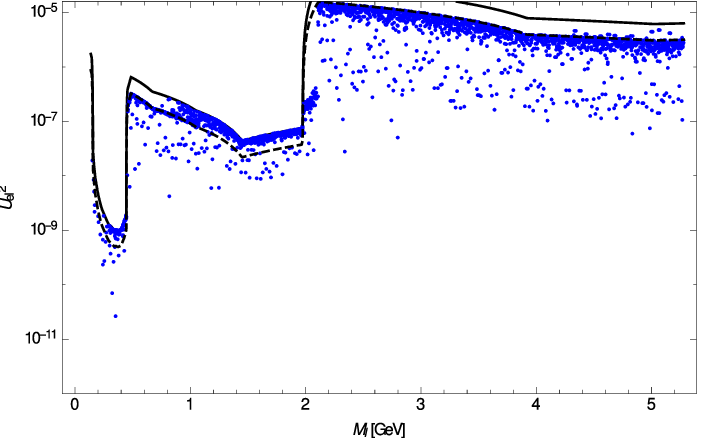}
\caption{\label{B-L_scan}
The blue dots represent the largest values for $U_{e I}^2$ found in a scan that is focused on the $B-L$ conserving region with $m_{\rm lightest}=0.23\,{\rm eV}$ and inverted hierarchy (scan II).
The solid black line is the direct search bound as published by the experimental collaborations. 
Values of $U_{e I}^2$ all the way up to the direct search bound can be realised in the entire mass range. However, for large $U_{e I}^2$ 
the constraint on $m_{ee}$ can only be avoided for comparably small values of the $B-L$-violating parameters in (\ref{BmLcons}). This implies that the two heavy neutrinos $N_I$ and $N_J$ with the largest $U_{e I}^2$ must have quasi-degenerate masses and equal mixings $U_{\alpha I}^2=U_{\alpha J}^2$, see (\ref{equalcouplings}).
If the mass splitting between $M_I$ and $M_J$ is smaller than the resolution of an experiment, the direct search bound applies to $U_{e I}^2+U_{e J}^2\simeq 2U_{e I}^2$.
That is, for mixings $U_{e I}^2$ above some critical value, the experimental bound is effectively twice as strong as the solid black line suggests, as indicated by the dashed black line.
Here we have roughly estimated the experimental resolution as 10 MeV. 
The plot shows that, if this is realistic, the critical value of $U_{e I}^2$ above which the dashed black line represents the appropriate direct search bound is $U_{e I}^2\sim 5\times 10^{-7}$. In a more precise future analysis, the actual mass resolutions of each individual experiment in each individual mass bin should be used to identify the correct upper bound precisely.
}
\end{center}
\end{figure} 
\item Near $M_I\sim m_W$, the direct search bounds from DELPHI and L3 become weaker than indirect bounds from LFV observables, which then provide the strongest constraints. Apart from lepton universality in meson decays, also the $\mu\rightarrow e \gamma$ decays and electroweak precision data are relevant in this mass region.
\end{itemize}

\subsection{Which indirect bounds matter?
}\label{LNVundLFVsec}
\paragraph{Lepton number and flavour violation} - 
The list in Sec.~\ref{DiscussionSmallLargeMixing} 
shows that there are many mass regions in which the combination of all constraints does not seem to give stronger upper bounds on $U_{\alpha I}^2$ than the direct search bounds alone.
This at first sight seems to contradict some earlier findings, which e.g. suggested the $0\nu\beta\beta$ decay alone could impose much stronger constraints on $U_{e I}^2$ than the direct searches at colliders \cite{Atre:2009rg,Blennow:2010th,Deppisch:2015qwa}.
The reason is that heavy neutrinos with large $U_{eI}^2$ and a non-degenerate mass spectrum can give a sizeable contribution to $m_{ee}$ \cite{LopezPavon:2012zg}, even in the region where leptogenesis is feasible \cite{Drewes:2016lqo,Hernandez:2016kel,Asaka:2016zib}.
The crucial point is that $0\nu\beta\beta$ decay is a LNV observable. 
In the seesaw limit $M_M\gg M_D$, LNV and LFV are closely related, as they originate at leading order from the same dimension 5 operator in (\ref{Leff}).  
For ``generic'' parameter choices or for $n=1$, all entries in the matrix $F M_M^{-1}F^T$ are of comparable size.
Then LNV observables usually allow to impose stronger bounds on $U_{\alpha I}^2$. This is the reason why the authors of \cite{Atre:2009rg,Blennow:2010th,Deppisch:2015qwa}, who studied the scenario with $n=1$, could derive very strong bounds on $U_{e I}^2$ from $0\nu\beta\beta$ decay searches.

However, in the model (\ref{L}), the scales of LNV and LFV are not identical for all parameter choices. For instance, total lepton number is conserved and neutrinos are Dirac particles if $M_M=0$, but individual lepton flavour numbers are still violated by neutrino oscillations.
Also in the seesaw limit $M_M\gg M_D$ under consideration here, all scenarios with $n>1$ contain parameter regions in which the total lepton number $L$ is approximately conserved (more precisely: $B-L$), see Sec.~\ref{MixingScenarios}. The observable LNV in these scenarios is suppressed by small $L$-breaking parameters in (\ref{BmLcons}). This is e.g. discussed in detail in \cite{LopezPavon:2012zg,Fernandez-Martinez:2015hxa}.
In the parametrisation (\ref{CasasIbarra2}) these can  be related to the splitting between two eigenvalues $|M_I-M_J|/(M_I+M_J)$ and  $e^{-|{\rm Im}\omega_{ij}|}$, i.e. the imaginary parts of the ``Euler angles''.
Hence, $0\nu\beta\beta$-decays can be strongly suppressed even if the $U_{e I}^2$ are comparably large \cite{Lopez-Pavon:2015cga,Helo:2015fba}, see \cite{Bernabeu:1987gr,Valle:1987gv,GonzalezGarcia:1988rw} for earlier discussions.  
Then LFV observables, which are generally not suppressed, come into play.
For $M_I\ll m_W$, the only relevant constraints from LFV observables come from lepton universality in meson and $\tau$-decays and CKM unitarity. 
The degree of saturation for these observables tends to be $S_i\sim 10^{-1}$ over wide ranges of $M_I$, which is mainly driven by the deviations of the observed rates from the SM prediction . Improving the bounds on lepton universality and CKM unitarity may therefore be an interesting target in order to further constrain the seesaw mechanism in the future.
For $M_I\sim m_W$, the constraint from $\mu\rightarrow e\gamma$ decays and electroweak precision data are also significant.

\begin{figure}
\center
\includegraphics[width=10cm]{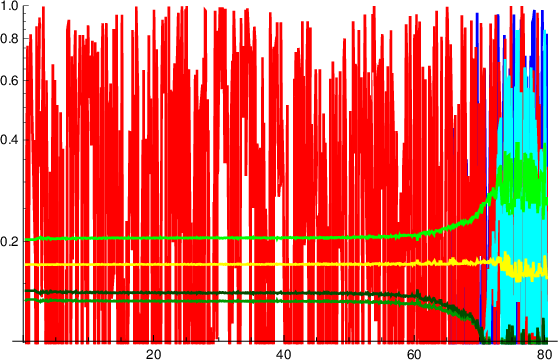}
\caption{A zoom into the lower left panel of Fig.~\ref{EsIH-1:fig}. The degrees of saturation for lepton universality (green lines) and CKM unitarity (yellow line) for the parameter choices that maximise $U_I^2$ are more or less constant in the mass range we consider. The value of $m_{ee}$ fluctuates, depending on whether approximate $B-L$-conservation is realised for the parameter choice found in the scan that maximises $U_I^2$. The situation is similar for the other benchmark scenarios.\label{ratiozoom}
}
\end{figure}

\paragraph{Most saturated bounds} - In our main scan, we randomised the parameters in (\ref{CasasIbarra2}), using flat probability distributions in $M_I$ and $\omega_{ij}$. With this measure, the vast majority of parameter choices do not correspond to approximate $B-L$ conservation, and the $0\nu\beta\beta$ bound tends to be saturated ($S_{0\nu\beta\beta}\gtrsim 1$). 
The approximately $B-L$-conserving regions in parameter space are very small and only consists of ``narrow funnels''. 
Near these funnels, the LNV observables are very sensitive to changes in the LNV parameters in (\ref{BmLcons}), which can be related to $e^{-|{\rm Im}\omega_{ij}|}$ and $|M_I-M_J|/(M_I+M_J)$.  
Whenever the scan happens to generate an approximately $B-L$-conserving parameter choice in a given $M_I$-bin, LNV observables can be avoided and LFV observables are the most saturated ones. 
For most values of $M_I$ these are lepton universality bounds, and a green dot appears in that mass bin in the upper panel of the corresponding figure. In all other cases, the $0\nu\beta\beta$ bound is the most saturated, and the dot indicating the largest active-sterile mixing in the upper panel is red.
The strong dependence of the LNV observables on ${\rm Im}\omega_{ij}$ and  $|M_I-M_J|/(M_I+M_J)$ near the parameter regions where $B-L$ is conserved allows to explain the large scattering in the values of $m_{ee}$  and the random distribution of green and red points, which are visible in the upper panels of figures \ref{ClNH-1:fig}-\ref{EsIH-2:fig}.
This is further illustrated in Fig.~\ref{ratiozoom}.

\paragraph{Radiative corrections to neutrino masses} -
As pointed out in section~\ref{MixingScenarios}, values of $U_{\alpha I}^2$ that significantly deviate from the ``naive seesaw'' expectation $U_{\alpha I}^2\sim (m_{\rm atm}^2+m_{\rm lightest}^2)^{1/2}/M_I$ can lead to large radiative corrections to the light neutrino masses $m_i$.
By using the radiatively corrected Casas-Ibarra parametrisation (\ref{CasasIbarra2}) in our main scan, we are guaranteed that the eigenvalues of $m_\nu=m_\nu^{\rm tree}+\delta m_\nu^{\rm 1loop}$ remain consistent with neutrino oscillation data even if one-loop corrections to $m_\nu$ are not small.
Consistency with the data can be realised in two ways. In the $B-L$-conserving limit, the symmetry guarantees that $m_\nu^{\rm tree}$ and $\delta m_\nu^{\rm 1loop}$ are small individually. In this case neutrino masses remain ``naturally'' small, and neutrino oscillation data does not impose an upper limit on $U_{\alpha I}^2$. 
Without a protecting symmetry, large $U_{\alpha I}^2$ can only be made consistent with small $m_i$ if there are accidental cancellations in $m_\nu$ that keep the $m_i$ small even if $m_\nu^{\rm tree}$ and $\delta m_\nu^{\rm 1loop}$ individually have larger eigenvalues. Such cancellations are generally thought to be fine tuned.
figures~\ref{ClNH-1:fig}-\ref{EsIH-2:fig} do not allow to distinguish these cases; by using (\ref{CasasIbarra2}), we do not prohibit or penalise accidental cancellations, and it is not possible to see how ``tuned'' the parameter choices that lead to comparably large mixings are.
Fig.~\ref{B-L_scan} shows that, for $M_I$ of a few GeV, values $U_{e I}^2> 5\times 10^{-7}$ can only be made consistent with experiment if there is a mass degeneracy $|M_I-M_J|/(M_I+M_J)<10^{-2}$. 
One could therefore argue that $U_{e I}^2\sim 10^{-7}$ is a reasonable estimate of the maximal active-sterile mixing angle that can be realised without tuning or a protecting symmetry. 


\subsection{Dependence on the neutrino mass spectrum}\label{subsec:massspectrum}

\paragraph{Choice of heavy neutrino mass range} - 
The choice of $M_{\rm max}$ appears to have no significant effect on the upper bounds on $U_{\alpha I}^2$. 
This can easily be understood qualitatively. The direct search bounds using particle kinematics are independent of the choice of benchmark scenario.
The indirect constraints on the mass and mixing of an individual $N_I$ from lepton number and flavour violation are affected by the properties of another $N_{J \neq I}$ most strongly if there are cancellations in (\ref{BranchingRatio}), (\ref{mee}) etc. This can most easily be realised if two of the $N_I$ have quasi-degenerate masses. 
If the mass splittings are large, the entries of $\Theta$ tend to correspond to the ``naive seesaw expectation''. 
Hence, the constraints on $U_{\alpha I}^2$ from most experiments (except neutrino oscillation data) are relatively independent of the properties of other RH neutrinos with masses $M_J\gg M_I$, and therefore independent of $M_{\rm max}$.

\paragraph{Absolute light neutrino mass scale} -
The absolute neutrino mass scale $m_{\rm lightest}$ has a significant effect on the combined constraints. 
\begin{itemize}
\item As already pointed out in Sec.~\ref{DiscussionSmallLargeMixing}, this parameter governs the lower bound on $U_I^2$ from neutrino oscillation data. If $m_{\rm lightest}$ is near the current upper cosmological limit, direct search experiments in the foreseeable future have a chance to reach this lower bound.
\item The absolute neutrino mass scale also affects the  parameter space for large  mixings. 
In the mass interval $M_I<m_D$ the dependence on the absolute neutrino mass scale is most visible in the bounds on $U_{\tau I}^2$ because this element is much less constrained by direct searches. The way how indirect constraints translate direct search bounds on $U_{e I}^2$ and $U_{\mu I}^2$ into an upper bound on $U_{\tau I}^2$ depends on $m_{\rm lightest}$. 
If the lightest neutrino is massless, then it is clear from (\ref{CasasIbarra2}) that $F$ has a vanishing eigenvalue, and therefore also $\Theta^\dagger\Theta$ has a vanishing eigenvalue.
Then there exists a direction in sterile flavour space (a linear combination of the $N_I$) that does not couple to active flavours, and neutrino masses $m_i$ are only generated by two linearly independent sterile flavour eigenstates. 
Similar to the $n=2$ scenario, this implies tight relations between $U_{\tau I}^2$ and $U_{e I}^2$, $U_{\mu I}^2$. 
These translate the direct search bounds on $U_{e I}^2$ and $U_{\mu I}^2$ into a bound on $U_{\tau I}^2$. It is dominated by the (weaker and hence more relevant) upper bounds on $U_{\mu I}^2$. 
If the lightest neutrino is not massless, then there is more freedom. However, the remaining indirect bounds prohibit arbitrarily large ${\rm Im}\omega_{ij}$, therefore the bounds on $U_{\mu I}^2$ still leave a visible imprint in the combined bounds on $U_{\tau I}^2$. The bounds on $U_I^2$ also depend on the absolute neutrino mass scale because weak bounds on $U_{\tau I}^2$ automatically imply weak bounds on $U_I^2$. 
\item By comparing Figs.~\ref{ElNH-1:fig} and \ref{EsNH-1:fig}, one can see that the choice of $m_{\rm lightest}$ also affects the allowed parameter space near $m_W$. The choice $m_{\rm lightest}=0$ leads to stronger constraints, probably for the same reasons as discussed in the previous point.
\end{itemize}

\paragraph{Light neutrino mass hierarchy} - 
The choice of active neutrino mass hierarchy does not appear to have a huge impact on the combined constraints on heavy neutrinos. For $m_{\rm lightest}=0$, the scan appears to find large $U_{e I}^2$ more efficiently for inverted hierarchy.
The difference between the hierarchies may become more relevant in the future, when the constraints on $m_{ee}$ and $m_{\rm lightest}$ become stronger. If only the light neutrino contribution to (\ref{mee}) is taken into account, it is well known that the allowed range of $m_{ee}$ for normal and inverted hierarchy differs if $m_{\rm lightest}<0.1$ eV.

\subsection{Implications for future experiments}
\paragraph{Neutrinoless double $\beta$ decay} - 
If only the light neutrino contribution\footnote{
It is important to distinguish the quantity $m_{ee}^\nu$ in (\ref{lightneutrinocontribution}) from the element $(m_\nu)_{ee}$ of the light neutrino mass matrix (\ref{activemass}).
} 
\begin{equation}
m_{ee}^\nu\equiv\sum_im_i (U_\nu)_{\alpha i}^2\label{lightneutrinocontribution}
\end{equation}
 to $m_{ee}$ in (\ref{mee}) is taken into account, it should not be possible to impose any constraints on neutrino properties from $0\nu\beta\beta$ decays: for both hierarchies, the range of values of $m_{ee}$ that are consistent with the cosmological limit on $m_{\rm lightest}<0.23$ eV \cite{Dell'Oro:2016dbc} lie below the current upper bound on $m_{ee}<0.2$~eV~\cite{Agostini:2013mzu}.
Hence, it seems surprising that this upper bounds imposes significant constraints on $U_{e I}^2$ even for $m_{\rm lightest}=0$. 
The reason is that both, radiative corrections and the contribution from $N_I$-exchange, can affect $m_{ee}$, as it has previously been pointed out in \cite{Lopez-Pavon:2015cga}.
To see this, one can use (\ref{SeesawConsistency}) to rewrite (\ref{mee}),

\begin{eqnarray}\label{meerewritten}
m_{ee}&=&\left|
m_{ee}^\nu
+f_A(\bar{M})\sum_IM_I\Theta_{e I}^2
+\sum_IM_I\Theta_{e I}^2[f_A(M_I)-f_A(\bar{M})]
\right|\\
&=&
\left|
[1-f_A(\bar{M})]m_{ee}^\nu
+(\delta m_\nu^{\rm 1loop})_{ee}f_A(\bar{M})
+\sum_IM_I\Theta_{e I}^2[f_A(M_I)-f_A(\bar{M})]
\right|.\nonumber
\end{eqnarray}
Here $\bar{M}$ is an arbitrary reference mass. 
The first term in the second line is always smaller than $m_{ee}^\nu$.
The second term is a radiative correction and usually neglected.
The third term can have either sign.
For $M_I\ll 100$ MeV, one can approximate $f_A(M_I)=1$. 
If this is true for all $M_I$, then one can immediately see that $m_{ee}=|(\delta m_\nu^{\rm 1loop})_{ee}|$. This is in principle interesting, but is disfavoured by cosmological data \cite{Hernandez:2014fha}. 
In the limit of superheavy $M_I$ one essentially recovers $m_{ee}=|m_{ee}^\nu|$ due to the suppression by the function $f_A$. 
However, for $M_I$ in the range we consider here, the second and third term can give a sizable positive contribution to $m_{ee}$.
This effect is strongest in the fine tuned scenarios where there are accidental cancellations in $m_\nu$ that keep neutrino masses small in spite of large $U_{\alpha I}^2$. 
This can be seen as follows. Since both terms are of order $\mathcal{O}[\Theta^2]$,\footnote{Recall that $\delta m_\nu^{\rm 1loop}$ is given by (\ref{RadiativeCorrectionExpression}).} they can only give a sizable contribution if there are some cancellations in $m_\nu$ that allow  values of individual elements $F_{\alpha I}$ to be much larger than the naive seesaw expectation $F_0$.
If these cancellations are protected by a symmetry ($B-L$ conservation), then the two heavy neutrinos with large mixings (let's say $N_2$ and $N_3$) must have degenerate masses ($|M_2-M_3|\simeq \mu_i\ll M_2,M_3\simeq M$), and the third one must have a very small mixing angle ($U_{1 \alpha}^2\ll U_{2 \alpha}^2, U_{3 \alpha}^2$), see section~\ref{MixingScenarios}. In this case all terms involving $N_1$ can be neglected because of the tiny $\Theta_{1 \alpha}\propto \epsilon_\alpha,\epsilon_\alpha'\ll 1$. 
Choosing $\bar{M}=(M_2+M_3)/2$, we can suppress the contributions from $N_2$ and $N_3$ to the third term due to the mass degeneracy. Finally, $\delta m_\nu^{\rm 1loop}$ is protected by the symmetry, meaning that the second term cannot be large.

On the other hand,
both, the second and third term on the right hand side of~(\ref{meerewritten}) can be unsuppressed if the cancellations in $m_\nu$ are accidental (or the $B-L$ breaking parameters are relatively large).
One would naively think that the suppression by $f_A$ implies that this is only possible for $M_I$ near 100 MeV. However, since $m_i\ll M_I$, it is easy to see that $M_If_A(M_I)\gg m_i$, and the contribution from sterile neutrino exchange can dominate when $f_A(M_I)M_I/m_i\gg\Theta_{e I}^2$. With a mass degeneracy of $10^{-3}$ and all ${\rm Im}\omega_{ij}\sim 3-5$, contributions from individual $N_I$ are much bigger than those from light neutrino exchange ($|f_A(M_I)\sum_IM_I\Theta_{e I}^2|/|m_{ee}^\nu|>10^4$) if the total sum $m_{ee}$ is comparable to the experimental bound.
This allows for $m_{ee}\gg |m_{ee}^\nu|$ and observable $0\nu\beta\beta$ decay in the entire mass range $m_\pi<M_I<m_W$ we consider.
In these scenarios, $U_{e I}^2$ is constrained from above by the bound on $m_{ee}$, and future experiments \cite{Agostini:2015dna} could see $0\nu\beta\beta$ decay in a region that appears forbidden if one uses the standard expression $|m_{ee}^\nu|$ instead of the full one~(\ref{meerewritten}) .
In contrast and as stated above, in the symmetry protected case,  searches for $0\nu\beta\beta$ decay are not able to constrain $U_{\alpha I}^2$.

\paragraph{Lepton universality} - For the parameter choices that maximise the $U_{\alpha I}^2$, the saturation  (\ref{SiDef1}) of the constraint on the quantities $\Delta r^\uppi_{\mu e}$ is around $\sim 0.2$ in the entire mass range we consider. Similarly, the bounds on  $\Delta r^K_{\mu e}$ and $\Delta r^\tau_{\mu e}$ tend to be saturated at a level $\sim 0.1$. This saturation is mainly driven by a discrepancy between the observed values and the SM predictions, which at least partly could be related to the existence of heavy neutrinos.
There are several experiments that can improve these constraints. For pions, these include PIENU at TRIUMF \cite{Ito:2015kvf}\footnote{While this work was being in progress the error bar on (\ref{ruppiexp}) has been reduced by almost a factor 2 \cite{Aguilar-Arevalo:2015cdf}.} and PEN at PSI \cite{Pocanic:2015rrw}.
For kaon decays, there are
E36 at JPARK \cite{Shimizu:2015xzf} and 
NA62 at CERN \cite{Parkinson:2014iza}.
The constraints may also be improved by the ORKA experiment at  FNAL \cite{Worcester:2013aje}. 
While none of these experiments is likely to rule out the SM for any of the parameter choices in our scan, deviations from the SM at a tentative level of significance are possible \cite{Asaka:2014kia}. 
Future searches for lepton universality violations therefore remain to be a promising approach to test the low scale seesaw.
This in particular applies to the approximately $B-L$ conserving limit, in which the (usually stronger) constraints from $0\nu\beta\beta$ decays can be avoided. 

\paragraph{Collider searches} - 
Most searches at colliders are focused on LNV processes, such as $pp\rightarrow W^*\rightarrow N\ell^\pm\rightarrow \ell^\pm\ell^\pm j j$ (for $M$ above the electroweak scale) or $B^\pm\rightarrow \mu^\pm N\rightarrow \mu^\pm\mu^\pm\pi^\mp$ (for $M\sim$ GeV).
The LNV is considered as ``smoking gun'' signals for heavy Majorana neutrinos because the only source of LNV in the SM, the sphaleron process, is strongly suppressed at temperatures below the electroweak scale. 
Hence, there is no SM background other than detector effects for LNV signals.\footnote{Let us recall that, in our conventions, the light neutrinos are massless in the SM, and that it is not clear at this stage whether light neutrino masses violate lepton number at all.} 
However, our results show that consistency of experimentally accessible $U_{\alpha I}^2$ with neutrino oscillation data and non-observation of neutrinoless double $\beta$ decay requires the implementation of an approximate lepton number conservation in the seesaw mechanism, which would effectively suppress all LNV signals.
Though this point had been noticed before \cite{Pilaftsis:1991ug,Gluza:2002vs,Gorbunov:2007ak,Kersten:2007vk,Gavela:2009cd,Ibarra:2010xw,Mitra:2011qr,Dev:2013oxa}, it has not been fully appreciated by the experimental collaborations so far.
Experiments that probe the mass range we consider with mixing angles $U_{\alpha I}^2>10^{-7}$ should therefore look for  signatures that do not rely on LNV, including LFV processes, displaced vertices or kinematic observables. 
The perspectives for such searches have been studied for hadron colliders
\cite{Helo:2013esa,Izaguirre:2015pga,Gago:2015vma,Dib:2015oka,Dib:2016wge}
as well as lepton colliders
\cite{Abada:2014cca,Blondel:2014bra,Graverini:2015dka,Antusch:2016vyf}.\footnote{The mass range $M_I>m_W$ has been subject to 
a larger number of studies for hadron colliders
\cite{Kersten:2007vk,BhupalDev:2012zg,Gago:2015vma,Das:2012ze,Anamiati:2016uxp,Dev:2015kca,Das:2015toa,Chen:2011hc} 
and lepton colliders
\cite{Banerjee:2015gca,Asaka:2015oia,Das:2016hof,Das:2012ze,Banerjee:2015gca,Antusch:2015mia,Antusch:2015gjw,Abada:2015zea}.
}

\subsection{Summary}
\paragraph{Mixing $U_{e I}^2$ with electron flavour} - 
Very strong constraints on $U_{e I}^2$ have been derived from neutrinoless double $\beta$ decay in the past \cite{Atre:2009rg,Blennow:2010th}
under the assumption that there is only one sterile neutrino ($n=1$). For $n>1$ they are much weaker because LNV signatures can be suppressed if there is an approximately conserved quantum number $B-L$. Without such symmetry, values of $U_{e I}^2$ that greatly exceed the bounds shown in \cite{Atre:2009rg,Blennow:2010th} can only be realised at the cost of considerable parametric tunings.

Radiative corrections $\delta m_\nu^{\rm 1loop}$ to the light neutrino masses are suppressed if $M_I/m_W\ll 1$. 
This suppression becomes inefficient for $M_I\gtrsim 10$ GeV. 
As a result, the combination of $0\nu\beta\beta$ decay and neutrino oscillation data prohibit large $U_{e I}^2$ unless there is either an approximate $B-L$ conservation or (fine tuned) accidental simultaneous cancellations in both, $m_{ee}$ and $m_\nu$. 
\begin{itemize}
\item For $M_I<m_K$, the upper bounds essentially coincide with the direct search bounds at $U_{e I}^2\lesssim 10^{-9}-10^{-8}$.
\item For $m_K<M_I\lesssim 1.5$ GeV, the direct search constraints are weaker ($U_{e I}^2\lesssim 10^{-7}-10^{-6}$). The upper bounds essentially coincide with these direct search bounds if there is an approximate $B-L$-conserving symmetry. Without a protecting symmetry, it is very difficult to make $U_{e I}^2>10^{-7}$ consistent with the observed $m_{ee}$ and $m_\nu$, see Fig.~\ref{B-L_scan}.
\item For $1.5 {\rm GeV} \lesssim M_I\lesssim m_D$, the upper bounds essentially coincide with the direct search bounds at $U_{e I}^2\lesssim 10^{-7}$. 
\item In a small mass region near $M_I\gtrsim m_D$, the combination of the  CHARM-bound on the quantity (\ref{ArtemReinterpret}) imposes a stronger constraint on $U_{e I}^2$ than the direct search alone. 
 \item For $m_D<M_I<m_B$ the direct search constraints are weaker ($U_{e I}^2\lesssim 10^{-5}-10^{-4}$). 
The upper bounds essentially coincide with these direct search bounds if there is an approximate $B-L$-conserving symmetry. Without a protecting symmetry, it is very difficult to make mixings $U_{e I}^2>10^{-7}$ consistent with the observed $m_{ee}$ and $m_\nu$, see Fig.~\ref{B-L_scan}.
\item For $m_B<M_I\ll m_W$, the situation is similar to the region $m_D<M_I<m_B$, but appears to be easier to circumvent the bound on $m_{ee}$ without a protecting symmetry because of the suppression of the $N_I$-contribution to (\ref{mee}) by the function $f_A$. It is also easier to realise larger $U_{e I}^2$ if $m_{\rm lightest}$ is larger.
\item For $M_I\lesssim m_W$, the direct search bound becomes weak, and indirect constraints from $\mu\rightarrow e\gamma$ searches, lepton universality and CKM unitarity dominate. For $m_{\rm lightest}=0$ and normal $m_i$-hierarchy, we found almost no allowed parameter choices with $U_{e I}^2>5\times10^{-5}$. In the other scenarios, it seems to be easier to realise mixings as large as $U_{e I}^2\sim 10^{-4}$, but we still found only few.  
\end{itemize}

\paragraph{Mixing $U_{\mu I}^2$ with muon flavour} - 
In almost the entire mass region we considered, the upper bound on $U_{\mu I}^2$  can be obtained by simply superimposing the known direct search bounds.
Large values of $U_{\mu I}^2$ can be made consistent with small neutrino masses $m_i$  if these are protected by an approximate $B-L$ symmetry or if there are accidental cancellations in $m_\nu$.
 In comparison to  $U_{e I}^2$, less parametric tuning is needed to reach large values of $U_{\mu I}^2$ without a protecting symmetry because cancellations only need to occur in $m_\nu$. In comparison, to achieve large values of $U_{e I}^2$ without protecting symmetry, one needs simultaneous cancellations in $m_\nu$ as well as $m_{ee}$.
This is clearly visible in the better convergence of the scan towards the upper direct search bounds. As pointed out in section~\ref{subsec:massspectrum}, the choice of $m_{\rm lightest}$ affects the convergence: It is excellent if the lightest neutrino is massless. If it is massive, the convergence is slightly worse.
There are only two mass regions in where the above statement does not apply:
\begin{itemize}
\item In a small mass interval near $M_I \sim m_D$, an upper bound on $U_{\mu I}^2$  (similar to $U_{e I}^2$) can be derived from the combination of the the CHARM bounds on the quantity (\ref{ArtemReinterpret}),  which extend to slightly higher masses than the NuTeV bounds on $U_{\mu I}^2$ itself, and all other constraints.
\item For $M_I$ near the W mass, the direct search bound becomes weak and indirect constraints from lepton universality, CKM-unitarity and $\mu\rightarrow e \gamma$ searches dominate. They do not appear to impose a strict upper bound which is stronger than that from CMS, but make it difficult to realise mixings larger than $U_{\mu I}^2\sim 10^{-4}$.\footnote{
In \cite{Antusch:2015mia}, much stronger bounds on $U_{\mu I}^2$ have been derived. Our results do not appear to be in contrast with those because they were derived under the assumption of the non-zero $U_{e I}^2$ preferred by the global scan in \cite{Antusch:2014woa}, which is found at a low statistical significance that does not affect the conditions we imposed in our scan.} 
\end{itemize}

\paragraph{Mixing $U_{\tau I}^2$ with tau flavour} -
For $n=2$ the  bounds on $U_{e I}^2$ and $U_{\mu I}^2$ strongly constrain the comparably
weak direct bounds on the mixing $U_{\tau I}^2$. 
This appears to already rule out the mass region below $M_I\simeq 240$ MeV if one takes the BBN bounds literally \cite{Asaka:2014kia}. 
The reason is that the parameter space for $n=2$ is sufficiently small that neutrino oscillation data prohibits large hierarchies between the mixings of a given heavy neutrino with different active flavours, see section \ref{sec:bbn}.
This is not the case for $n=3$: The upper bound on $U_I^2$ lies above the lower bound from BBN even for $M_I<240$ MeV because $U_{\tau I}^2$ can be comparably large.
Hence, we cannot deduce a lower bound on $M_I$ within the mass intervals we studied for the general $n=3$ scenario. However, we can still obtain non-trivial constraints on $U_{\tau I}^2$.
\begin{itemize}
\item In the entire mass region $M_I<m_D$, combining all direct and indirect search constraints does allow to impose stronger upper bounds $U_{\tau I}^2$ than any individual measurement in all scenarios.
These are orders of magnitude stronger than the known direct constraints and mainly come from the combination of neutrino oscillation data and direct search bounds on $U_{e I}^2$ and $U_{\mu I}^2$. For $M_I<m_K$ they can be as strong as $U_{\tau I}^2<10^{-8}$, for $m_K<M_I<m_D$ they are weaker ($U_{\tau I}^2<10^{-7}-10^{-6}$).
\item For $M_I>m_B$ the behaviour is qualitatively the same as for $U_{\mu I}^2$.
\end{itemize}

\section{Conclusions}\label{sec:conclusions}
The properties of heavy sterile neutrinos in the minimal seesaw model can be constrained by combining various experimental and cosmological data sets, as illustrated in Fig.~\ref{SummaryPlot}.
In the present work, we focused on the case with three heavy neutrinos and Majorana masses $M_I$ between the pion and W boson masses.
Our analysis extends previous studies, which were mostly performed under the assumption that there are only one or two heavy neutrinos, and includes updated experimental data from neutrino oscillation experiments, direct searches at colliders and fixed target experiments, searches for rare lepton decays, electroweak precision data, neutrinoless double $\beta$ decay and lepton universality.
In spite of the fact that the combination of different observables imposes considerable constraints on the model parameters, we find that they do not impose much stronger bounds on the magnitude of the heavy neutrino mixings $U_{\alpha I}^2$. 
This is in stark contrast to the scenario with $n=2$, where the combination of known direct and indirect constraints leads to much stronger bounds on the magnitude of the $U_{\alpha I}^2$ than one may expect from considering them individually.

One reason is that scenarios with several heavy neutrinos contain parameter space regions in which lepton number is approximately conserved, which allows to circumvent all constraints from lepton number violating observables. This in particular applies to neutrinoless double $\beta$ decay and radiative corrections to light neutrino masses.
Because of this, the combined upper bound on the mixings $U_{e I}^2$ and $U_{\mu I}^2$ of heavy neutrinos with electron and muon flavour are dominated by the direct search constraints for most choices of $M_I$.
The combined direct and indirect constraints on $U_{\tau I}^2$ in the region $M_I<m_D$, on the other hand, are orders of magnitude stronger than the direct search bounds alone. 
Another reason for the weaker bounds in scenarios with three or more heavy neutrinos lies in the \emph{seesaw relation}, which constrains the mixing angles of heavy neutrinos with individual light flavours due to the requirement to explain light neutrino oscillation data. In scenarios with only two heavy neutrinos this relation does not only imply that the lightest active neutrino must be massless, it also allows to impose lower bounds on the mixings of sterile neutrinos with individual active flavours. Due to the larger parameter space, these lower bounds are much weaker with three heavy neutrinos and can be entirely avoided if the lightest neutrino is massless. A lower bound can, however, still be imposed on the sum $U_I^2=U_{e I}^2+U_{\mu I}^2+U_{\tau I}^2$ of the mixings with different active flavours.

Mixing angles larger than $U_{\alpha I}^2\sim 10^{-7}-10^{-6}$ generally require a parameter tuning to avoid large neutrino masses and observable neutrinoless double $\beta$ decay unless there is a protecting symmetry (approximate $B-L$-conservation). 
An interesting consequence is that 
the combined constraints from direct searches, neutrino oscillation data and
neutrinoless double $\beta$ decay can be a factor 2 stronger than the upper bound on $U_{\alpha I}^2$ from direct searches quoted in the literature: To avoid large neutrino masses, two of the heavy neutrinos $N_I$ and $N_J$ must have degenerate masses. If these cannot be resolved experimentally, then the direct search bound should be applied to the sum $U_{\alpha I}^2+U_{\alpha J}^2$. 
Since the allowed parameter space for mixings larger than than $U_{\alpha I}^2\sim 10^{-6}$ consists only of rather specific regions in which the violation of $B-L$ is suppressed, it can probably be considerably constrained (or even excluded) if one imposes additional requirements (besides explaining the observed neutrino masses). Examples for such additional requirements could be successful low scale leptogenesis or a theoretical prejudice on the flavour structure of the Yukawa coupling and Majorana mass matrices from model building.

There are several existing and upcoming collider experiments which are capable of probing significant fractions of the allowed parameter space.
The NA62 experiment may close the gap between the upper bound on $U_{e}^2$ and $U_{\mu}^2$ from experiment and the lower bound from BBN for $M_I$ below the kaon mass for scenario A, thus excluding part of the mass range we consider. 
The proposed SHiP experiment at CERN could probe most of the parameter space for $M_I$ below the D meson mass. A part of this region will also be accessible to a SHiP-like experiment using the DUNE beam. If the mass $m_{\rm lightest}$ of the lightest SM neutrino is near the cosmological upper bound, a realistic improvement of the current SHiP design could reach down to the lower bound on $U_I^2$ from neutrino oscillation data.
SHiP could also improve the sensitivity of past searches in the mass range between the D meson and B meson masses by 2-4 orders of magnitude.
Heavy neutrinos with larger masses can be searched for in gauge boson decays at LHC. A powerful lepton collider, such as the ILC or the proposed FCC-ee, or the CEPC would, however, be several orders of magnitude more sensitive than the LHC.
A discovery at any of these experiments would be of great importance, as it could  unveil the origin of neutrino masses. 

The vast majority of past searches has been focused on 
LNV processes, 
which are viewed ``smoking gun'' signals for heavy Majorana neutrinos. However, our results confirm the result of earlier studies that consistency of experimentally accessible $U_{\alpha I}^2$ with neutrino oscillation data and non-observation of neutrinoless double $\beta$ decay requires the implementation of an approximate lepton number conservation in the seesaw mechanism, which would effectively suppress all LNV signals.
Experiments that probe mixing angles $U_{\alpha I}^2>10^{-7}$ should therefore look for  signatures that do not rely on LNV, including LFV processes, displaced vertices or kinematic observables (e.g. ``peak searches''). 

Apart from direct searches in these collider or fixed target experiments, indirect probes that can pick up the traces of heavy neutrinos will considerably gain sensitivity in the future.
This includes future neutrinoless double $\beta$ decay searches, searches for violation of lepton universality in the decays of mesons and $\tau$ leptons, rare LFV decays and muon-to-electron conversion in nuclei.

In addition to neutrino oscillations, low scale leptogenesis is one key motivation to search for heavy neutrinos with masses below the electroweak scale. Previous studies that aimed to identify the parameter region where leptogenesis is possible  did not incorporate all bounds self-consistently. Direct search constraints were simply superimposed on the projection of the leptogenesis region on the mass-mixing plane after the completion of the parameter scan.
It would be interesting to perform a parameter scan that consistently incorporates all constraints we discuss here and the requirement to explain the observed BAU. This may reduce the viable leptogenesis parameter space in comparison to previous estimates.
Such a study would be crucial to evaluate the perspectives for future collider searches to unveil the origin of matter in the universe.

\section*{Acknowledgements}
We would like to thank Nicola Serra, Andrei Golutvin, Elena Graverini, Takehiko Asaka, Artem Ivashko, Robert Shrock and Artur Shaikhiev for useful discussions.
We are also grateful to Jacobo Lopez-Pavon, Pilar Hernandez and Enrique Fernandez-Martinez for pointing out the importance of the approximate $B-L$ conservation in the context of radiative corrections to the seesaw relation and neutrinoless double $\beta$ decay, which was missed in the first version of this paper.
This work was supported by the Gottfried Wilhelm Leibniz program of the Deutsche Forschungsgemeinschaft (DFG) and by the DFG cluster of excellence Origin and Structure of the Universe.

\bibliographystyle{apsrev}
\bibliography{all}
\end{document}